\documentclass[manuscript]{aastex}
\usepackage{color}

\shorttitle{Comparison of mm-wave and X--ray emission}
\shortauthors{Monje et al.}

\begin{document}


\title{Comparison of Millimeter-wave and X--ray Emission\\ in Seyfert Galaxies}

\author{R. R. Monje, A. Blain and T. Phillips}
\affil{
Division of Physics, Mathematics and Astronomy, California Institute of Technology, 1200 E. California Blvd, Pasadena, CA 91125-4700}

\begin{abstract}
We compare the emission at multiple wavelengths of an extended Seyfert galaxy sample, including both types of Seyfert nuclei. We use the Caltech Submillimeter Observatory (CSO) to observe the CO $J=2-1$ transition line in a sample of 45 Seyfert galaxies and detect 35. The galaxies are selected by their joint soft X--ray (0.1--2.4 keV) and far--infrared ($\lambda$ = 60--100 $\mu$m) emission from the ROSAT/IRAS sample. Since the CO line widths (W$_{CO}$) reflect the orbital motion in the gravitational potential of the host galaxy, we study how the kinematics are affected by the central massive black hole, using the X--ray luminosity. A significant correlation is found between the CO line width and hard (0.3--8 keV from Chandra and XMM--Newton) X--ray luminosity for both types of Seyfert nuclei. Assuming an Eddington accretion to estimate the black hole mass (M$_{BH}$) from the X--ray luminosity, the W$_{CO}$--L$_X$ relation establishes a direct connection between the kinematics of the molecular gas of the host galaxy and the nuclear activity, and corroborates the previous studies that say that the CO is a good surrogate for the bulge mass. We also find a tight correlation between the (soft and hard) X--ray and the CO luminosities for both Seyfert types. These results indicate a direct relation between the molecular gas (i.e. star formation activity) of the host galaxy and the nuclear activity. To establish a clear causal connection between molecular gas and the fueling of nuclear activity, high-resolution maps ($<$~100 pc) of the CO emission of our sample will be required and provided by the forthcoming ALMA observatory.

\end{abstract}


\section{Introduction}

Seyfert galaxies are classified into two types based solely upon their optical spectrum properties. Type 1 Seyferts (Sy1s) are those with very broad H I, He I, and He II emission lines. The forbidden lines [O III], [N II] and [SII], though narrower than the very broad permitted lines, are still broader than the emission lines in most starburst galaxies. Type 2 Seyferts (Sy2s) have permitted and forbidden lines with approximately the same full width half maximum (FWHM), similar to the FWHMs of the forbidden lines in Seyfert 1s, but do not present a broad line feature \citep{Ost00}. Evidence now exists that the main differences in the optical spectra are due to Seyfert nuclei being surrounded by a torus of dusty, obscuring gas a few parsecs from the center. In Sy1s the orientation of the torus axis is close to the line of sight allowing direct observations of the inner broader region associated with the accretion disk of the active galactic nucleus (AGN). In type 2 nuclei, the orientation of the torus shields the nucleus from direct view and only the more extended narrow line clouds are observed. Several studies analyzing the properties of Seyfert nuclei galaxies have reported a difference in the infrared properties between the Seyferts types, where type 2 nuclei present nearly an order of magnitude enhancement of their infrared emission from their disk with respect to those with type 1 nuclei \citep{ede87,mai95}. These results can be easily explained in terms of an elevated rate of star formation of massive stars in galaxies with type 2 nuclei. In addition, \cite{mai97} found through their $^{12}$CO (1--0) observations of a large sample of Seyfert galaxies, no significant difference in the total amount of molecular gas as a function of the Seyfert nuclear type. Therefore, they concluded that the total amount of molecular gas is not responsible for the enhanced star-forming activity in Seyfert 2 hosts. Contrary to those observations, \cite{hec89} and \cite{cur00}, using observation of the CO $J = 1-0$ line, have found that type 2 Seyferts do indeed have a higher molecular gas content than Seyfert 1s. \cite{cur00} found that for Seyferts with far-infrared (FIR) luminosity (L$_{FIR}$) $\approx 10^{10} L_{\odot}$, the CO/FIR luminosity ratio in type 2 is at least three times that in type 1 sources. They suggested that this molecular gas may be related in some indirect way to the nuclear material hypothesized to obscure the broad line region in type 2 Seyferts.
In this paper, we study the molecular gas content of Seyfert types by comparing the luminosities of a higher excitation line (CO $J=2-1$) to the far-infrared and X--ray luminosities of a sample of 35 active galaxies. The sources were selected by their soft X--ray (0.1--2.4 keV) and FIR emission from a ROSAT/IRAS sample generated originally by \cite{bol92} and modified by \cite{con98}, who used new VLA images of all the objects to eliminate uncertainties coming from the original ROSAT and IRAS positions.
Molecular gas is important not only to support the star formation activity in the different Seyfert types, but also to understand the powering mechanism of the nuclear activity in active galactic nuclei. The source of the accreted gas is still unclear. Most studies suggest that the massive black holes are fed by infalling interstellar gas, which is mostly in molecular form in the central kiloparsec of spiral galaxies. Single dish observations of CO emission can not give a direct answer to the molecular gas-nuclear activity relation, since any relation between the fueling (through CO luminosity) and the massive black hole data, such as X--ray luminosity (assuming most of the X--ray luminosity in Seyfert galaxies comes from the accretion disk around the massive black hole sitting in the center of the galaxy) must be searched for in the central few 100 pcs, of nearby AGN. However, \cite{yam94} and \cite{kaw05} reveal an interesting close relation between L$_{CO}$ and L$_X$ in Seyfert 1s at low and high redshift, respectively. 
In this paper, we study the properties of an extended Seyfert galaxy sample including both types of Seyfert galaxies, from the kinematics and multi-wavelength luminosity comparison point of view. In Section 2, observations and results in terms of line intensities, correlation coefficients and linear fits are presented. In Section 3, we summarize the results and discuss the interpretation of the CO line width (W$_{CO}$) - L$_{X}$, L$_{CO}$ - L$_{FIR}$ and L$_{CO}$ - L$_{X}$ relations.

\section{Observations and Results}

We have used the 10.4 m Caltech Submillimeter Observatory (CSO) telescope to observe the CO $J=2-1$ transition line towards the center of  45 Seyfert galaxies and have detected 35. Table \ref{tbl-1} lists the sources parameters obtained from NED\footnote{See http://nedwww.ipac.caltech.edu/} and Hyperleda\footnote{See http://leda.univ-lyon1.fr/}. The observations were mostly done during the winter and spring of 2009, with system noise temperatures of typically 300--400 K. Pointing was checked regularly on planets with typical pointing errors less than 5$\arcsec$ during one night. The CSO beam size for the $^{12}$CO $J=2-1$ line (with rest frequency\footnote{See http://physics.nist.gov/cgi-bin/micro/table5/start.pl} equal to 230.538 GHz) is approximately 30$\arcsec$ and the beam efficiency is 0.69. Peak intensities in T$^{*}_{A}$ scale are shown in Table \ref{tbl-2}. T$^{*}_{A}$ is converted to main beam brightness scale, T$_{mb}$, by dividing the T$^{*}_{A}$ by the beam efficiency. We have applied the necessary corrections, cold spillover and beam coupling, to those sources with size similar to, or smaller than the beam. This gives T$^{*}_{A}$/$\beta$, where $\beta$ is the spillover and beam coupling coefficient. %$\beta$ is 1 for sources bigger than the beam.
Source sizes have been determined by their optical extent. Since most of our sources are bigger than the beam, we assume extended sources, i.e., $\theta_{beam} \leq \theta_{source}$. Therefore, no beam dilution correction has been applied in our temperature calculations. We used a 1 GHz band width FFTS back-end with 8192 channels and a 1 GHz band-width acousto-optical spectrometer (AOS) with 2048 channels.\\
The molecular line emissions observed for our sample are shown in Figure \ref{spectra}. A first-order polynomial is used to correct the baseline. A 3$\sigma$ detection has been assumed to calculate the upper limit for the non-detected sources.

\subsection{CO line width and the galaxy inclination} \label{bozomath}

Figure~\ref{Wco-sini} shows that the CO line widths correlate with the galaxy inclination. This result shows that the sources have been correctly identified by Condon et al (1998), in spite of the large pointing errors of IRAS and ROSAT. The line widths were measured at 20\% of the peak intensity since it gives a more robust measurement of the maximum rotation velocity of the disk as shown by \cite{ho07}. The galactic disk inclination angles of these Seyferts were estimated from the Hyperleda database. The CO line width has been also corrected by the blue luminosity (W$_{CO}$/L$^{1/4}$) to avoid selection effects, since brighter sources rotate faster on average \citep[][which establish that the luminosity, L, is proportional to the 4$^{th}$ power of the rotational velocity, V, i.e. L $\approx$ V$^{4}$]{tul77}. We notice that when applying the luminosity correction the correlation improves slightly and reduces the spread. \\
Figure~\ref{Wco-sini} shows that face-on galaxies with smaller inclinations tend to have narrower velocities than edge-on with larger inclinations, following $\Delta V_{CO} \approx 2V_{MAX}\cdot$ sin(i). This is an unsurprising result since the CSO beam samples CO emission on scales of few kpcs (see Table~2). As molecular disks are roughly coplanar with stellar disks on these scales, a rough correlation is expected. 

The inclination--corrected CO line width--redshift correlation has been checked for selection effects of more distant objects being more luminous. In our sample the inclination--corrected CO line width values are distributed around an average value of 500 km~s$^{-1}$ for any redshift, which corresponds to a typical rotational velocity value for spiral galaxies.

\subsection{CO line width and the X--ray luminosity}

We run a simple statistical approach to our investigation of the CO kinematics. For all the Seyferts with detected CO emission, we examine possible correlation between the inclination--corrected CO line width and the soft (ROSAT, 0.1--2.4 keV) and hard (Chandra or XMM--Newton, 0.3--8~keV) X--ray luminosity: the data is listed in Table~\ref{tbl-2}. The hard X--ray luminosities in Table~\ref{tbl-2}, are corrected for absorption using the X--ray fitting software from HEASARC\footnote{See http://heasarc.gsfc.nasa.gov}. For those type 1 (un~obscured) AGNs with no available hard X--ray data, we extrapolate the ROSAT (0.5--2 keV) into Chandra (0.3--8 keV) X--ray fluxes, by using the tool PIMMS \footnote{See http://heasarc.gsfc.nasa.gov/Tools/w3pimms.html}, applying the net count rate, a power law spectrum with index $\Gamma$ = 1.6 and an estimate of the Galactic neutral hydrogen column density from \cite{kal05}. The correlation between X--ray power and CO kinematics has not been examined before, to the best of the authors' knowledge. Figure~\ref{Wco-lx} shows the soft X--ray (0.1--2.4 keV) (left) and hard (0.3--8 keV)(right) luminosity, L$_X$, versus the corrected CO line width, W$_{CO}$, in a log--log plot. For the least-squares fit and slope calculations the bisector of the ordinary least-squares (OLS) regression described in \cite{iso90}, has been used. Table~\ref{tbl-3} lists the results of the L$_{X}$--W$_{CO}$ correlation together with the $\chi^{2}$ value and probabilities. The results from Figure~\ref{Wco-lx} show a significant correlation between the CO line width and both soft and hard X--ray luminosity for type 1 Seyfert galaxies, with a correlation coefficient of 0.67 and 0.62 and a null probability of 0.12 and 0.32, respectively. In most type 2 Seyferts the observed 0.1--2.4 keV flux is a mixture of emission from the host galaxy, reflected X--ray emission from the AGNs and, if the absorbing column density is low enough, some transmitted intrinsic flux. The soft X--ray photons, originating within a radius close to the nucleus ($\leq$1 pc), are absorbed by the dusty torus and only the fraction that is scattered toward our line of sight escapes from above or/and below the torus and can be detected, leading to the reduction in observed X--ray flux from Seyfert 2s. In consequence, the soft X--ray luminosity is fainter and does not present a significant correlation with the CO line width. 

The observed CO line width reflects the dynamics of the molecular gas in the inner parts of the galaxy. Therefore, we use the CO line widths to estimate the galaxy dynamical masses in this region. For a rotating disk of radius R, the dynamical mass enclosed within R can be determined by the following equation \citep{sol97},
\begin{equation}
M_{dyn}\approx \frac{R\cdot \Delta V^{2}}{2 \cdot sin^{2}(i)\cdot G} ,
\end{equation}
Here we assume that the gas emission comes from a rotating disk of outer radius \textit{R} (in units of kiloparsecs) observed at an inclination angle \textit{i}. We use a radius, R, equal to the telescope beam FWHM at each source, see Table 2.

At the same time, X--ray luminosity in Seyfert galaxies is related to the black hole mass (BHM) \citep{gra01}. Given the bolometric luminosity for an AGN, we can estimate the minimum mass of the black hole by using the \textit{Eddington limit} relation. In some cases the true bolometric luminosity cannot be calculated easily, as in the case of Seyfert 2s where the optical and ultraviolet (UV) radiation is highly obscured by the dusty torus. However, we can estimate the black hole mass from the X--ray luminosity. First, by assuming that an important fraction of the bolometric luminosity is emitted in the X--ray range by the central source. Second, the AGN luminosity, \textit{L}, can be estimated from the luminosity in any given band \textit{b}, $L_{b}$, by applying a suitable bolometric correction $f_{bol,b} = L/L_{b}$. We will use the bolometric corrections calculated by \cite{mar04}. \cite{mar04} use in their calculations the intrinsic luminosity which is the total luminosity directly produced by the accretion process, i.e. the sum of the optical-ultraviolet and X--ray luminosities radiated by the accretion disc and hot corona, respectively. We also assume that the X--ray bolometric correction is the same in type 1 and type 2 Seyfert galaxies. As expected from Figure \ref{Wco-lx}, the dynamical and back hole masses present a significant correlation shown in Figure \ref{masses}. 

\subsection{Relation between the CO luminosity with FIR and X--ray luminosities}

Figure \ref{Lco-Lfir} shows the strong correlation between the far-infrared luminosity$\footnote{The FIR luminosity, in the wavelength range $40\mu m < \lambda < 120 \mu m$ in $\left[ W \cdot m^{-2}\right]$, is estimated using the relation \citep{hel88},
$L_{FIR}=1.26\times 10^{-14}(2.58\cdot S_{60\mu m}+S_{100\mu m})$. }$, L$_{FIR}$, and the CO line luminosity, L$_{CO}$, of the Seyfert 1s (circles) and Seyfert 2s (triangles) from our sample, in a log--log plot, to confirm the relation observed between far-infrared and CO line luminosity \citep{ric84,you86,hec89,san91,rig96}. The strong correlation is interpreted as the result of the link between the amount of molecular gas and the rate of star formation. The FIR emission for both types of galaxies is very similar with a mean ratio of L$_{FIR}$(Sy1)/L$_{FIR}$(Sy2) $\approx$ 0.88. This result suggests two different possibilities. First, that the FIR emission comes from dust re-radiation of starlight from regions outside the torus, where both classes of Seyferts have similar properties. Second, the FIR in Seyfert galaxies can be powered by the AGN, i.e. non-thermal radiation coming from the nucleus torus, but in that case, the torus should have a similar covering fraction and optical depth to obtain the same emission from both Seyfert 2s as in 1s.

Studying the relation between the CO and FIR luminosities can give us a better understanding of the origin of FIR emission in Seyfert galaxies, and the relationship between the AGN activity, the star formation and the interstellar medium (ISM), via the molecular gas content. Taking the average value of $L_{CO}/L_{FIR}$ for both Seyfert types we obtain the following result,
\begin{equation}
 \frac{L_{CO}}{L_{FIR}}(Sy1)\approx 1.33 \pm 0.13 \cdot \frac{L_{CO}}{L_{FIR}}(Sy2) \approx 3.21 \times 10^{-8} K \cdot km \cdot s^{-1}\cdot kpc^{2} \cdot  L^{-1}_{\odot},
\end{equation}
There is not a significant difference in the total molecular gas content between the two Seyfert types. This result is consistent with the one obtained by \cite{mai97}, establishing that the total amount of molecular gas is not responsible for the enhanced star-formation activity observed in Seyfert 2 hosts \citep{gon93}. 

Another interesting observational result from this survey, is the strong correlation between the X--ray and the CO luminosity, see Figure \ref{Lco_Lx}, with a null probability of $1.83 \times 10^{-5}$ and a correlation coefficient of 0.79. A least-squares fit yields to a fitted relation of \textit{Log(L$_{CO}$)=1$\cdot$Log(L$_{X}$)-33.43} when considering both Seyfert types of our sample. To rule out the possibility of a correlation driven by the luminosity distance (D$^2$), we study the correlation between L$_{CO}$/L$_{FIR}$ and L$_X$/L$_{FIR}$, and between L$_{CO}$/L$_X$ and L$_X$, in both cases we will get rid of the distance dependence if any, and find a significant correlation in both cases.% These results indicate that the LCO-LX correlation is real and we decide to leave teh plots in Figure 6 in more conventional units.}

\section{Summary and Discussion}

We have analyzed the gas properties towards the galaxy nuclei of 18 Seyfert 1 and 17 Seyfert 2 galaxies, taken from the ROSAT/IRAS sample generated by \cite{con98}. The CO line is a molecular gas tracer and provides information about the kinematics of the inner galactic disks through the line width. We investigate the CO kinematics and find that the CO line width moderately correlates with the host galaxy inclination for both types of Seyferts (Figure 2). This may initially suggest that the CO is coplanar with the galactic disk. However, \cite{hec89} showed that the CO emission may also be correlated with the torus. Studies on the relative orientation of a radio jet and the host galactic disk in Seyfert Galaxies, e.g. \cite{sch02}; \cite{nag99}, have shown that the torus is not coplanar with the galactic disk. The combination of these studies suggests that a small, though not negligible, fraction of the CO emission is coming from the nuclear regions and not solely from the galactic disk. We examine the relation between the CO kinematics and the X--ray power and find a significant correlation between the CO line widths and the hard X--ray luminosity both AGNs types. Based on the assumption of Eddington accretion, we analyze the correlation between the dynamical mass calculated from the CO line width and the black hole mass calculated from the L$_{X}$ for Seyfert galaxies. The measured correlation corroborate the recent studies that suggest that the inclination--corrected CO line width is a surrogate for the bulge velocity dispersion of the host galaxies \citep{shi06,wu07}, and the black hole--bulge relation obtained with this assumption \citep{fer00}. As shown in Figure 4, the Eddington limit approximation in our calculations, underestimate the M$_{BH}$ by at least a factor of 10 in comparison to the M$_{BH}$-M$_{bulge}$ relation for normal galaxies where $\left\langle M_{BH}/M_{bulge}\right\rangle$ $\sim$ 0.001 \citep{mar03}.  In order to get a better understanding of the M$_{BH}$-W$_{CO}$ relation for Seyfert galaxies, we have to rely in more accurate measurements of the M$_{BH}$ and higher resolution images of CO luminosity, in future studies.

\cite{yam94} studied the X--ray (0.5--4.5 keV) and CO luminosity relation in a sample of 13 Seyfert 1s and 5 quasars and found a significant correlation. They suggest a scenario where the star formation activities directly control the mass accretion rate at the central black hole, concluding that the more powerful monsters live in the more actively star-forming host galaxies. We analyze the L$_{CO}$--L$_{X}$ relation with a more extended sample using both Seyfert types and the X--ray bands 0.1--2.4 keV and 0.3--8 keV and find similar correlations between the X--ray and CO luminosities for both Seyfert types. Two possible scenarios can be considered in the nuclear activity--total molecular gas content relation. First, circumnuclear star-formation activity could drive gas into the innermost galactic regions to feed a black hole, thus creating a star formation/active galactic nuclei connection \citep{nor88,hec89}. The CO emission can be related to the infalling interstellar gas feeding the accretion disk of the massive black holes, which is mostly in molecular form located in the central kiloparsecs. Second, the more powerful X--rays from the central black hole could provide additional heat to the molecular cloud, enhancing the excitation of CO molecules in the inner region (few pc) enhancing the total CO luminosity. However, since the CO emission observed in our sample with the CSO, typically comes from inner regions of the galaxy disks (2 kpc - 20 kpc), as determined from the available interferometer data  \citep[e.g.][]{tay99, isr09, cas08}, while most of the X--ray luminosities likely come from the central engine region (a few pc--100pc, at most). To establishing a clear causal connection between molecular gas in the inner 1 kpc and the fueling of nuclear activity will require higher resolution maps of the CO emission. The forthcoming ALMA (Atacama Large Millimeter Array) observatory will provide the necessary spatial resolution \citep{mai08}, which will enable to obtain much accurate kinematics an extended maps of the distribution and dynamics of the molecular gas.

\acknowledgments

The authors would like to thank the CSO staff for their support during observations. We thank the anonymous referees for the valuable suggestions to improve the manuscript. We are grateful for interesting discussion with Nick Scoville, Martin Emprechtinger and Tom Bell. The CSO is founded by the National Science Foundation under the contract AST-08388361. This research has made use of the NASA/IPAC Extragalactic Database (NED) (which is operated by the Jet Propulsion Laboratory, California Institute of Technology, under contract with the National Aeronautics and Space Administration)

{\it Facilities:} \facility{CSO}, \facility{ROSAT}, \facility{IRAS}, \facility{Chandra} .

\appendix

\section{CO Line Luminosity}

We have calculated the CO line luminosity using the integrated line intensity (T$\Delta$V), the intensity (I) and the flux (S$_{CO}$). The calculated parameters for each source are listed in Table \ref{tbl-2}. \\
To convert from integrated intensity in units of K $\cdot$ km/s into line intensities $erg \cdot cm^{-2} \cdot s^{-1} \cdot sr^{-1}$ we use the Rayleigh-Jeans relation,
\begin{equation}
I\left[erg \cdot cm^{-2} \cdot s^{-1} \cdot sr^{-1}\right] = \frac{2 k_{B} \cdot \nu^{3}}{c^{3}}\cdot \frac{T \Delta V \left[K \cdot km/s\right]}{B_{eff}}  \times 10^{6},
\end{equation}
where c is the speed of light, k$_{b}$ is the Boltzmann constant, B$_{eff}$ is main beam efficiency, and  v is the line frequency in GHz. \\
Flux in $Jy \cdot km/s$ are derived as follows,
\begin{equation}
S[Jy \cdot km/s] = I \cdot \frac{c}{\nu} \cdot \Omega_{b}\cdot \left({\frac{\nu_{0}}{\nu}}\right)^{2} \times 10^{20},
\end{equation}
where $\Omega_{b}$ is the solid angle in \textit{sr}. For a Gaussian beam, the solid angle is given by,
\begin{equation}
\Omega_{b}= 1.1333 \cdot B^{2} \cdot \frac{1}{206265^{2}}~~sr
\end{equation}
where B is the half power beam width in arcsec. \\
We calculate the CO luminosities using the formula in Solomon et al 1997,
\begin{equation}
L_{CO} = (c^{2}/2\textit{k})\cdot S_{CO}\cdot \Delta V \cdot \nu^{-2}_{obs}\cdot D^{2}_{L} \cdot (1+z)^{-3},
\end{equation}
where D$_{L}$ is the luminosity distance $\footnote{For all the sources, we obtained the luminosity distance using the web site calculator of http://www.astro.ucla.edu/~wright/CosmoCalc.html, using a cosmology of $H_{0} = 77 km\cdot s^{-1} \cdot Mpc^{-1}$, $\Omega_{M} = 0.27$ and $\Omega_{V}=0.73$.}$ in Mpc.




\clearpage
\begin{figure*}[!htb]
\centering
\vspace{1cm}

%\plotone{/figure1a.eps}\hspace{.1cm}
\includegraphics[scale=.22, angle = -90]{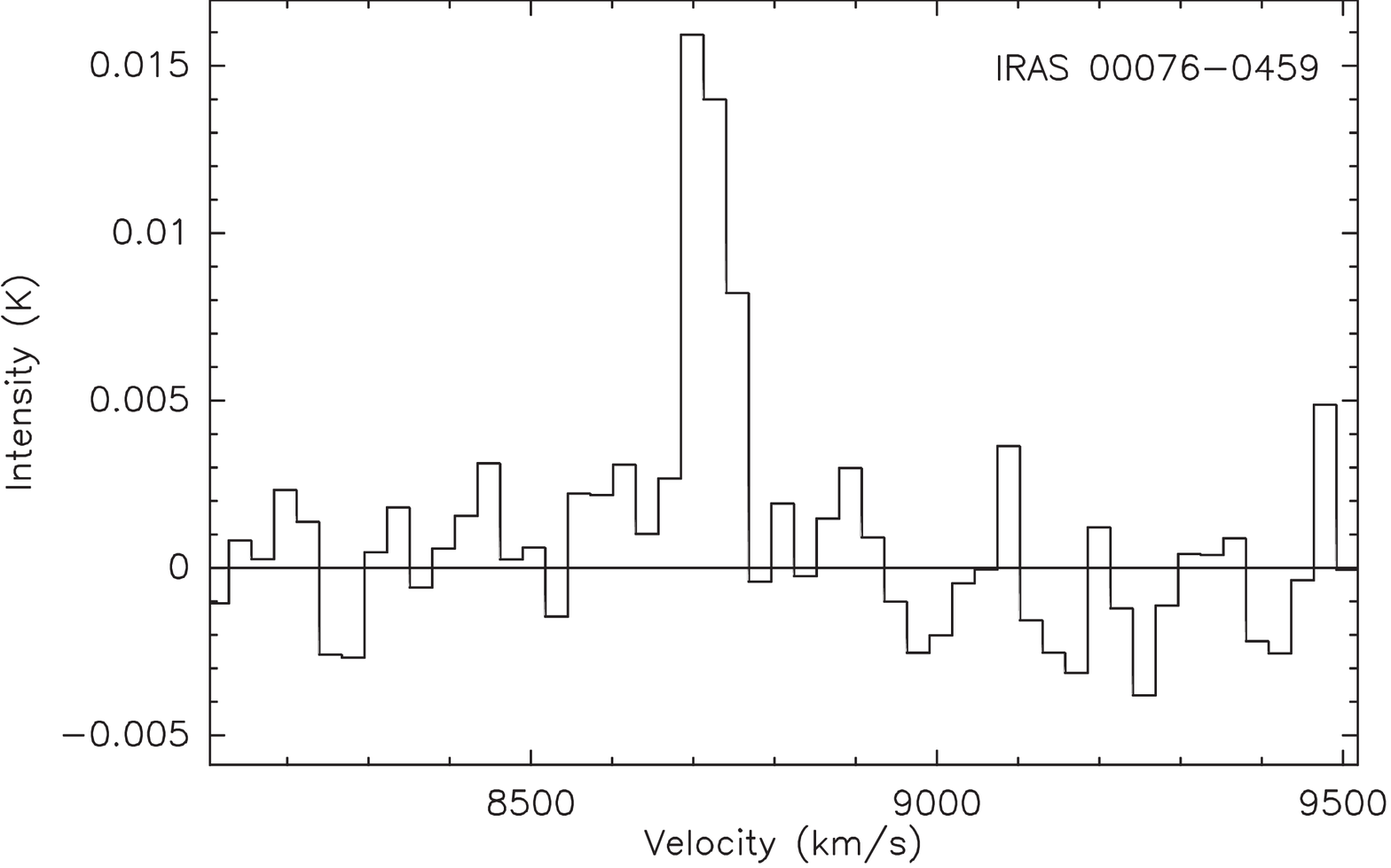}\hspace{.1cm}
\includegraphics[scale=.22, angle = -90]{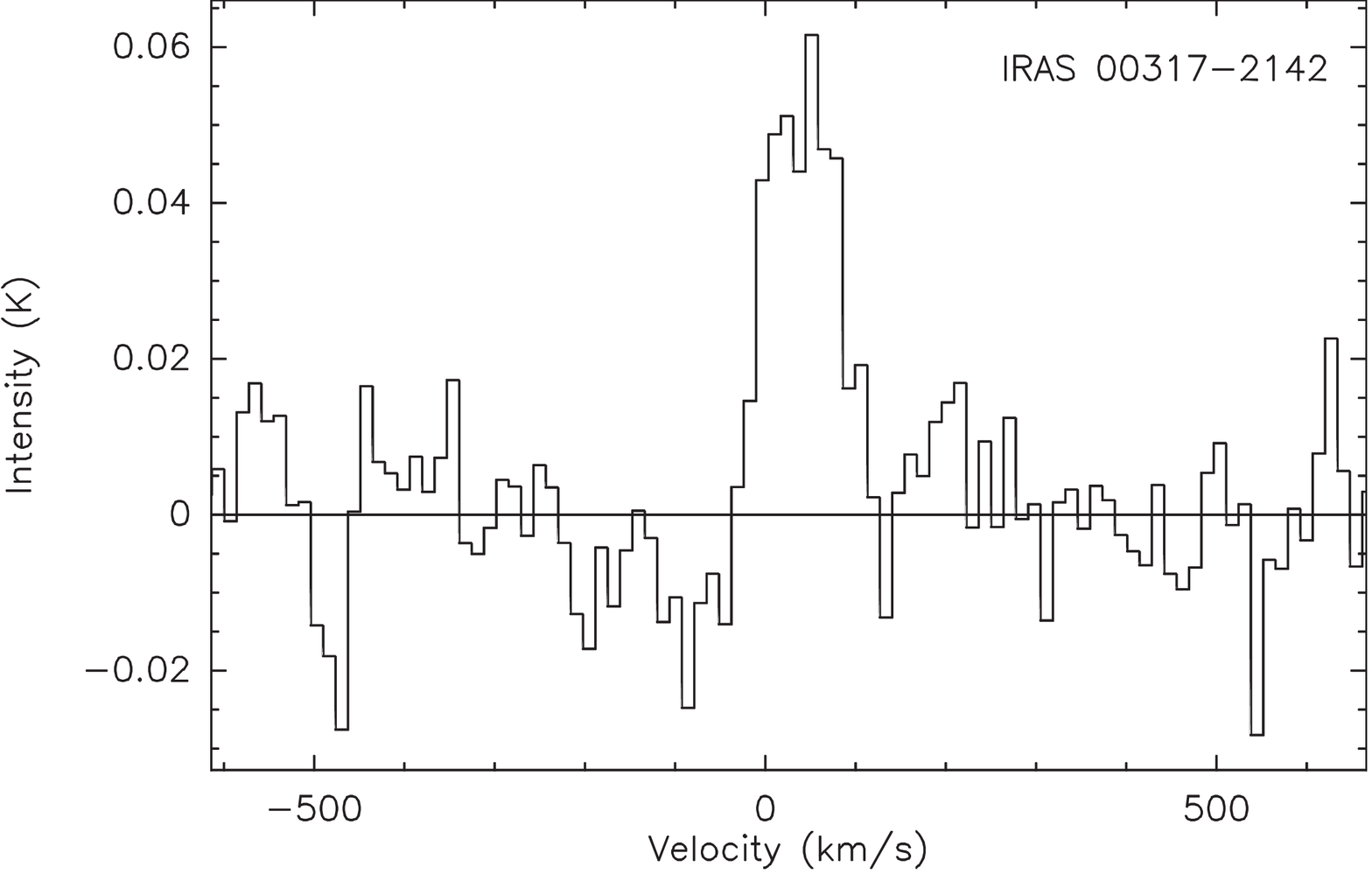}\hspace{.1cm}
\includegraphics[scale=.22, angle = -90]{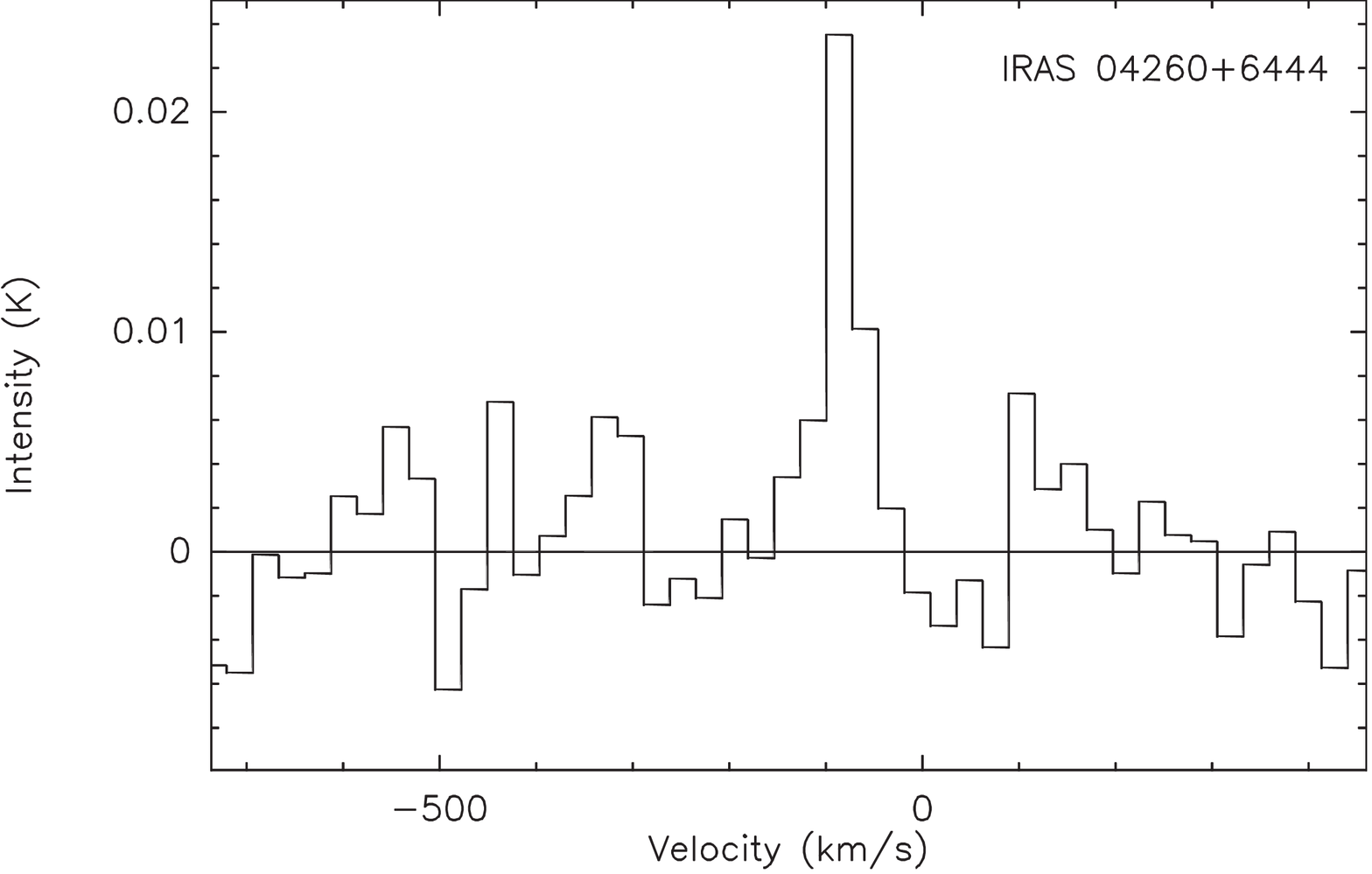}\\[20pt]
\includegraphics[scale=.22, angle = -90]{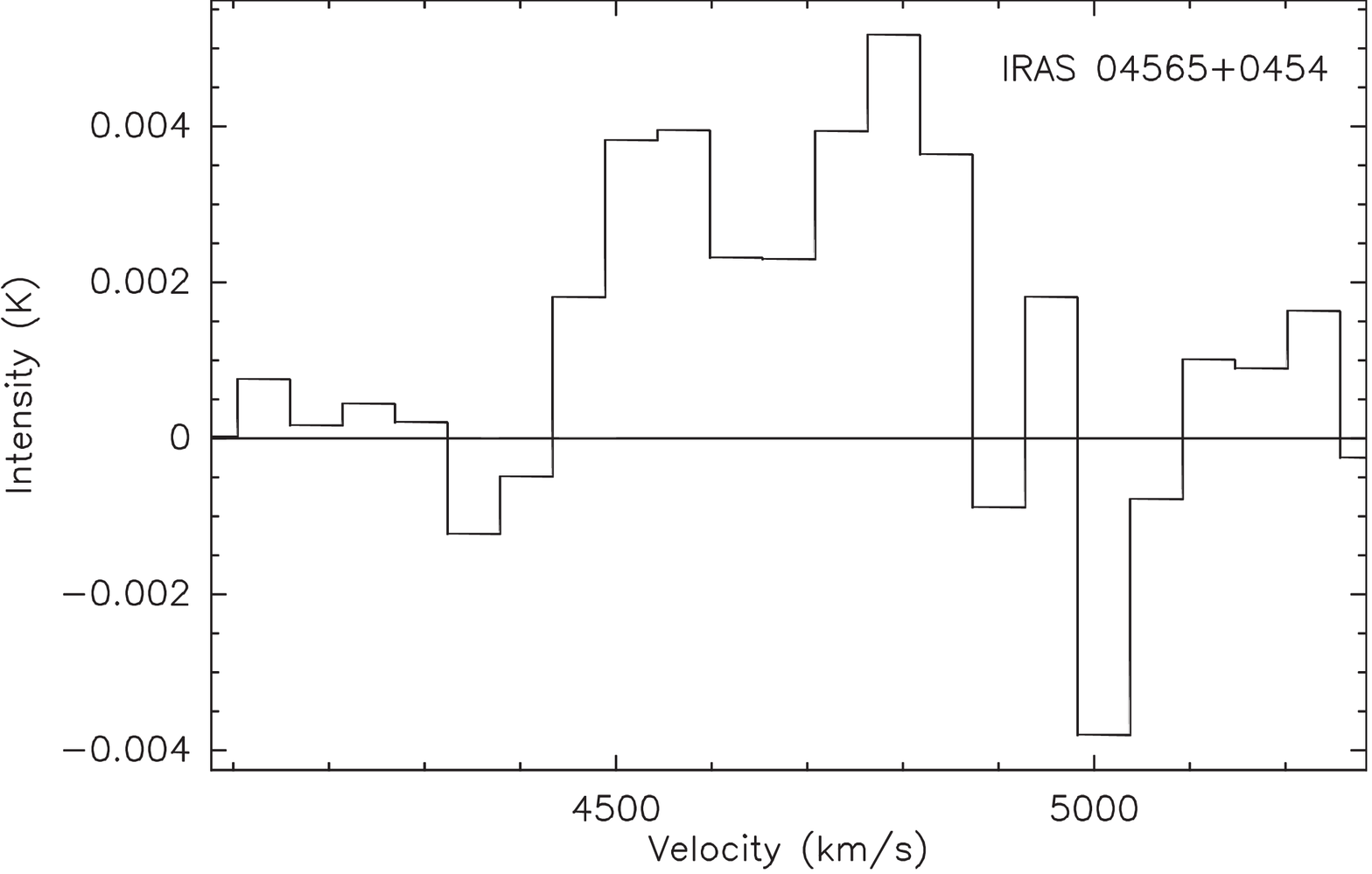}\hspace{.1cm}
\includegraphics[scale=.22, angle = -90]{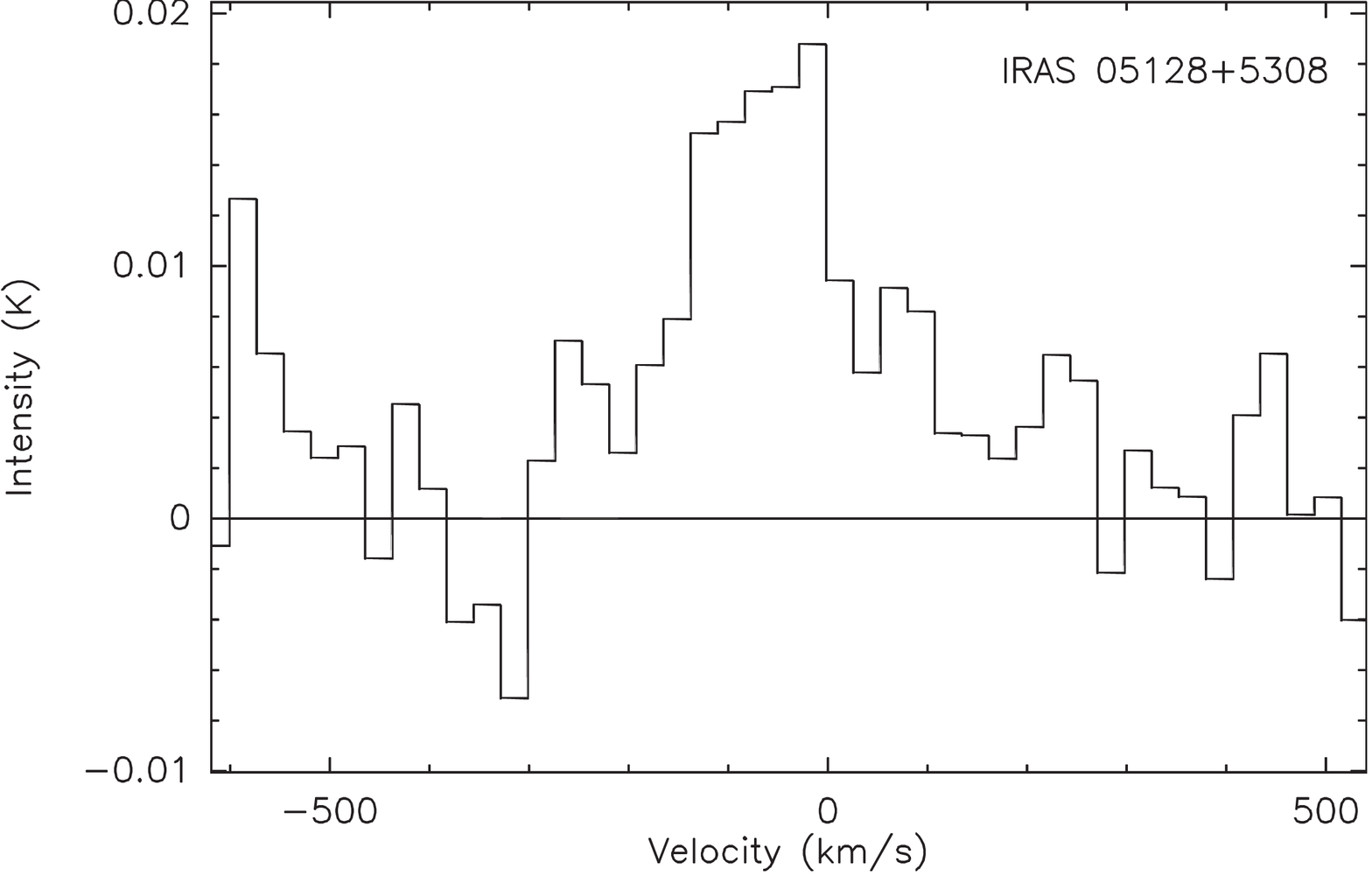}\hspace{.1cm}
\includegraphics[scale=.22, angle = -90]{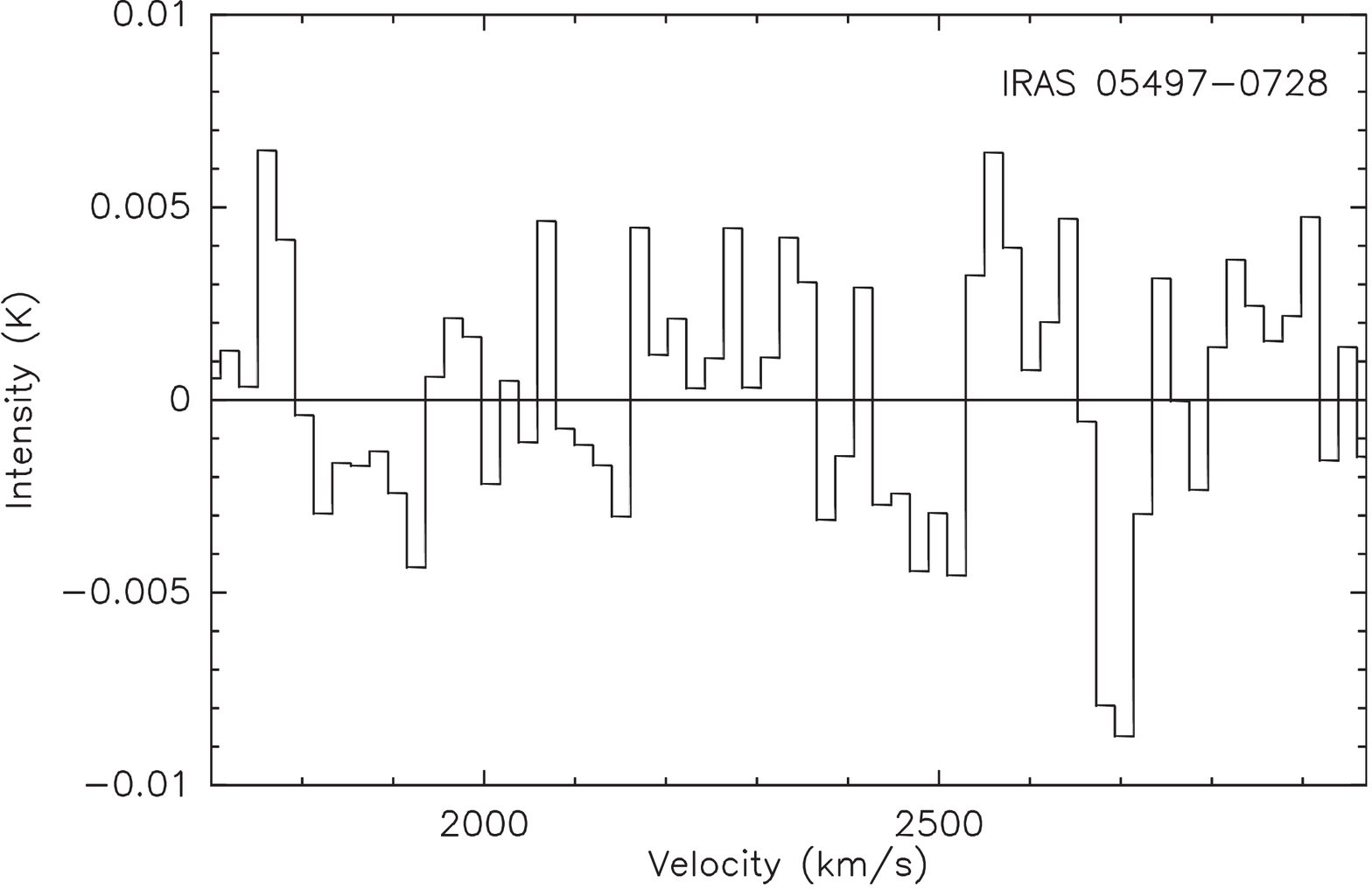}\\[20pt]
\includegraphics[scale=.22, angle = -90]{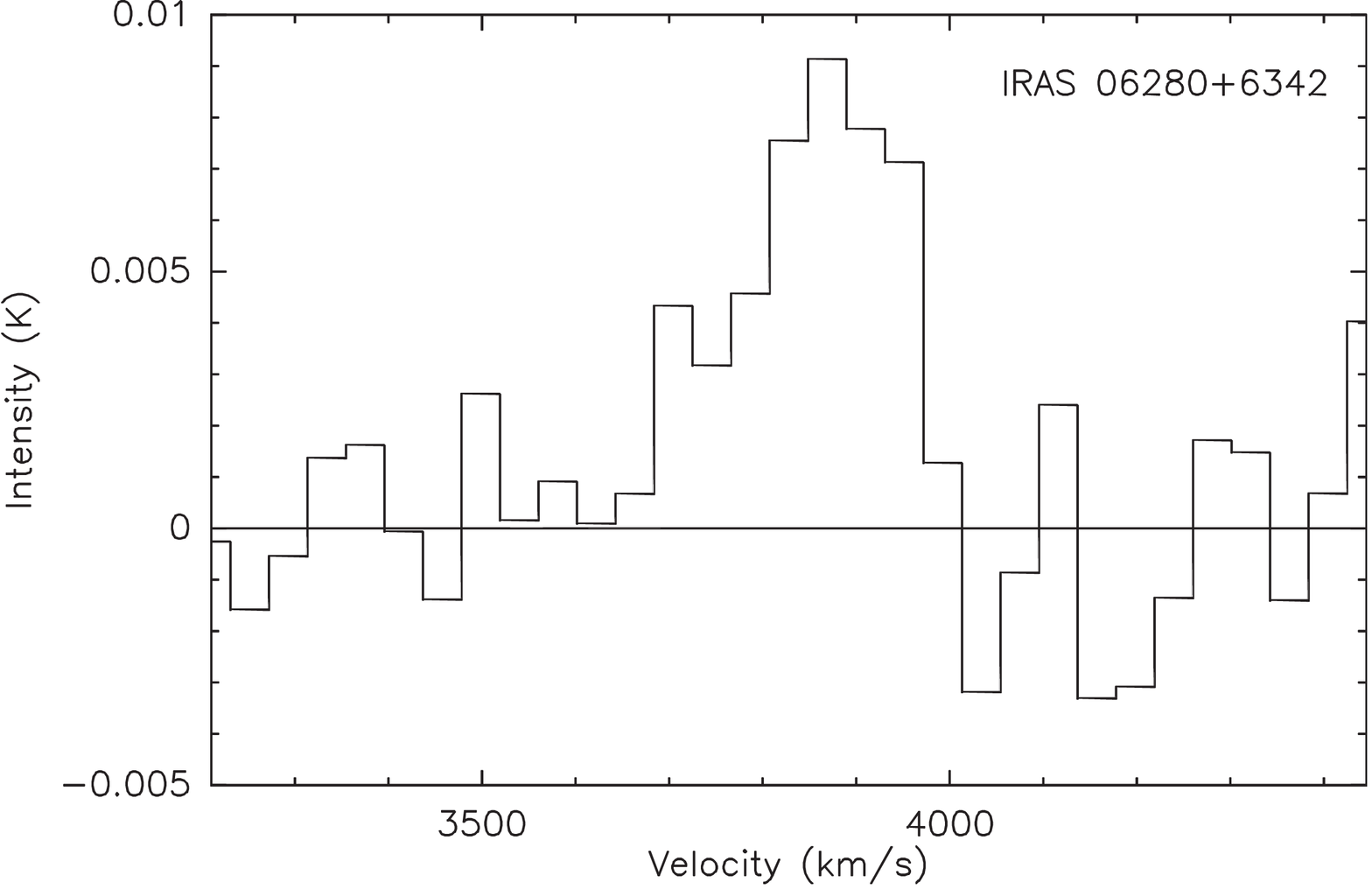}\hspace{.1cm}
\includegraphics[scale=.22, angle = -90]{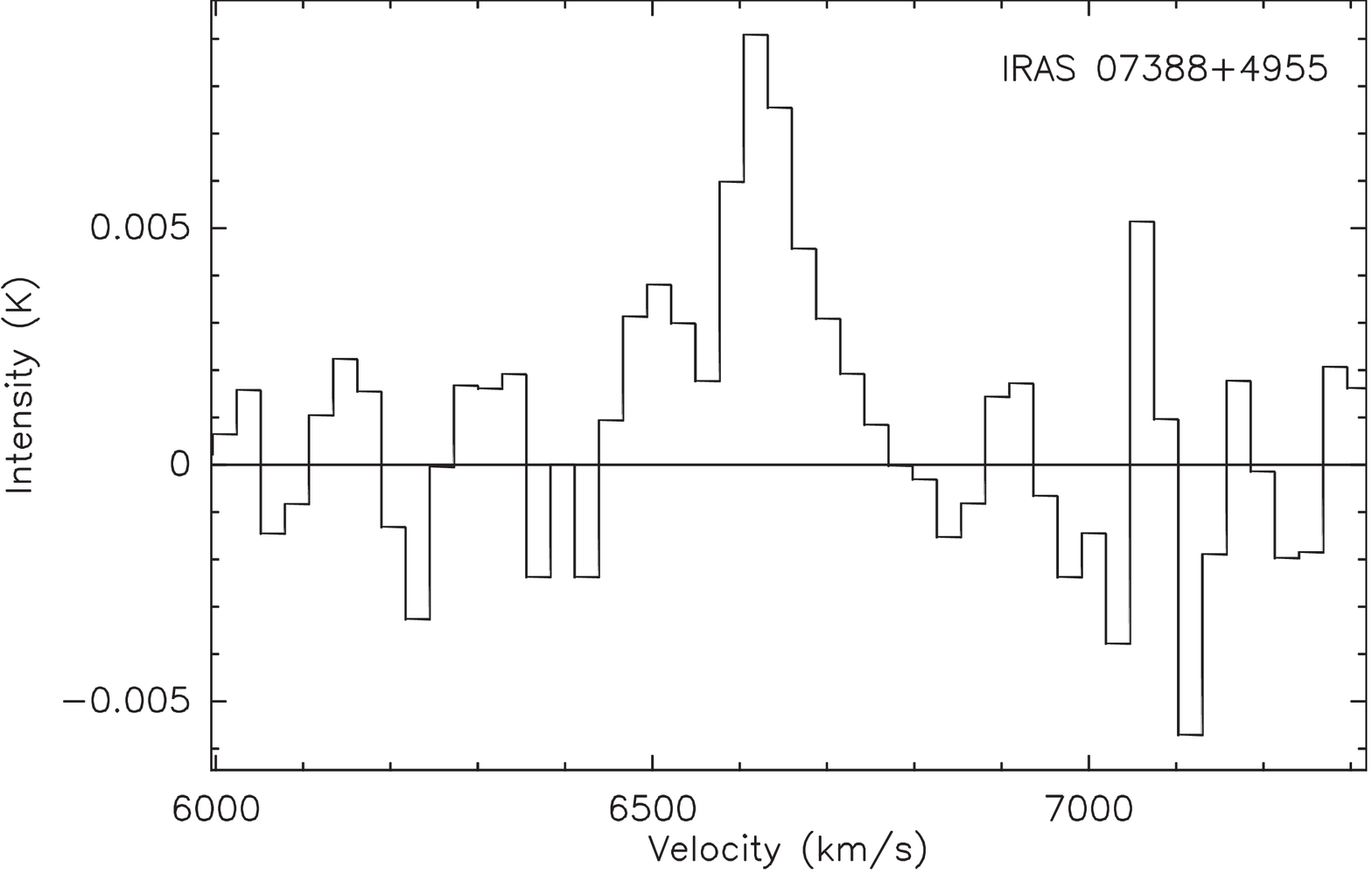}\hspace{.1cm}
\includegraphics[scale=.22, angle = -90]{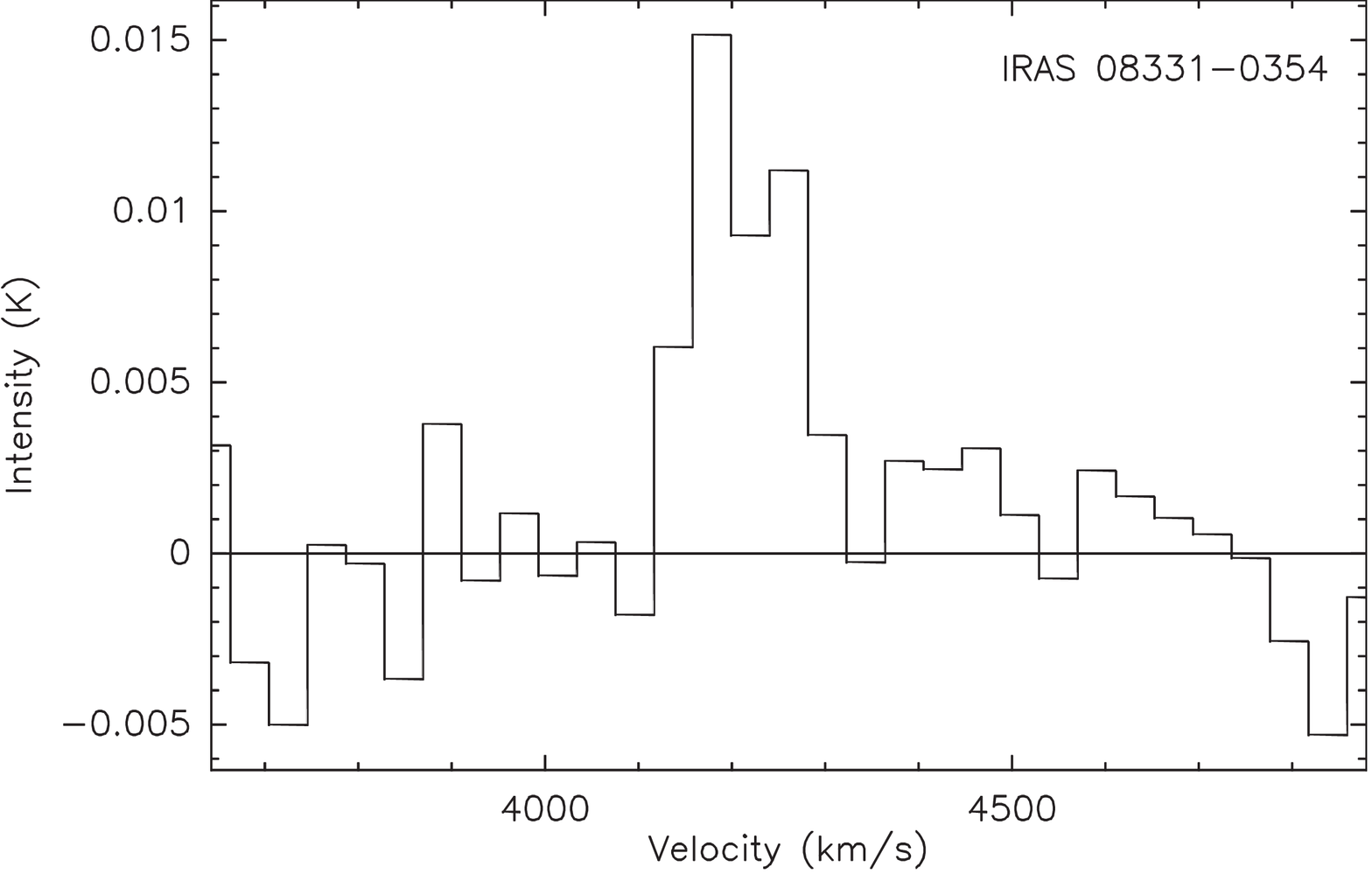}\\[20pt]
\includegraphics[scale=.22, angle = -90]{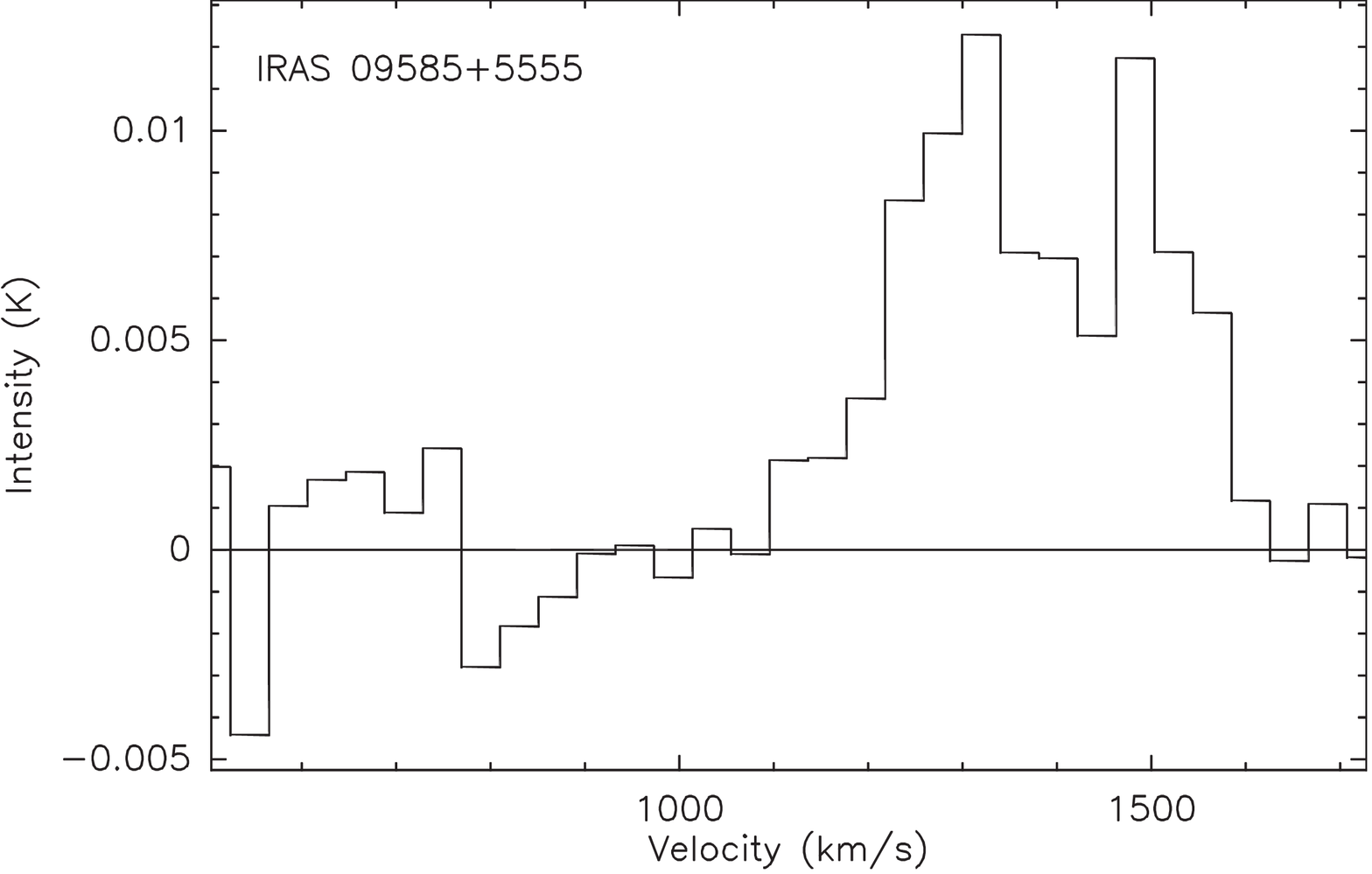}\hspace{.1cm}
\includegraphics[scale=.22, angle = -90]{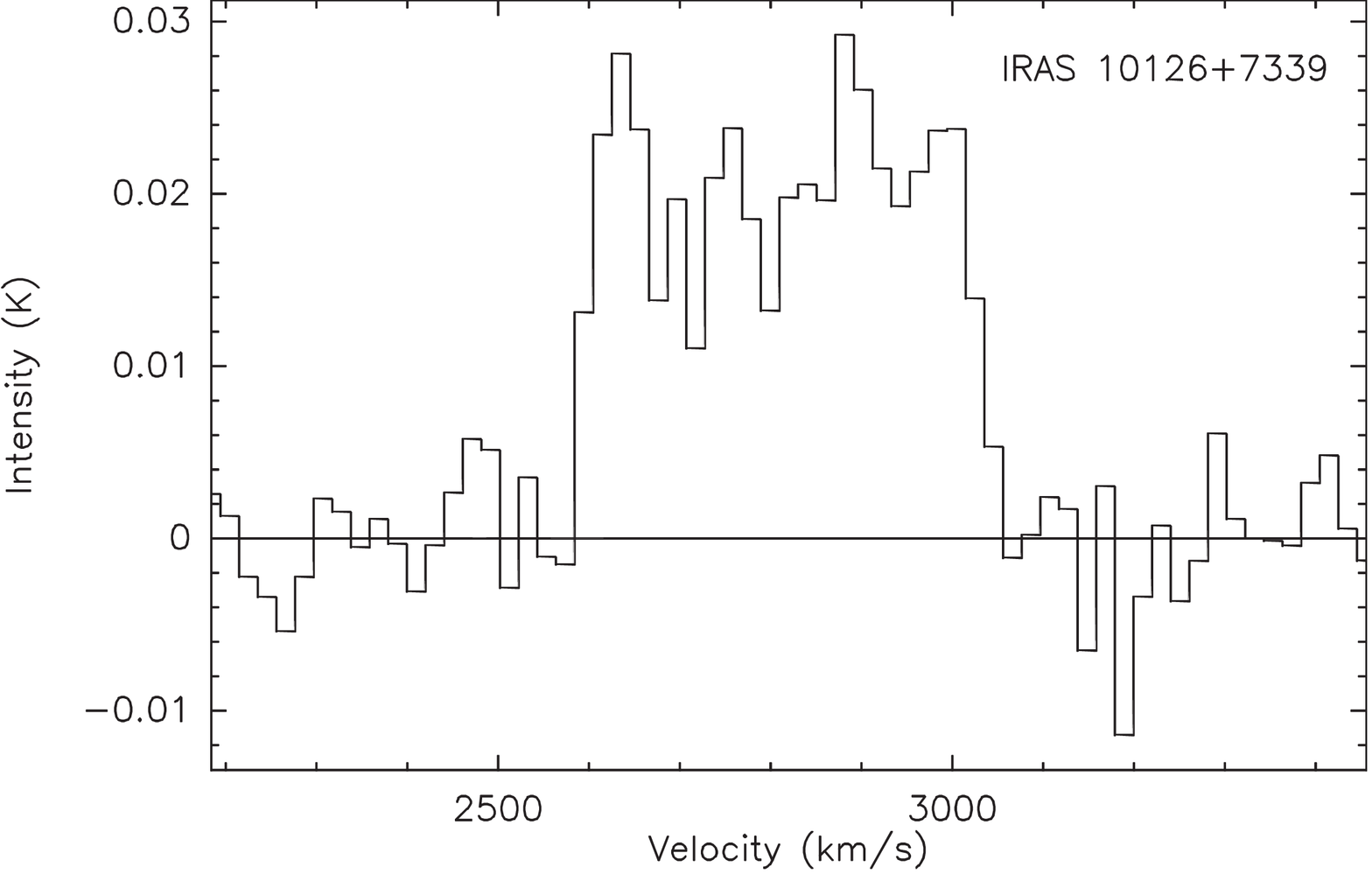}\hspace{.1cm}
\includegraphics[scale=.22, angle = -90]{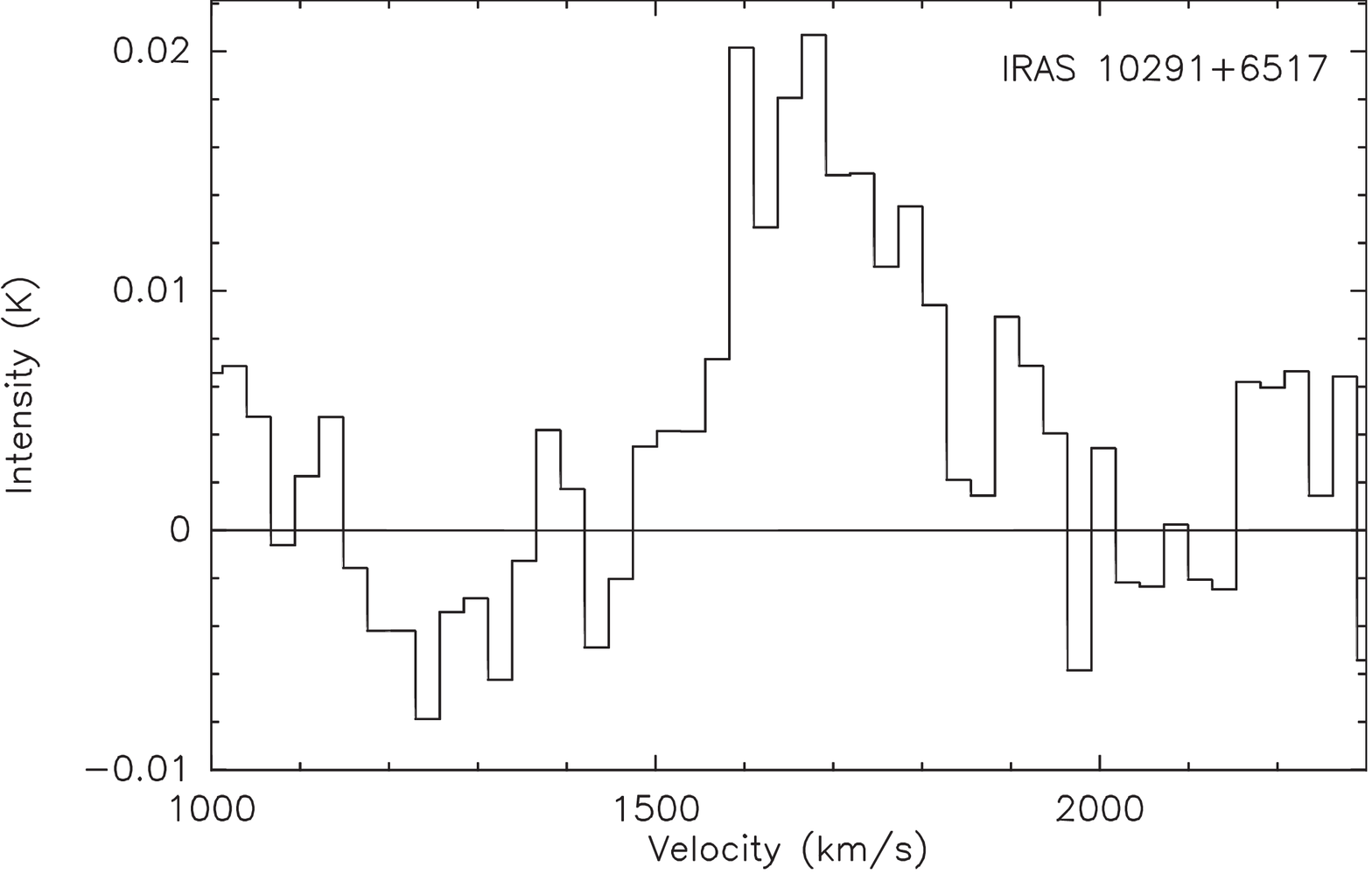}\\[20pt]

\includegraphics[scale=.22, angle = -90]{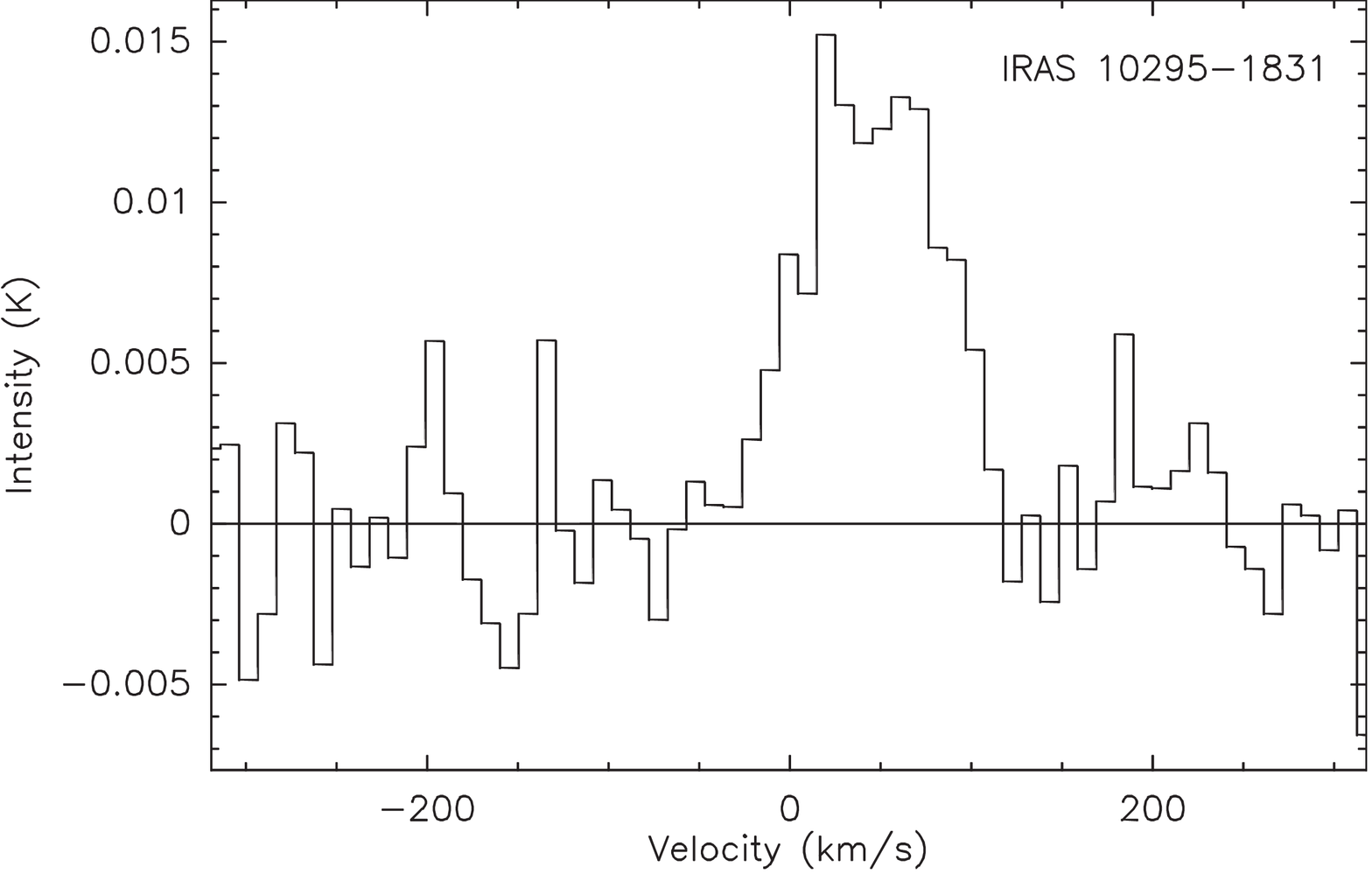}\hspace{.1cm}
\includegraphics[scale=.22, angle = -90]{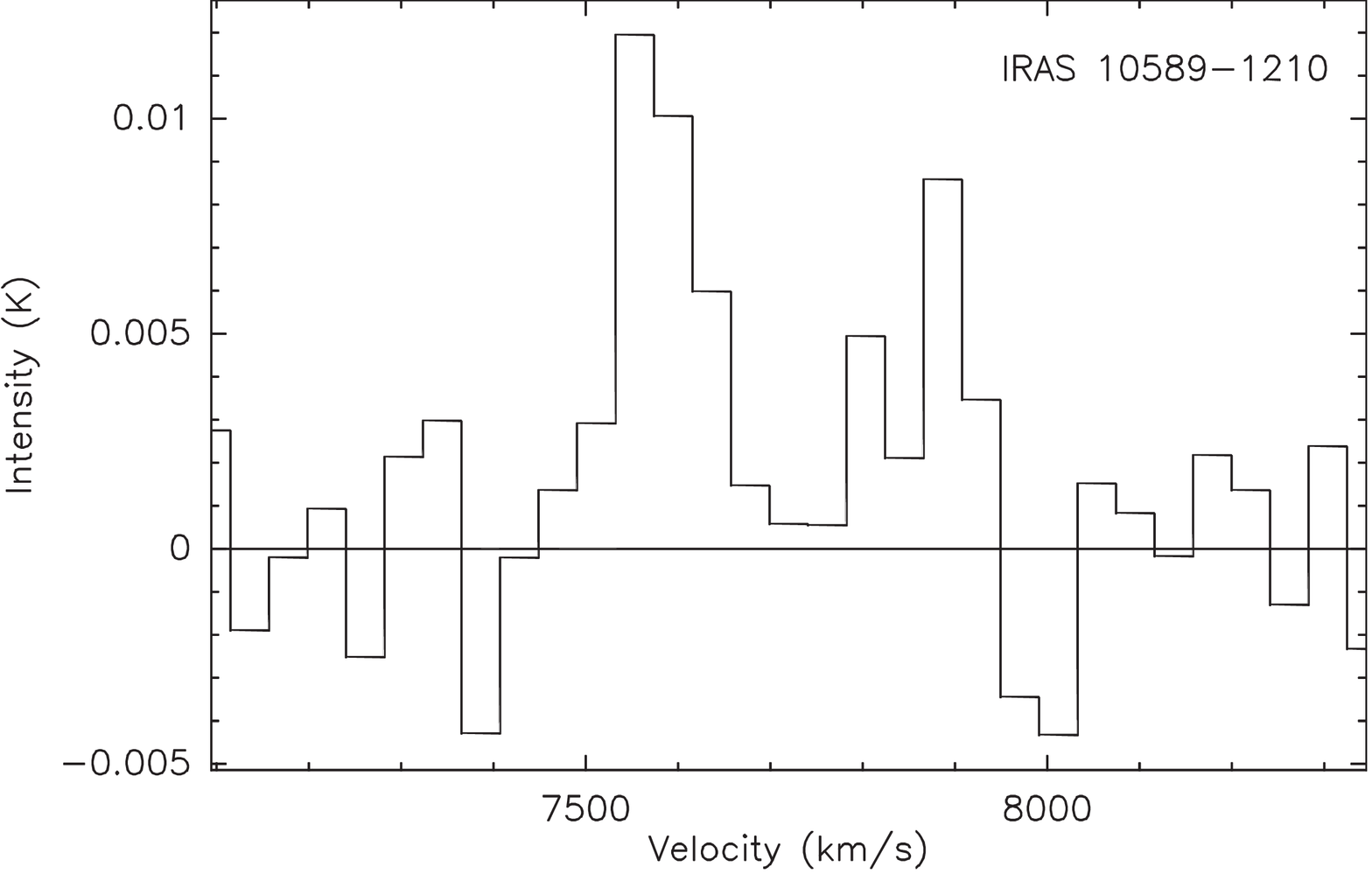}\hspace{.1cm}
\includegraphics[scale=.22, angle = -90]{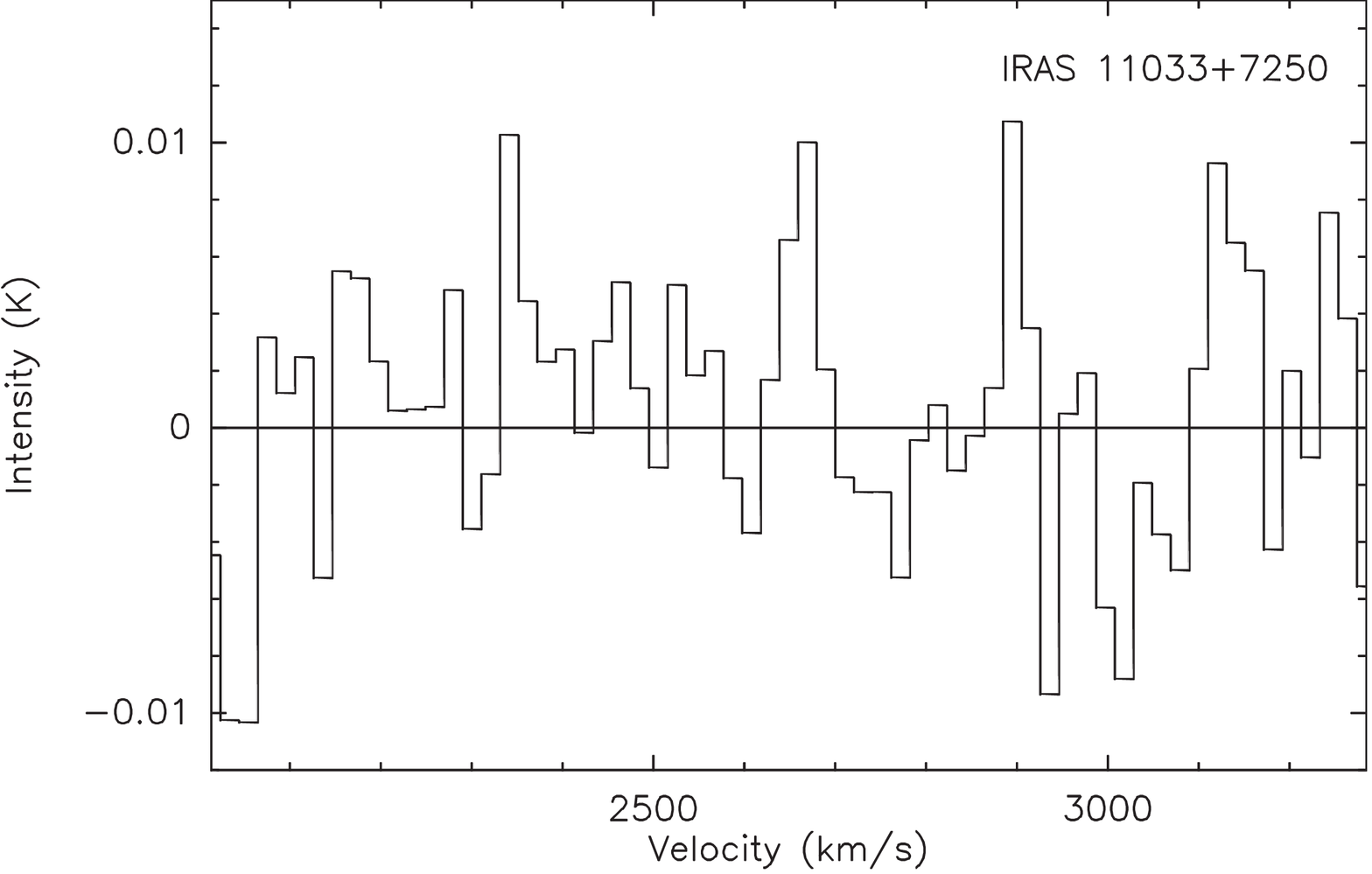}

\caption{\textit{CO $J$=2--1} observations of our sample. The intensity scale is $T^{*}_{A}$ in K. }
\label{spectra}
\end{figure*}

\begin{figure*}[!htb]
\setcounter{figure}{0}
\centering
\vspace{1cm}
\includegraphics[scale=.22, angle = -90]{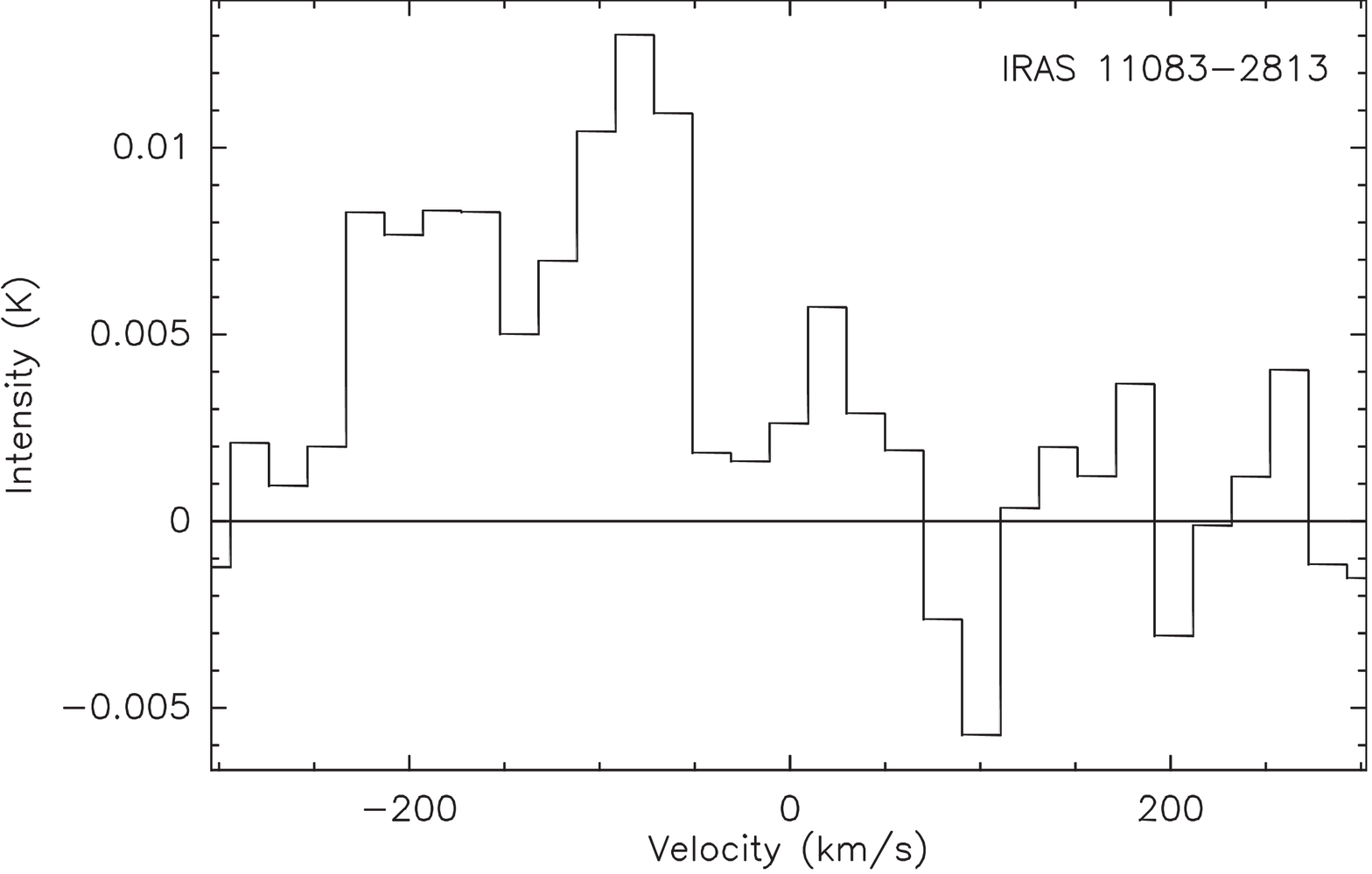}\hspace{.1cm}
\includegraphics[scale=.22, angle = -90]{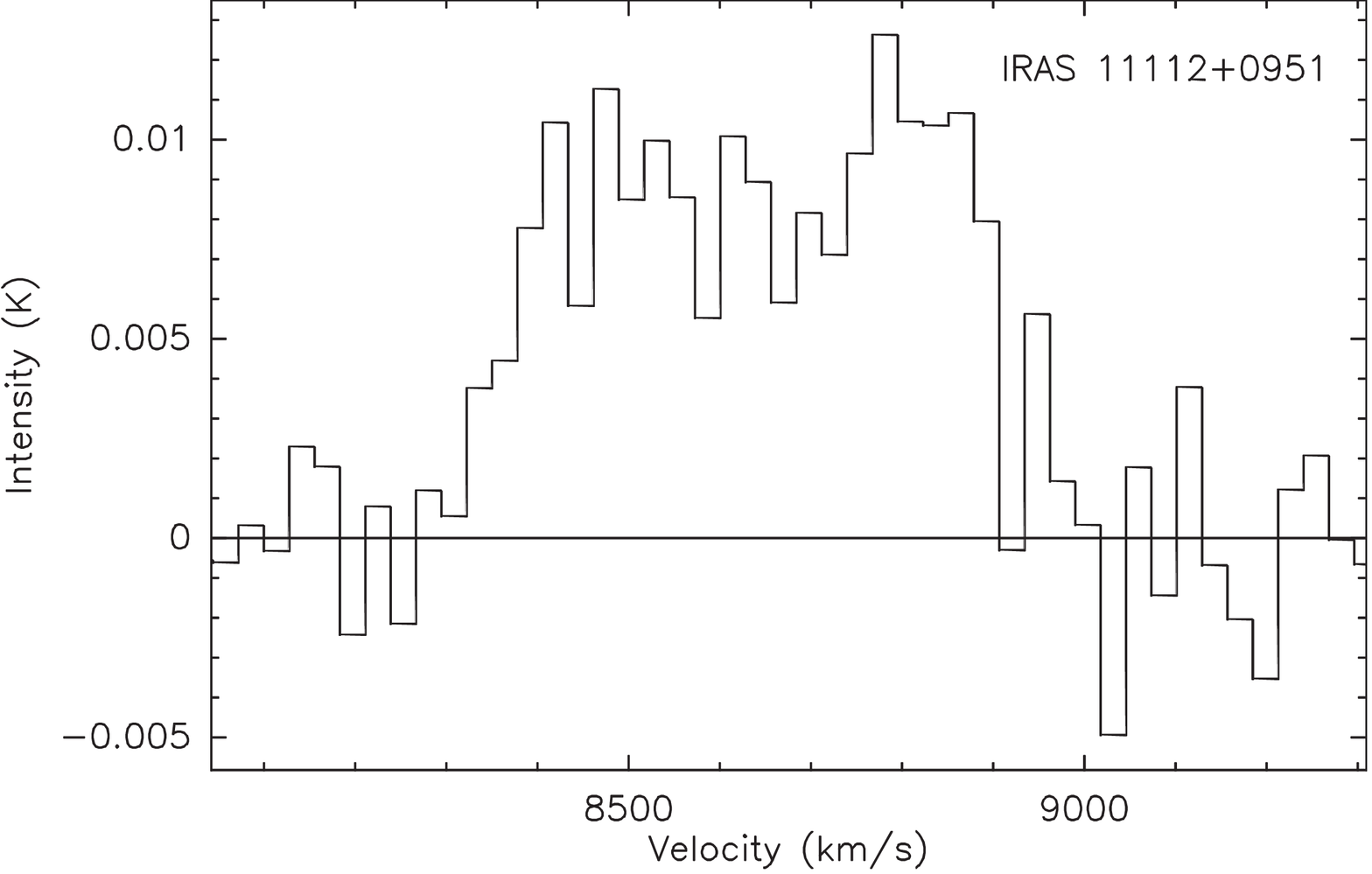}\hspace{.1cm}
\includegraphics[scale=.22, angle = -90]{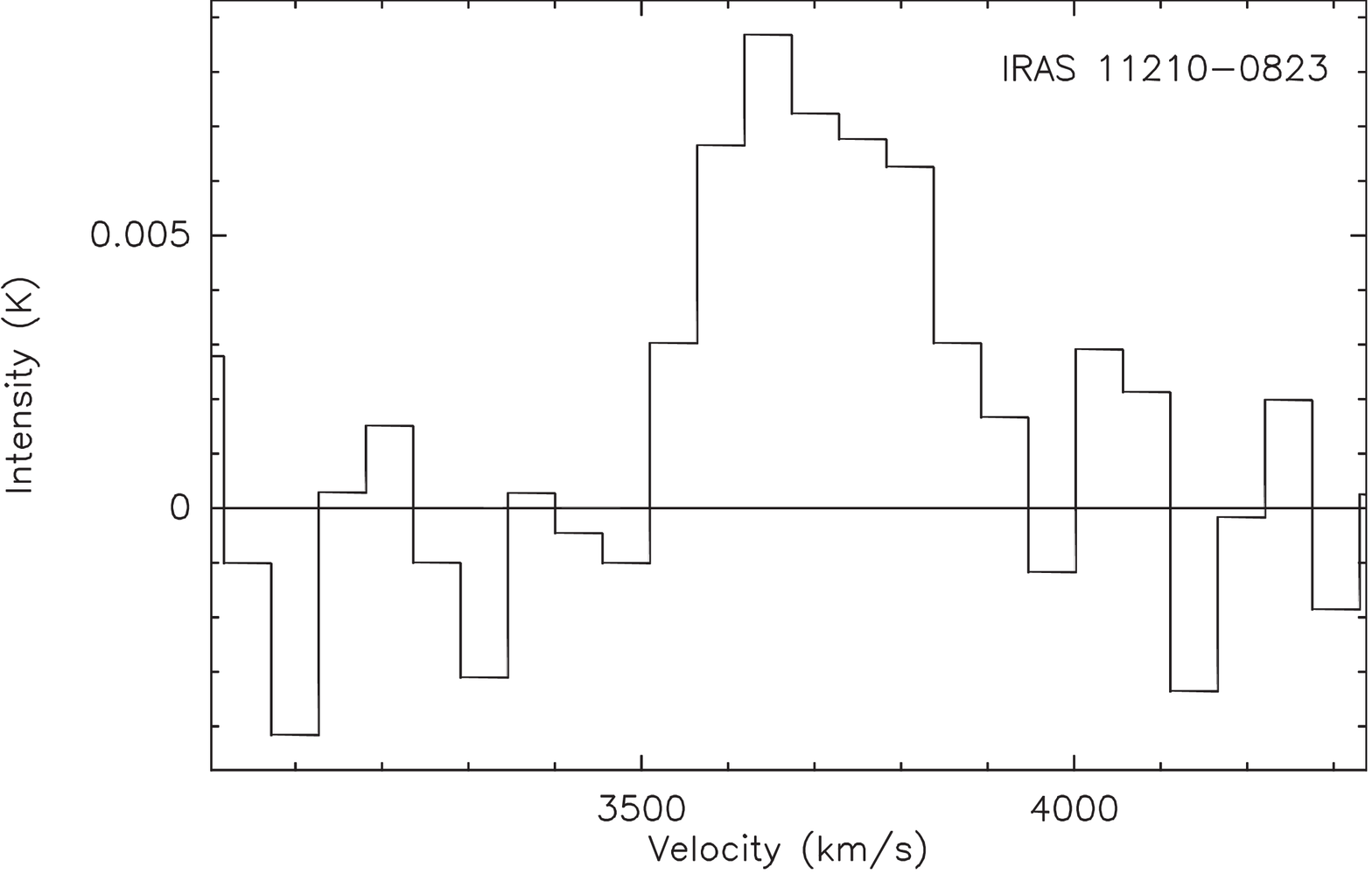} \\[20pt]

\includegraphics[scale=.22, angle = -90]{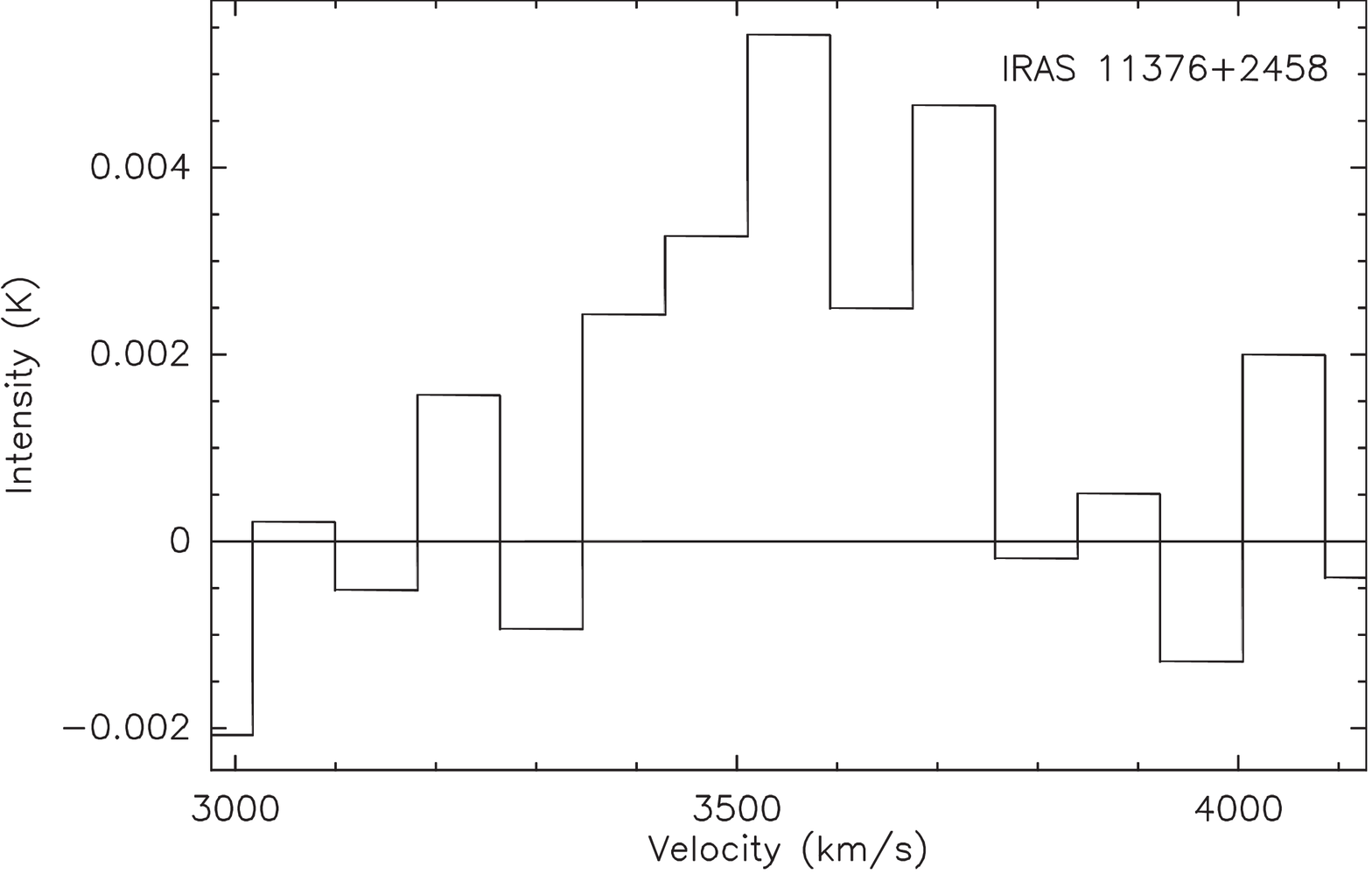}\hspace{.1cm}
\includegraphics[scale=.22, angle = -90]{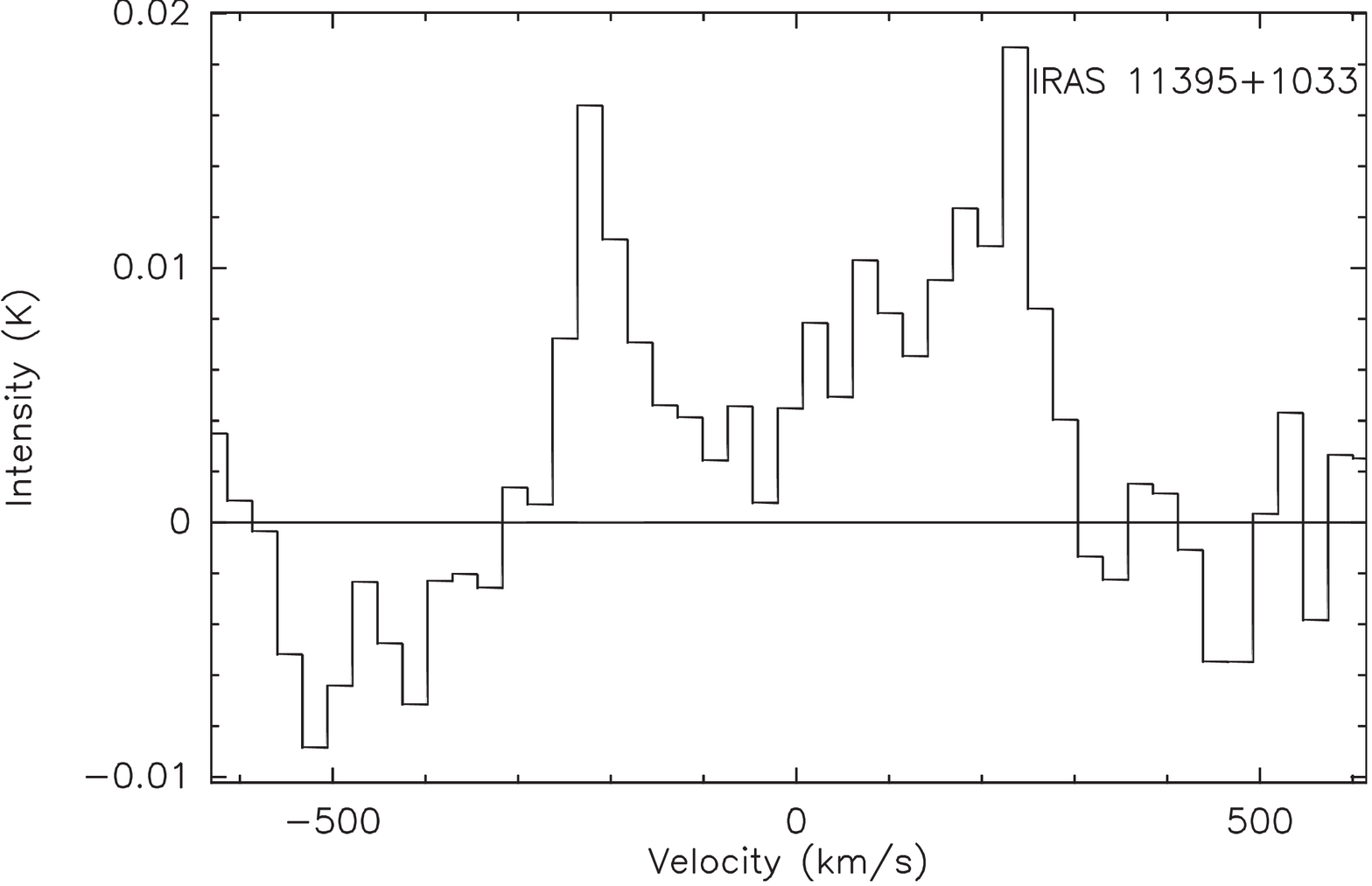}\hspace{.1cm}
\includegraphics[scale=.22, angle = -90]{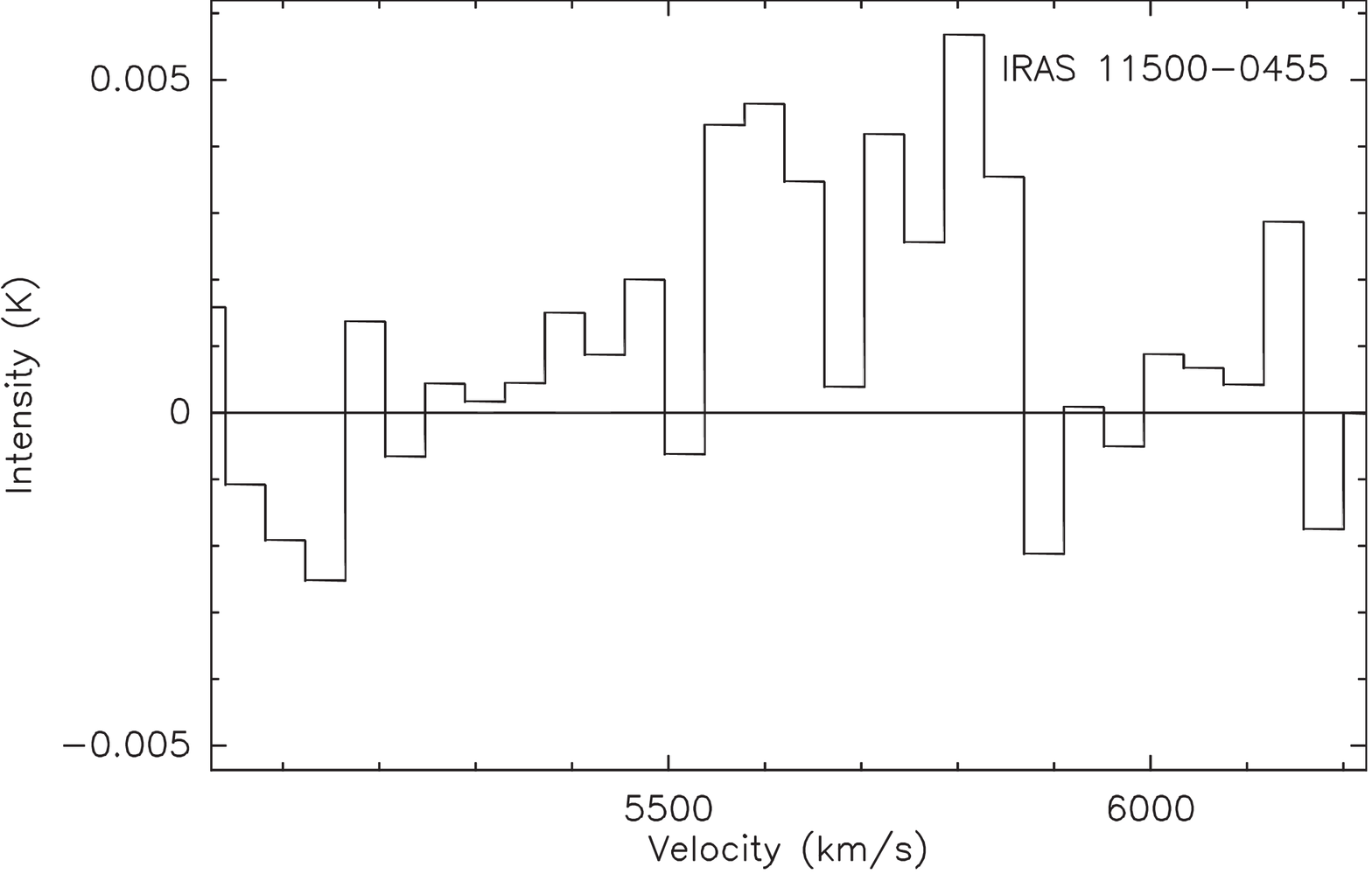}\\[20pt]

\includegraphics[scale=.22, angle = -90]{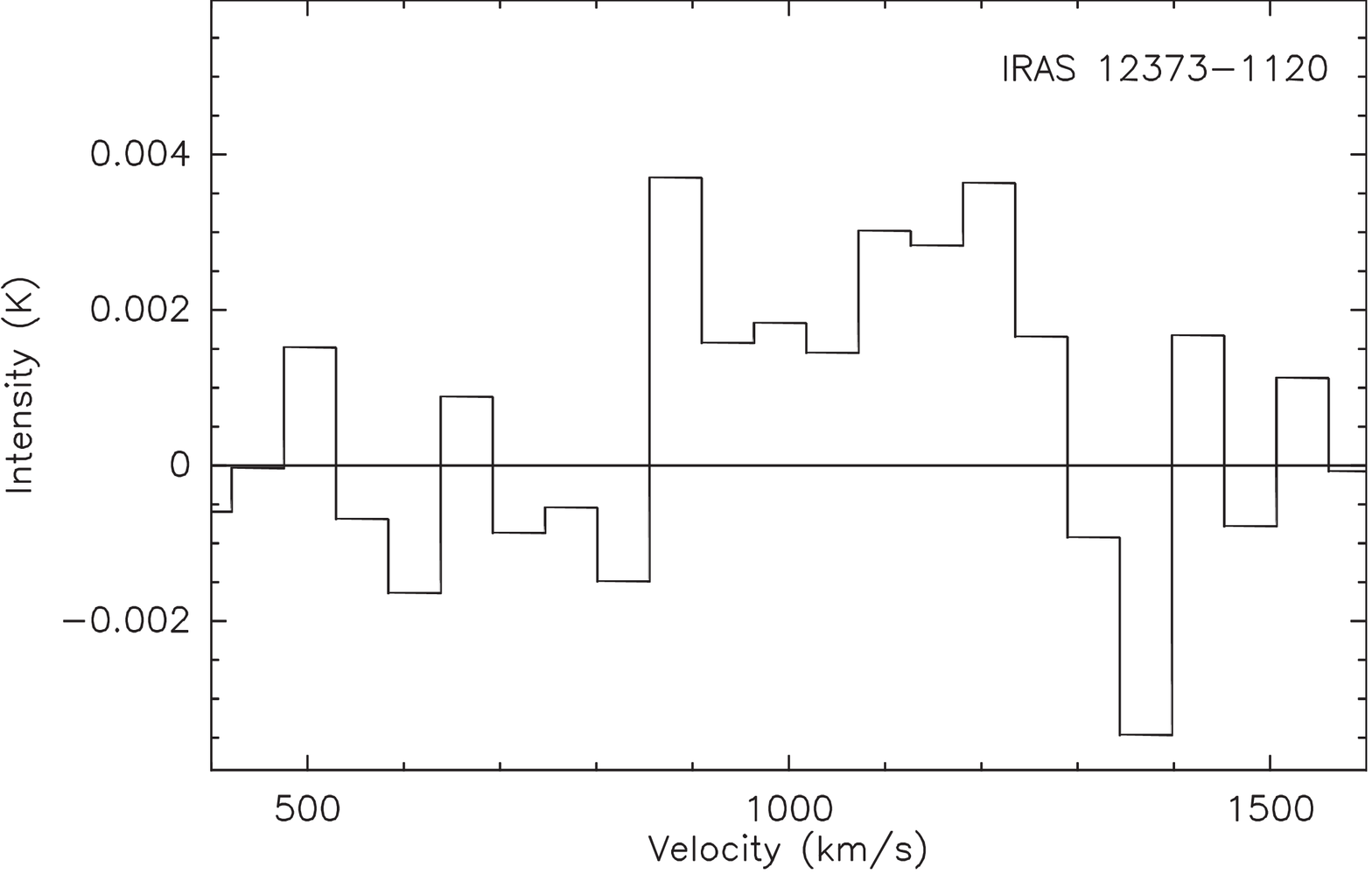}\hspace{.1cm}
\includegraphics[scale=.22, angle = -90]{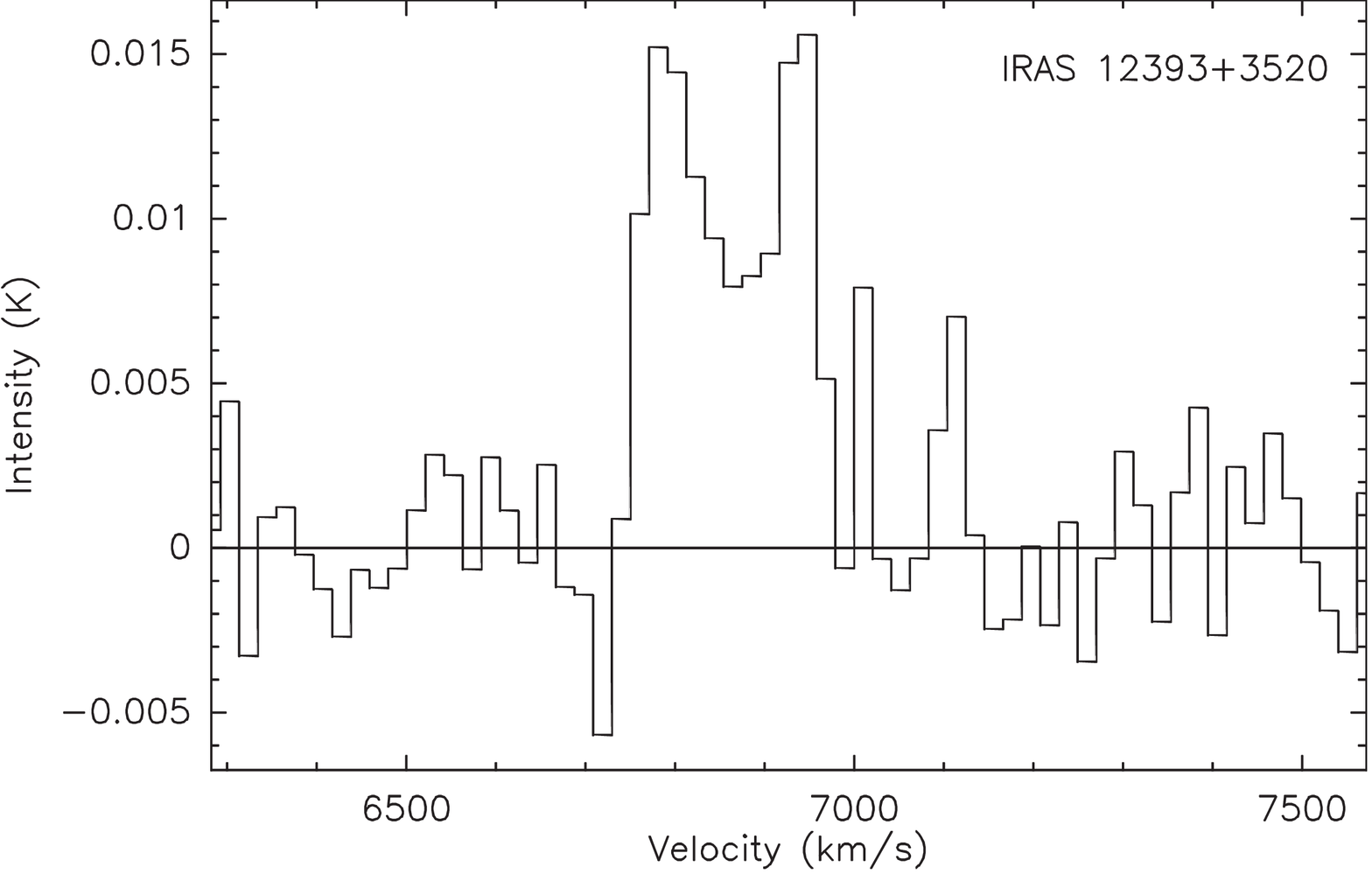}\hspace{.1cm}
\includegraphics[scale=.22, angle = -90]{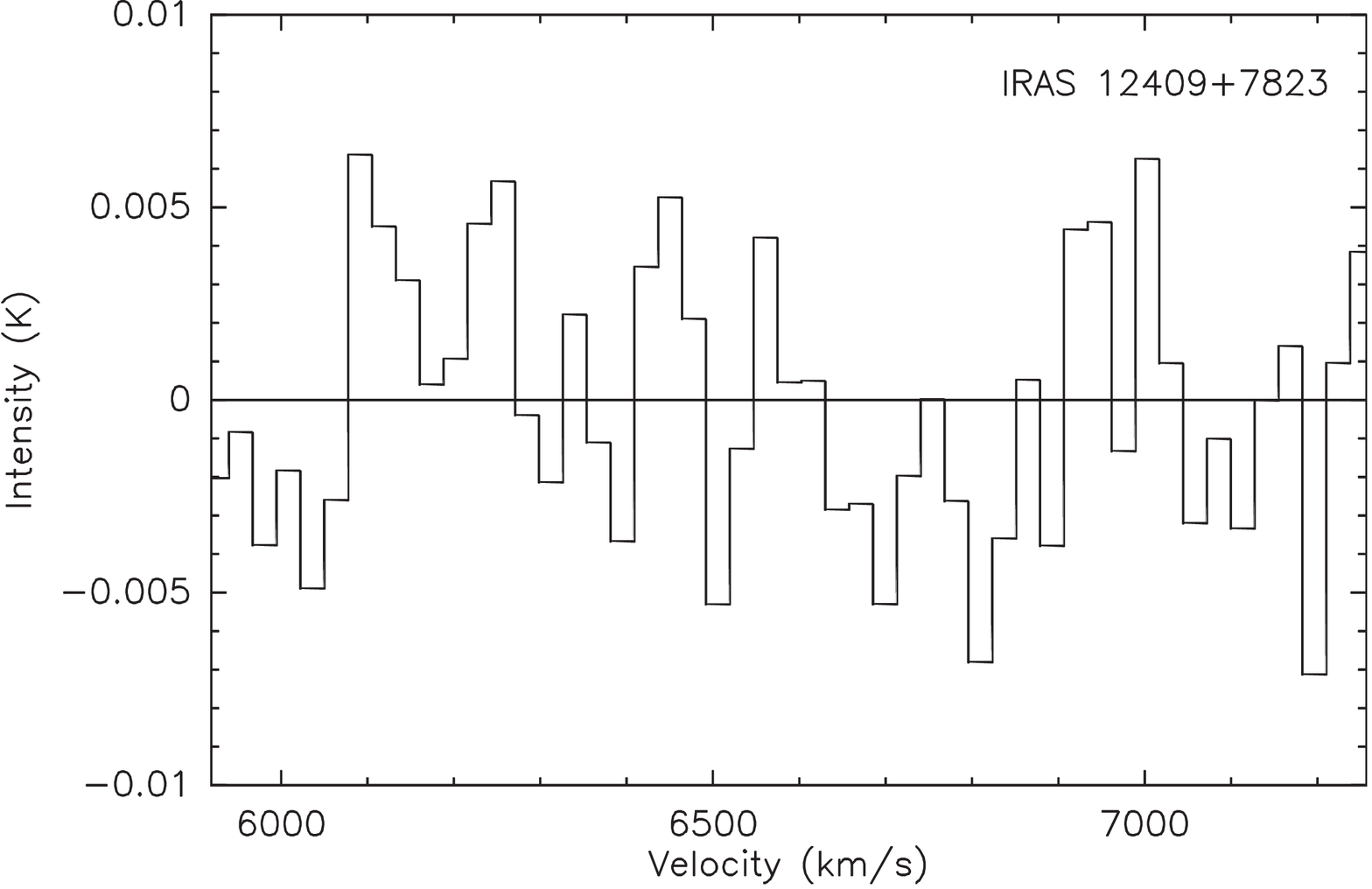}\\[20pt]

\includegraphics[scale=.22, angle = -90]{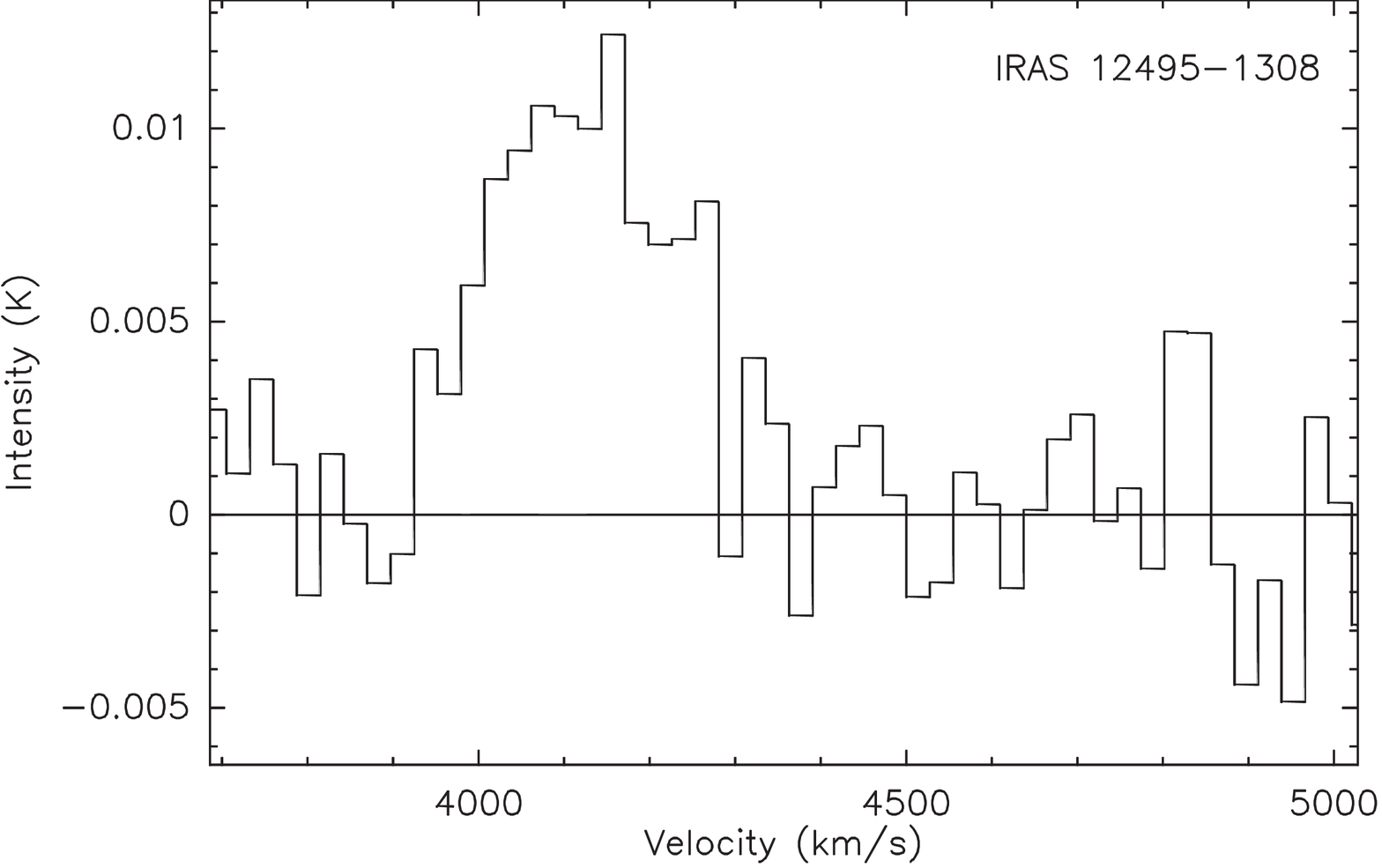}\hspace{.1cm}
\includegraphics[scale=.22, angle = -90]{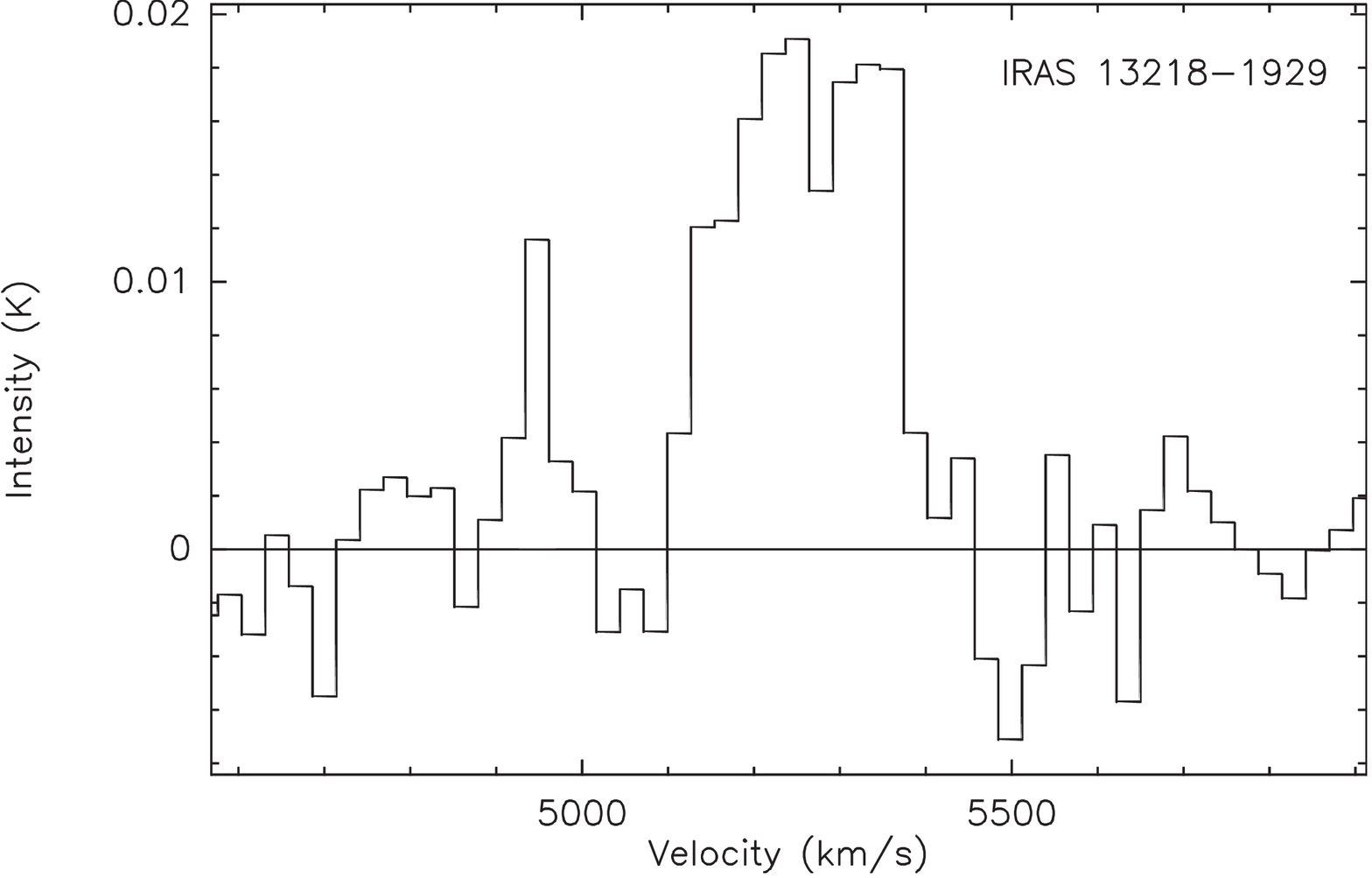}\hspace{.1cm}
\includegraphics[scale=.22, angle = -90]{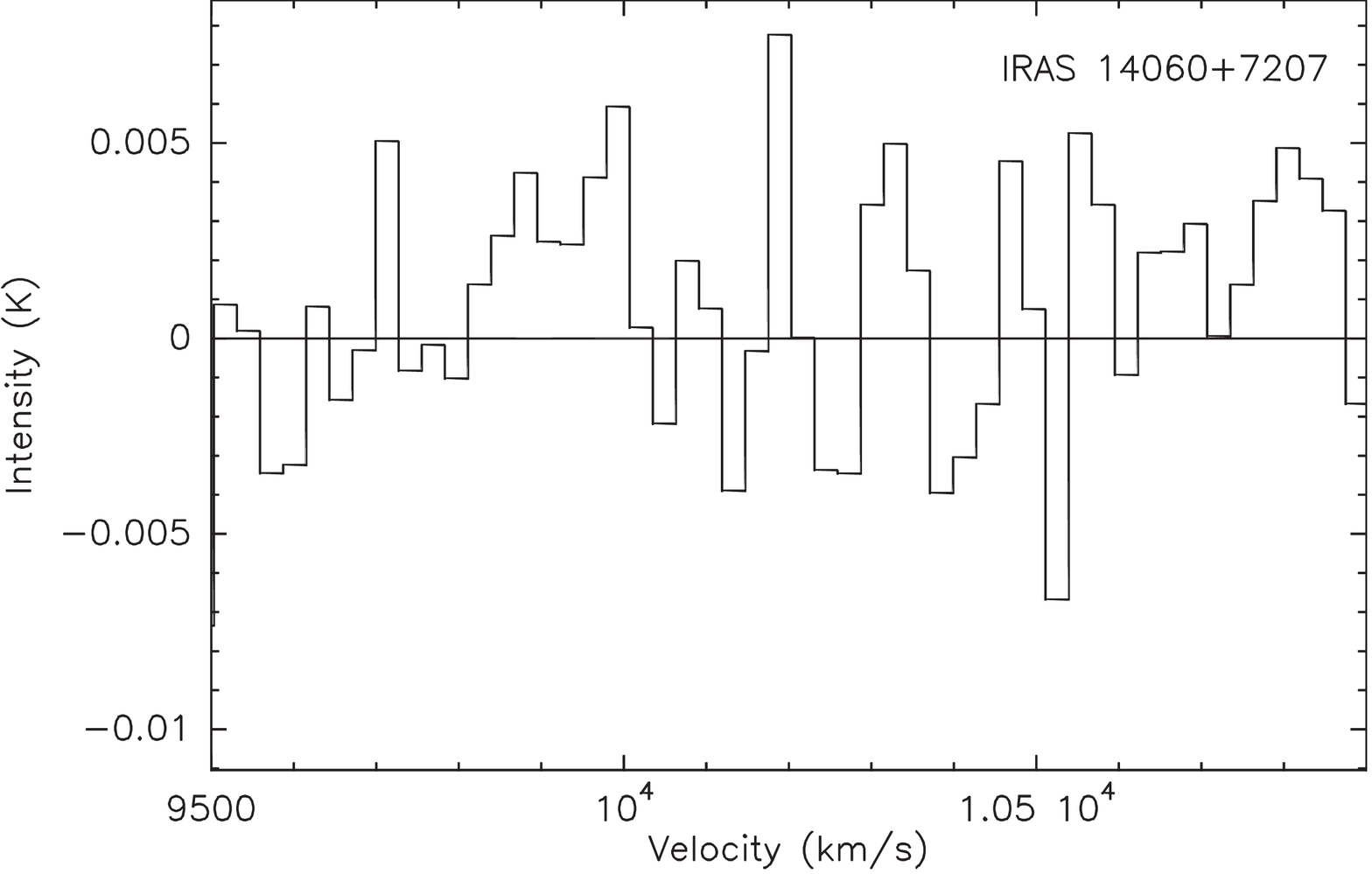}\\[20pt]

\includegraphics[scale=.22, angle = -90]{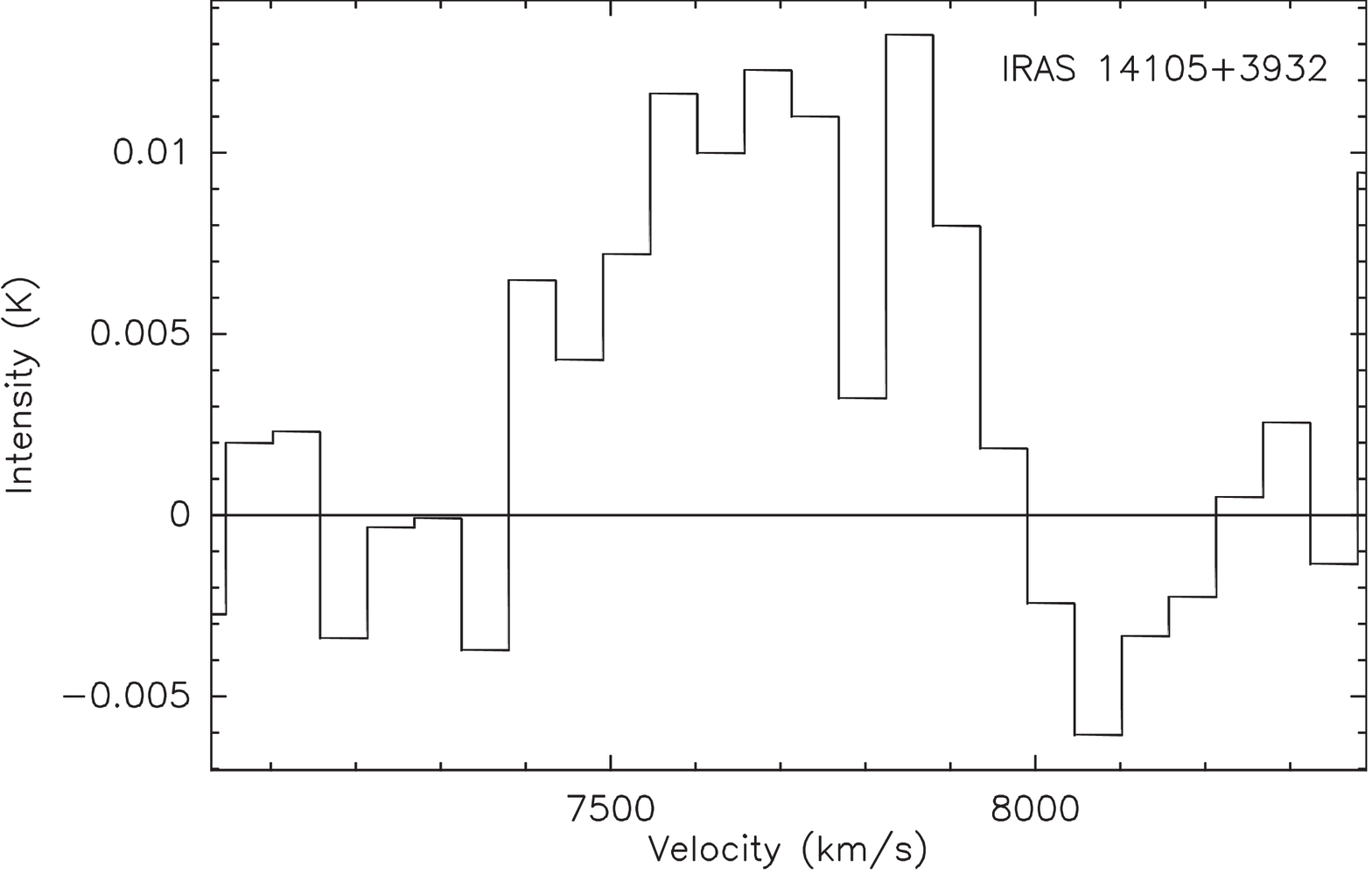}\hspace{.1cm}
\includegraphics[scale=.22, angle = -90]{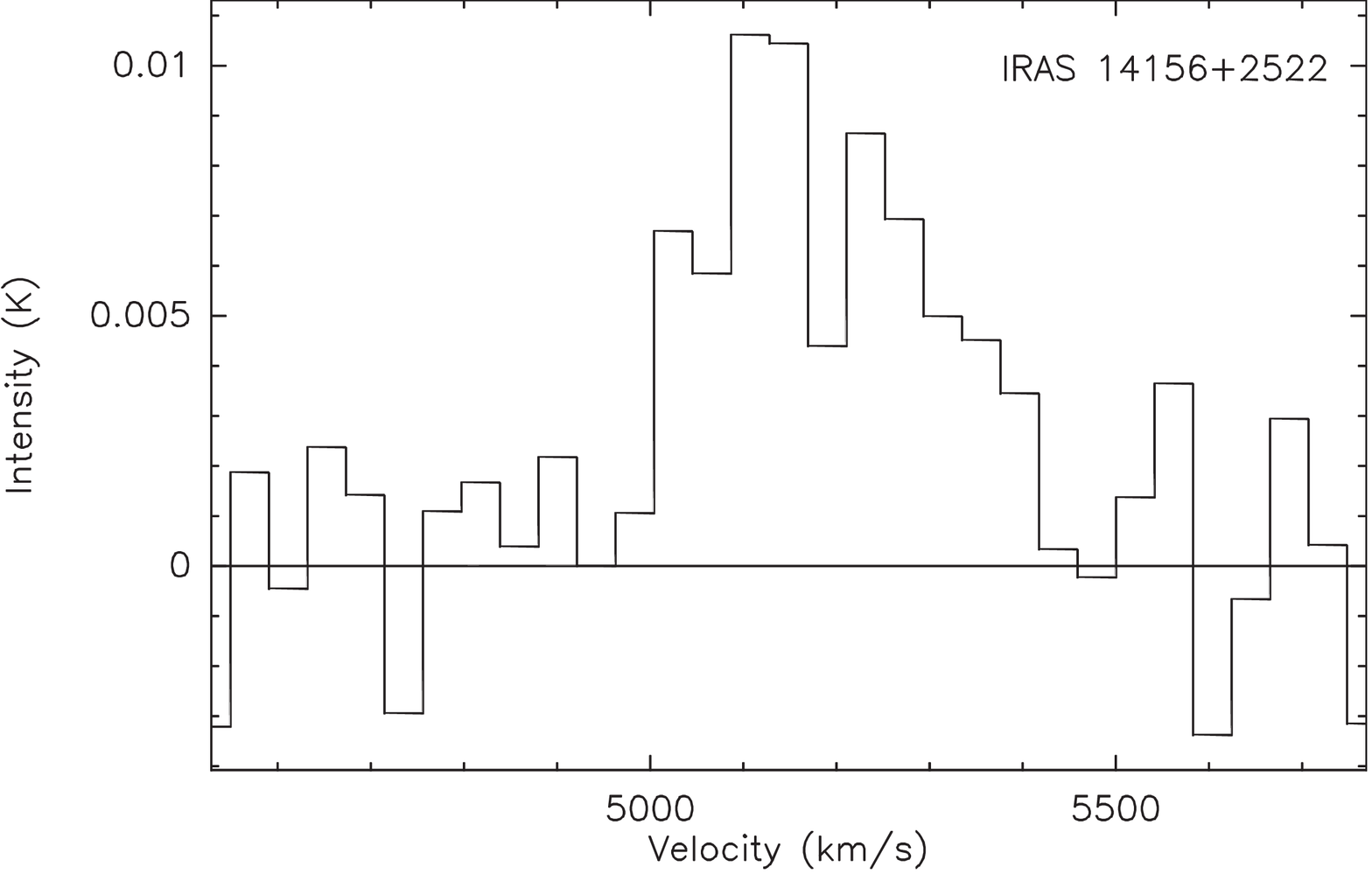}\hspace{.1cm}
\includegraphics[scale=.22, angle = -90]{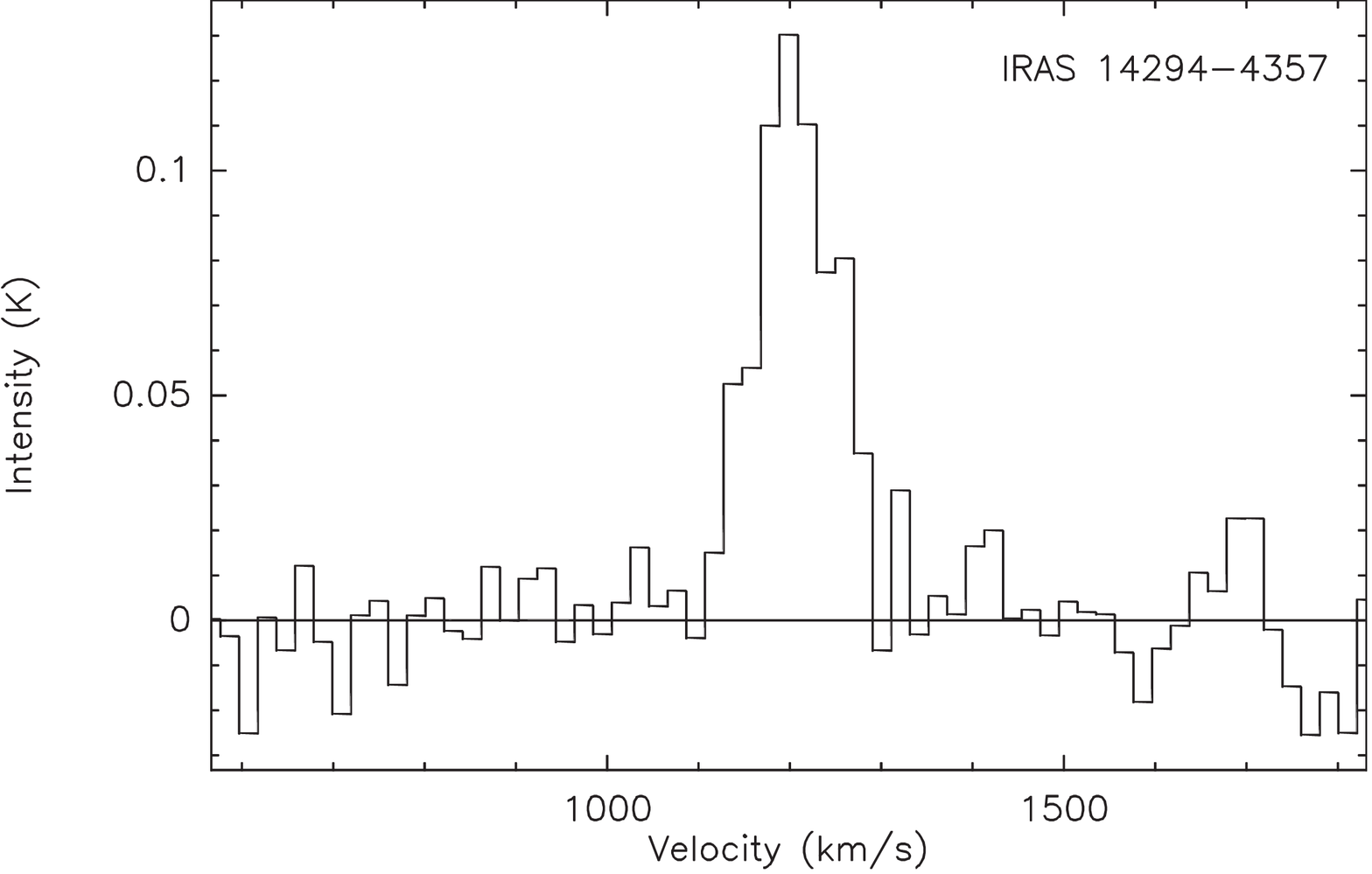}
\caption{continued}
\end{figure*}

\clearpage
\begin{figure*}[!htbp]
\setcounter{figure}{0}
\centering
\vspace{1cm}
\includegraphics[scale=.22, angle = -90]{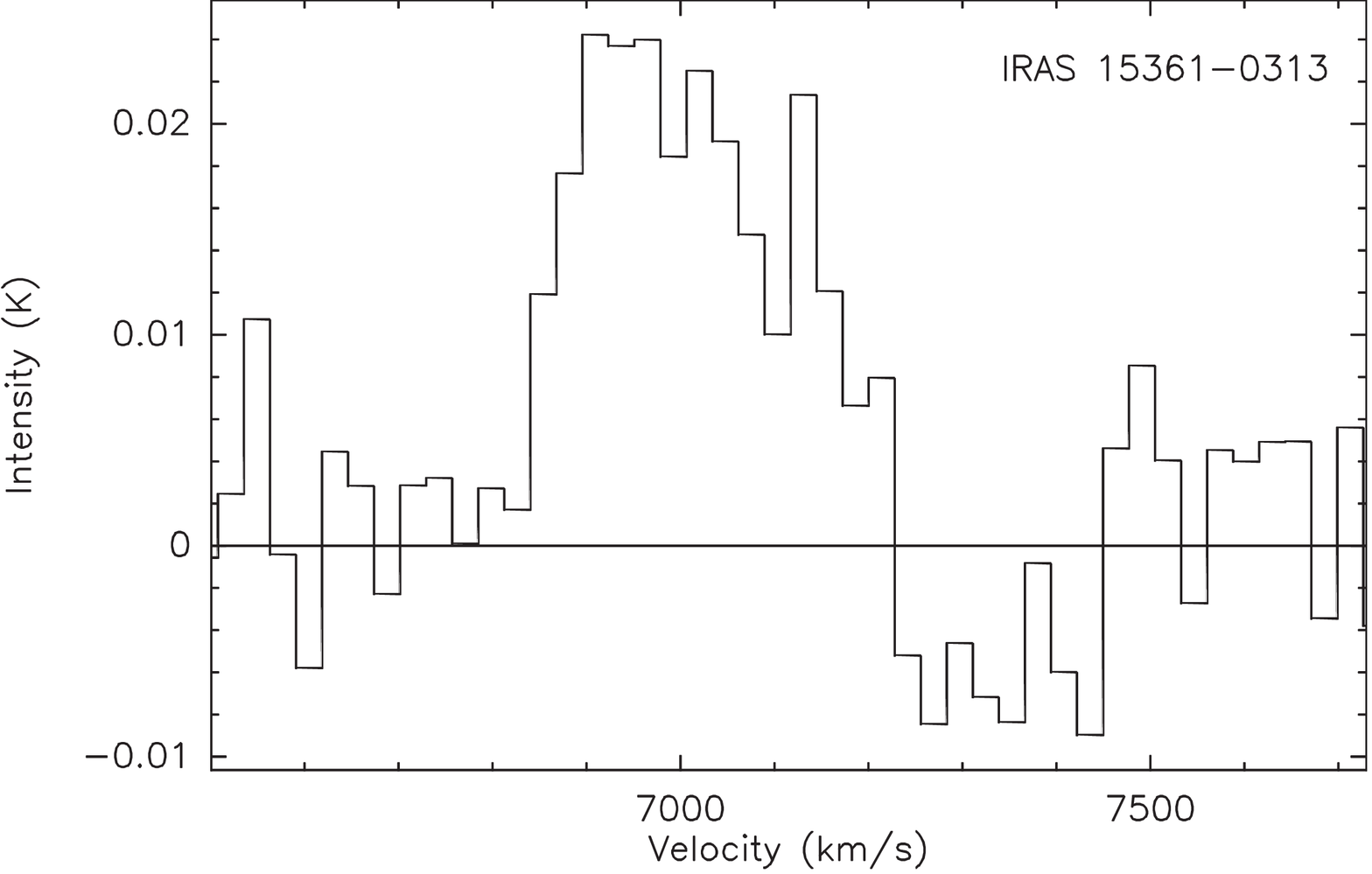}\hspace{.1cm}
\includegraphics[scale=.22, angle = -90]{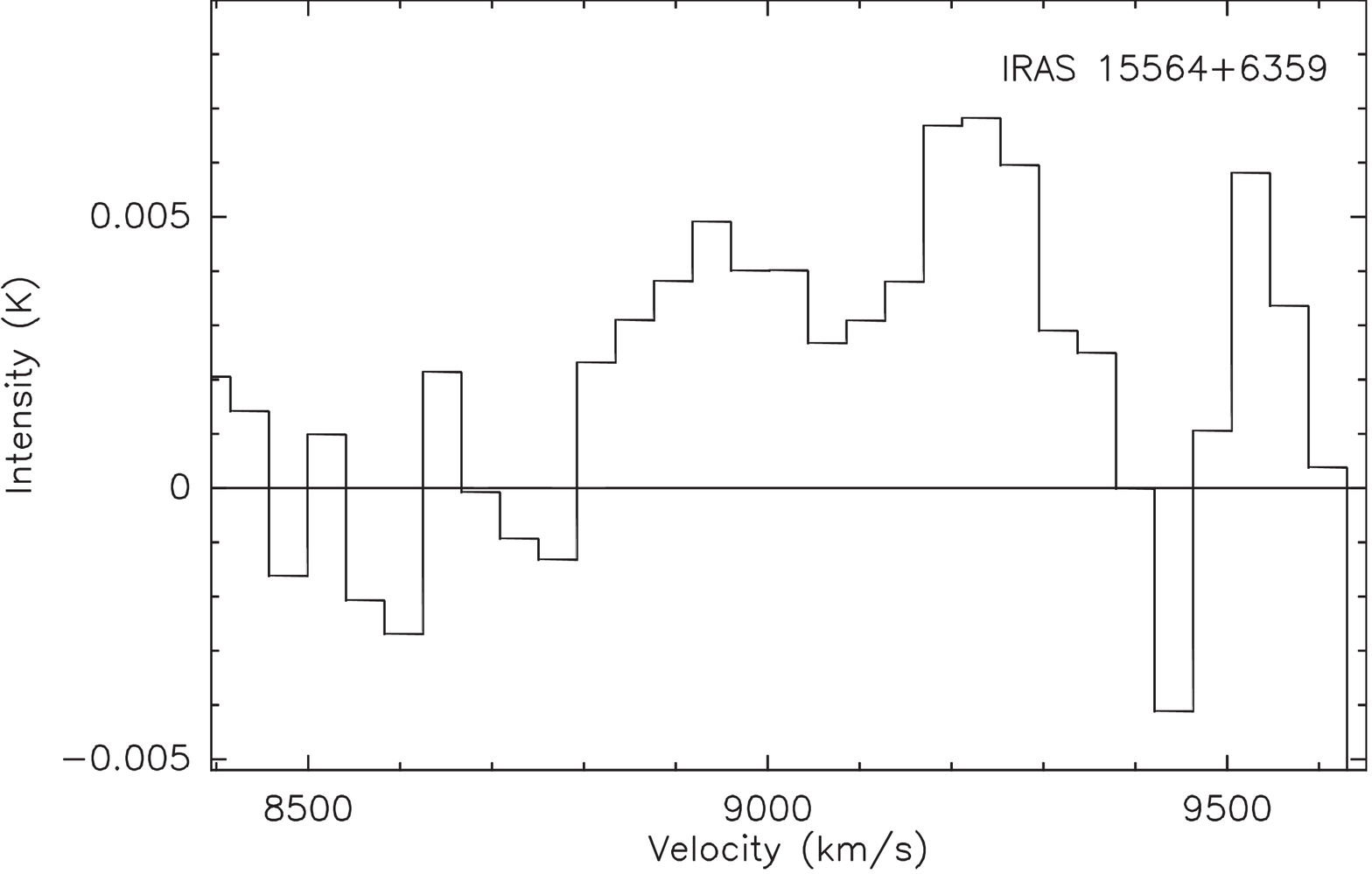}\hspace{.1cm}
\includegraphics[scale=.22, angle = -90]{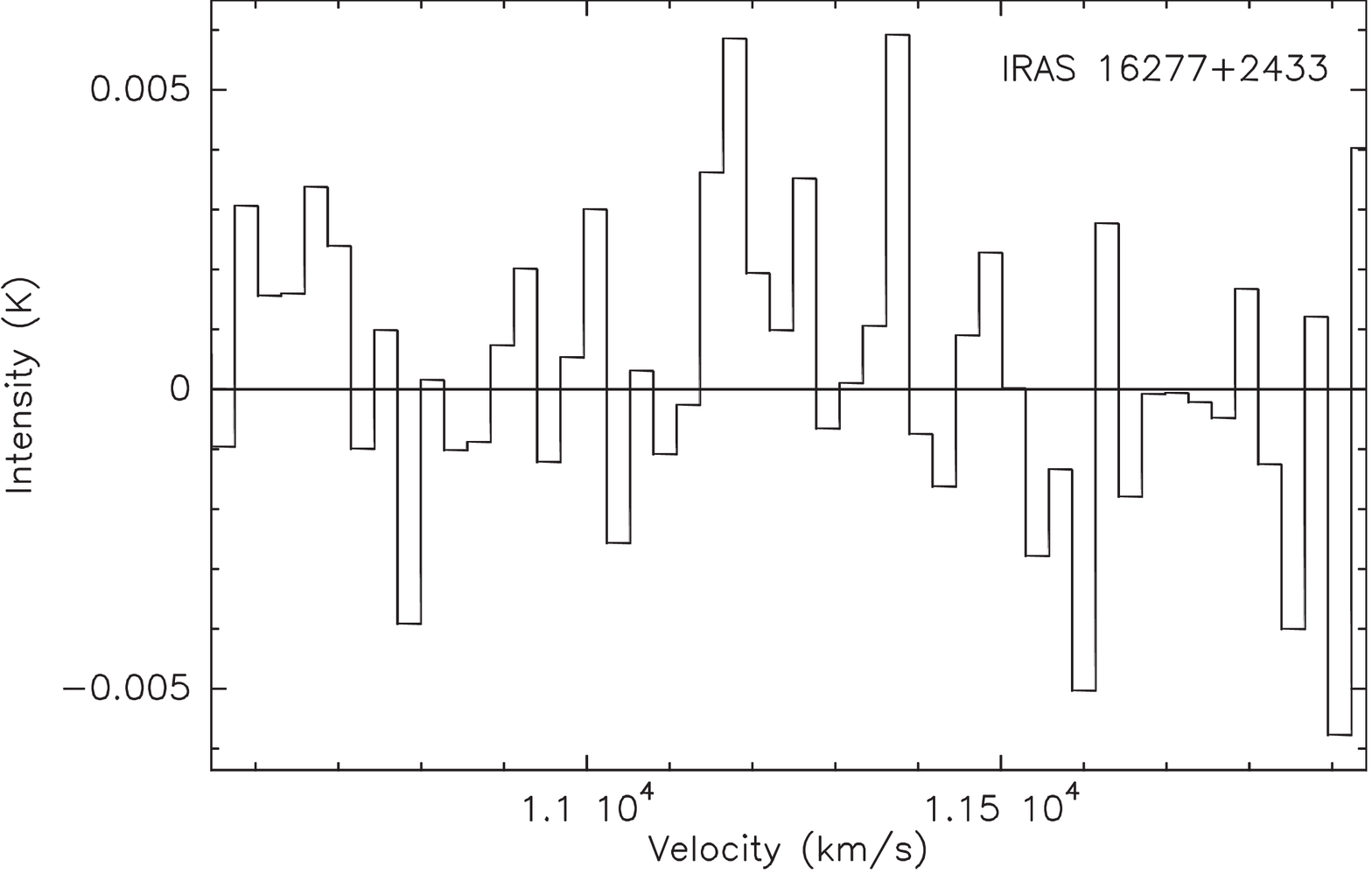}\\[20pt]

\includegraphics[scale=.22, angle = -90]{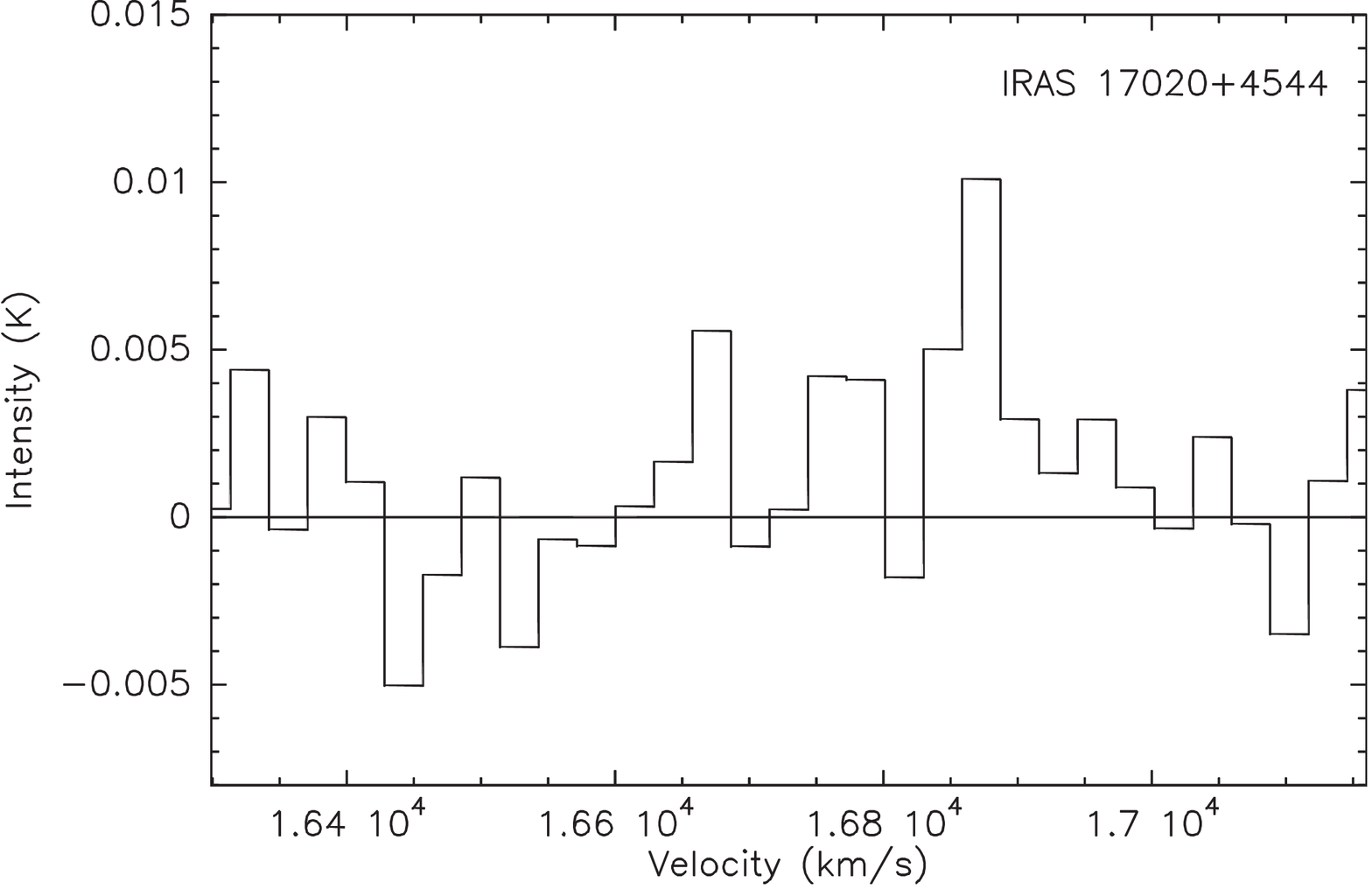}\hspace{.1cm}
\includegraphics[scale=.22, angle = -90]{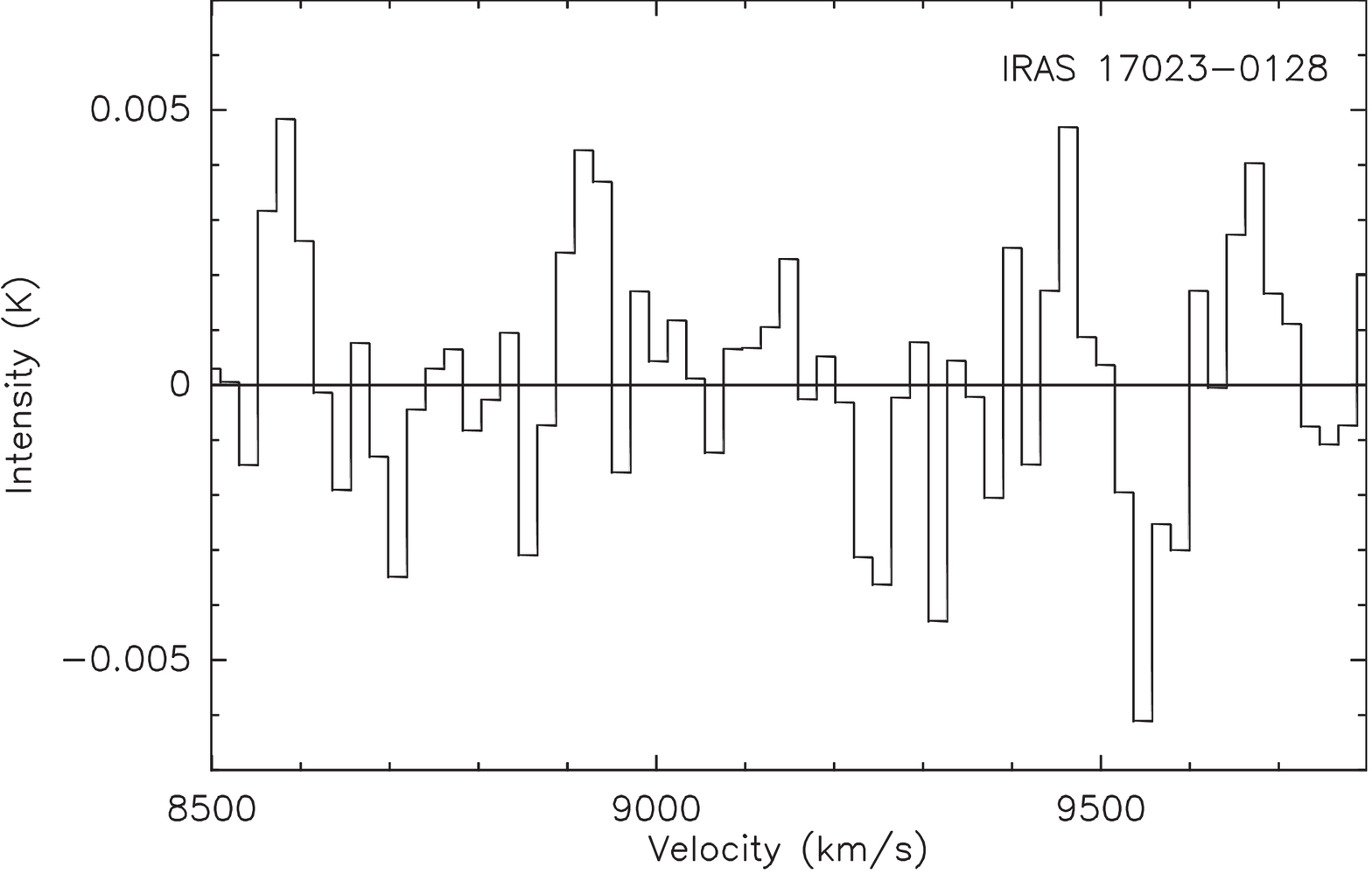}\hspace{.1cm}
\includegraphics[scale=.22, angle = -90]{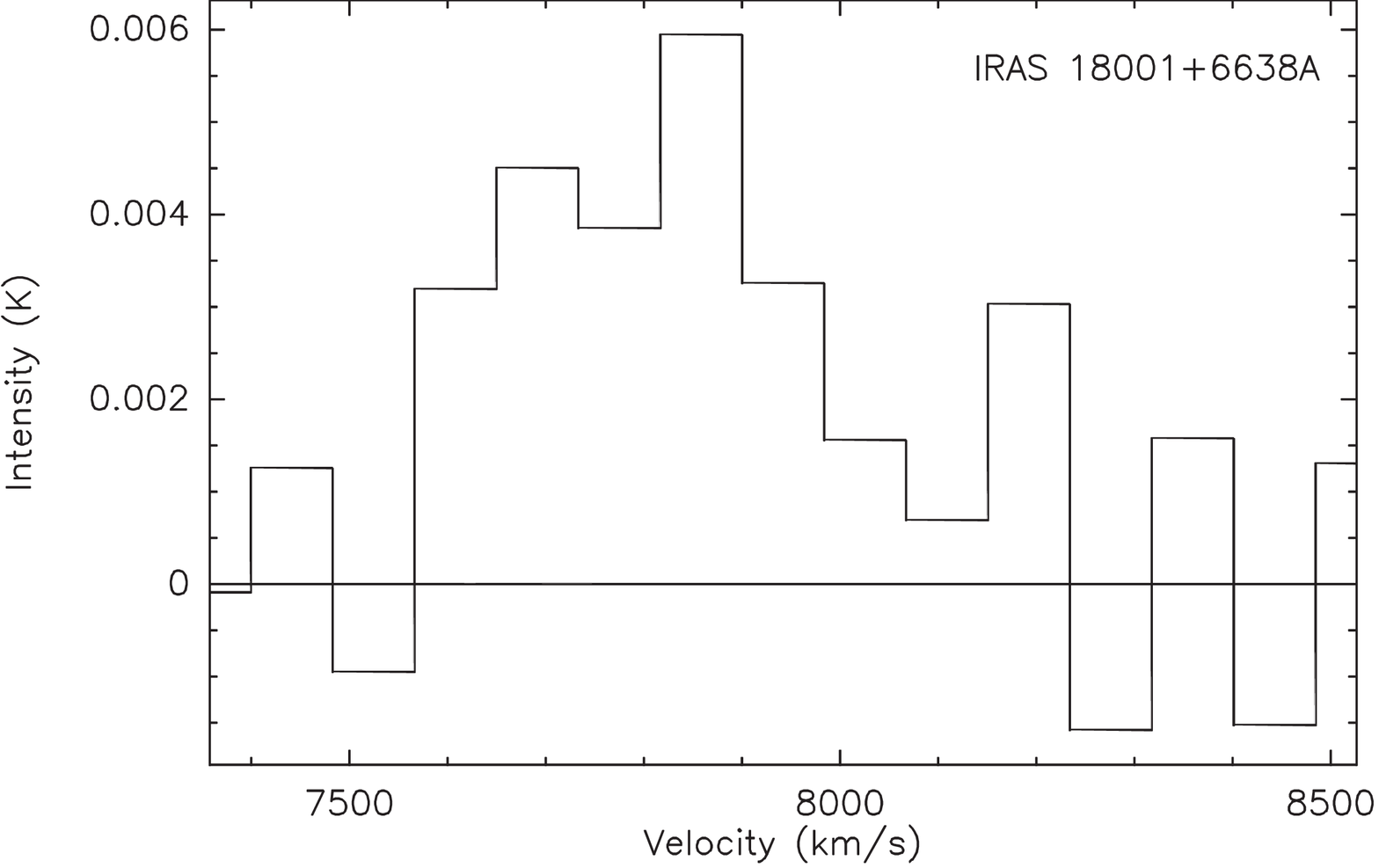}\\[20pt]

\includegraphics[scale=.22, angle = -90]{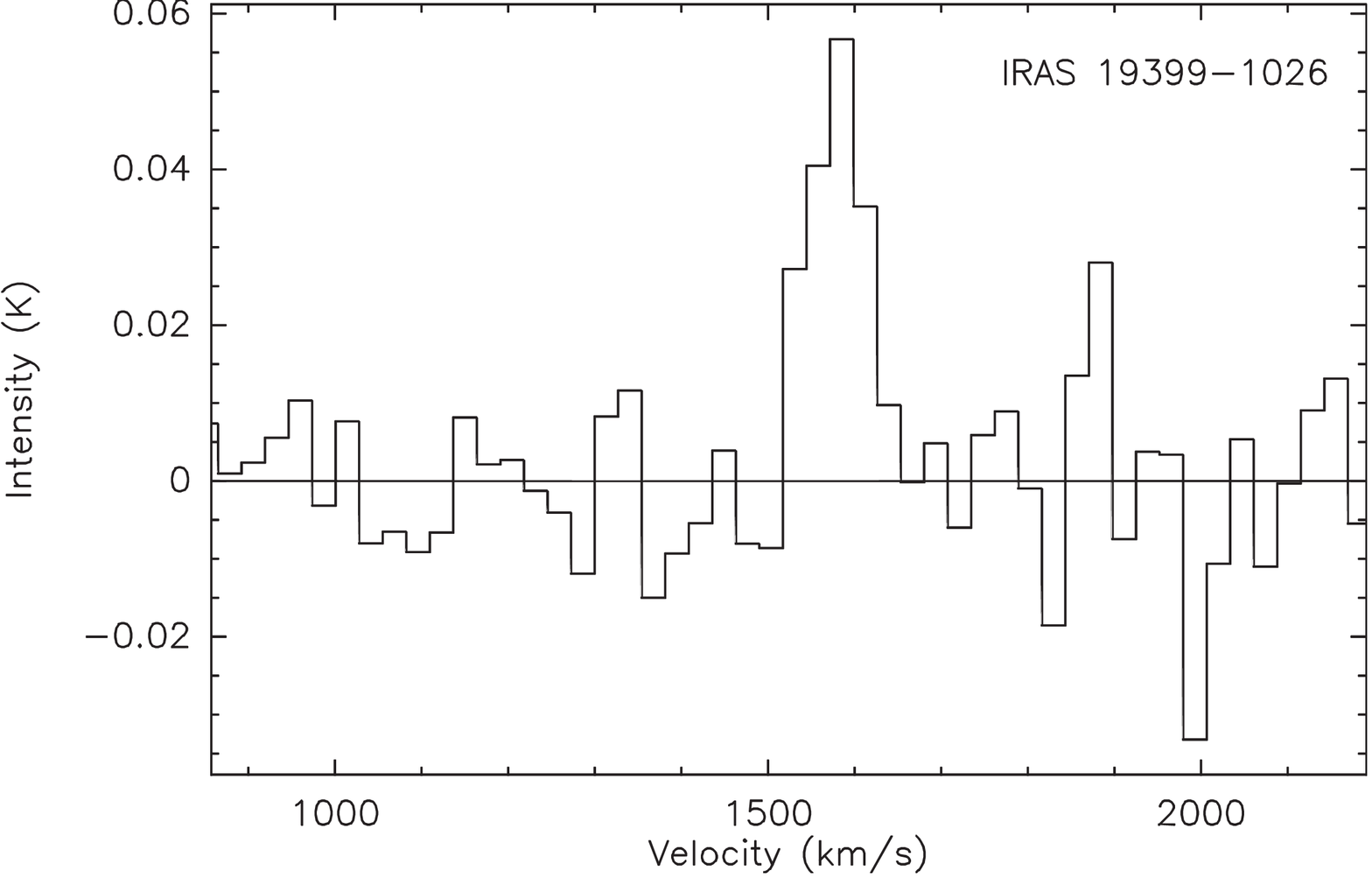}\hspace{.1cm}
\includegraphics[scale=.22, angle = -90]{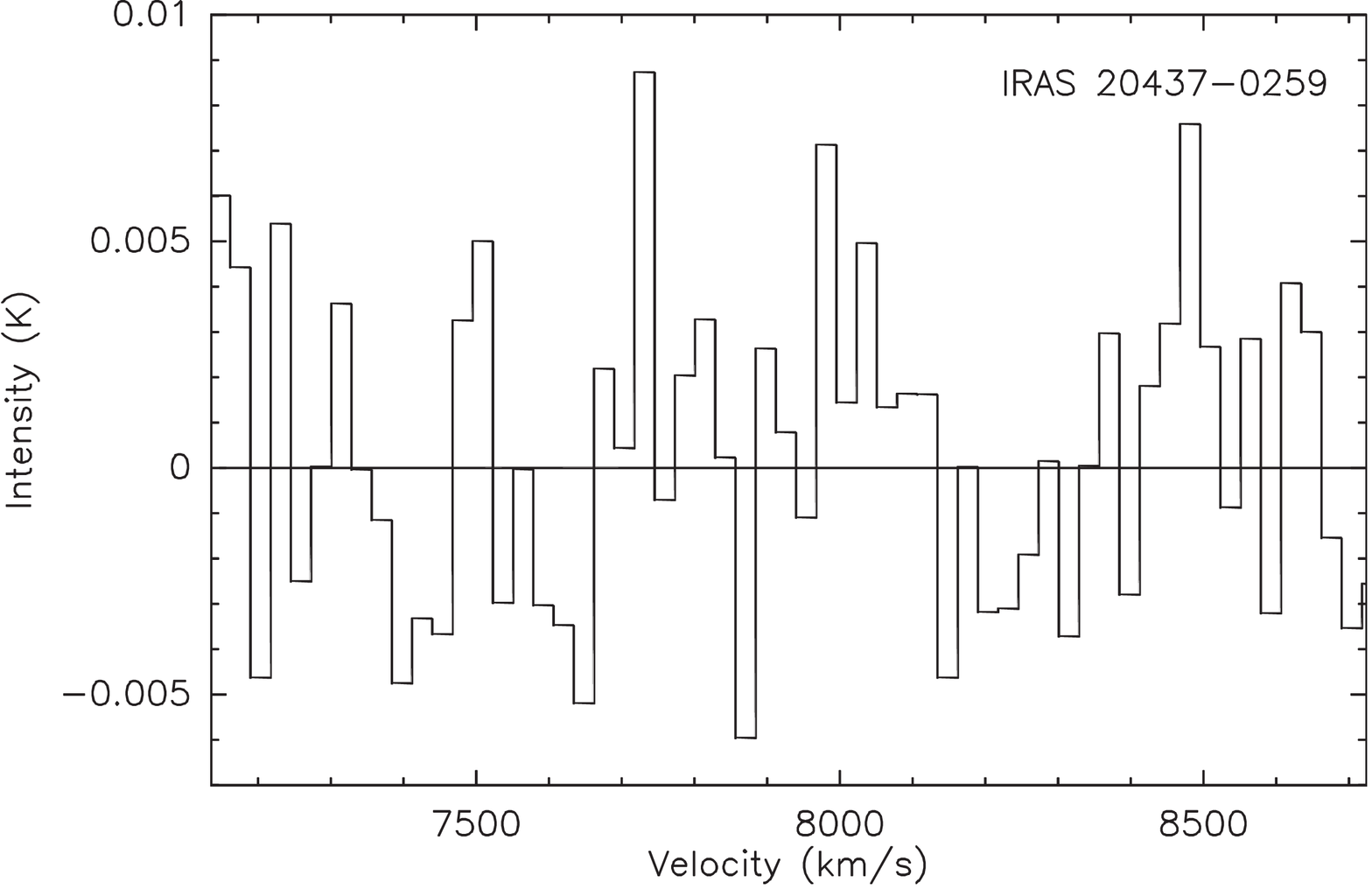}\hspace{.1cm}
\includegraphics[scale=.22, angle = -90]{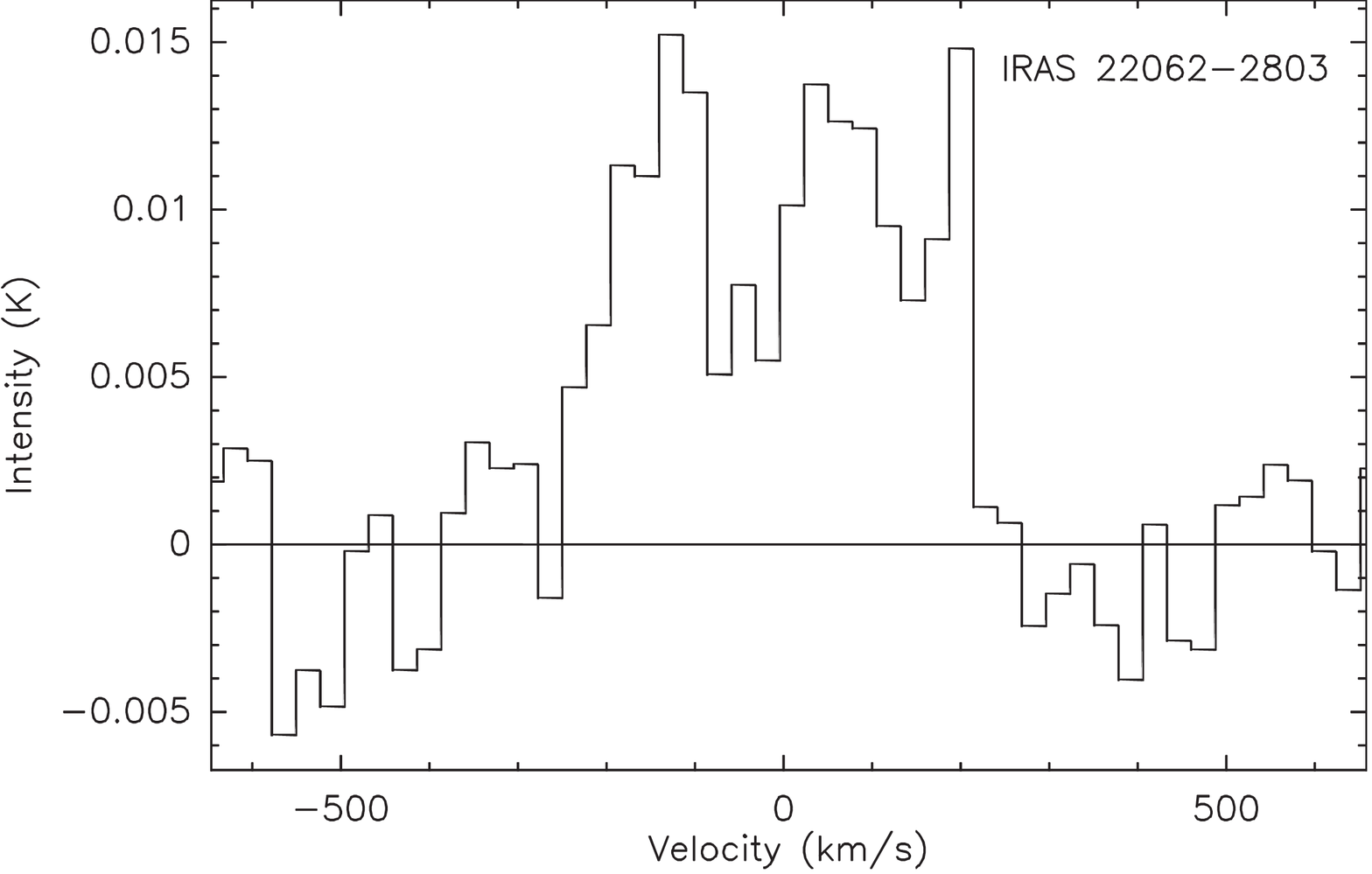}\\[20pt]

\includegraphics[scale=.22, angle = -90]{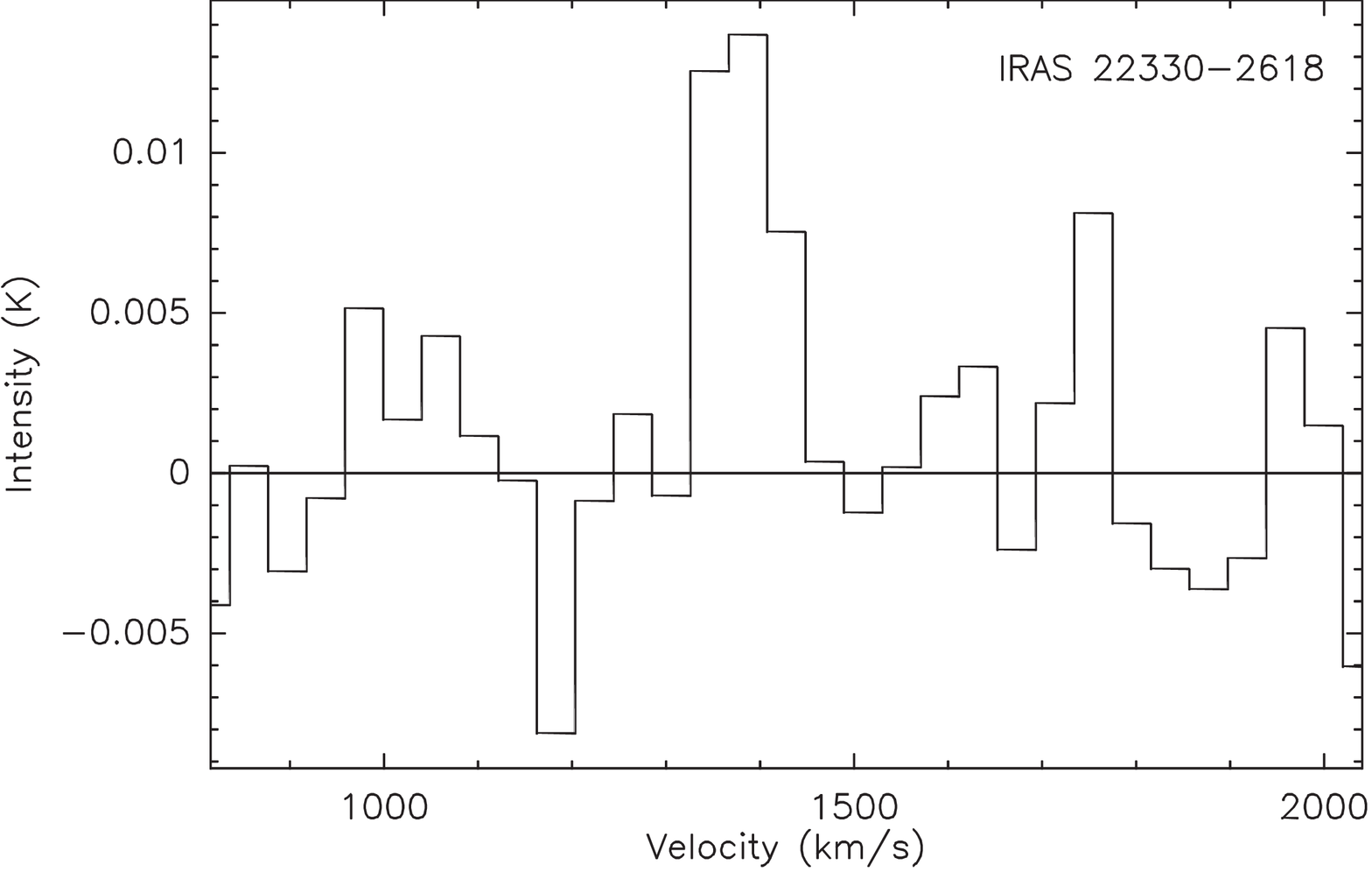}\hspace{.1cm}
\includegraphics[scale=.22, angle = -90]{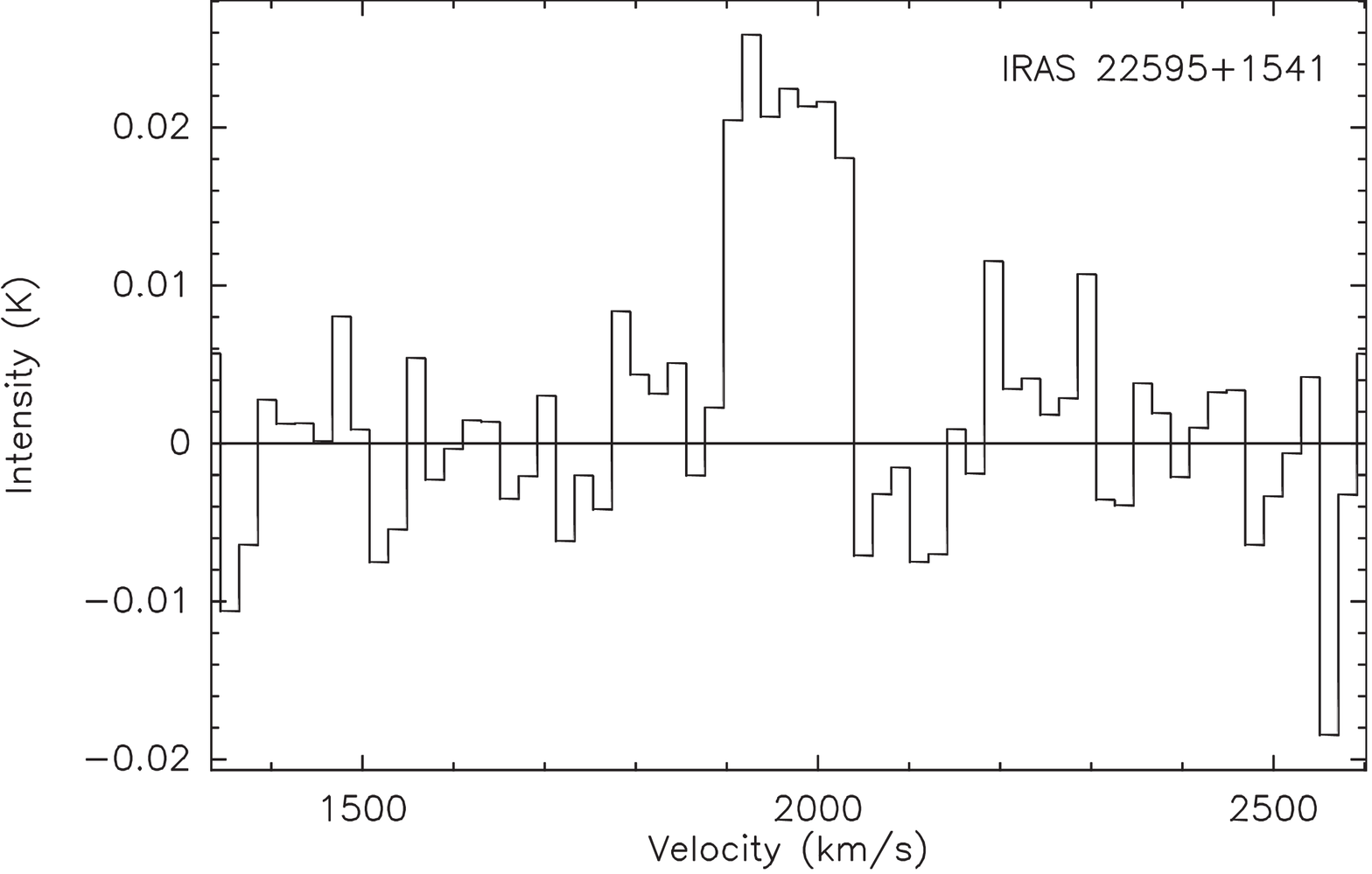}\hspace{.1cm}
\includegraphics[scale=.22, angle = -90]{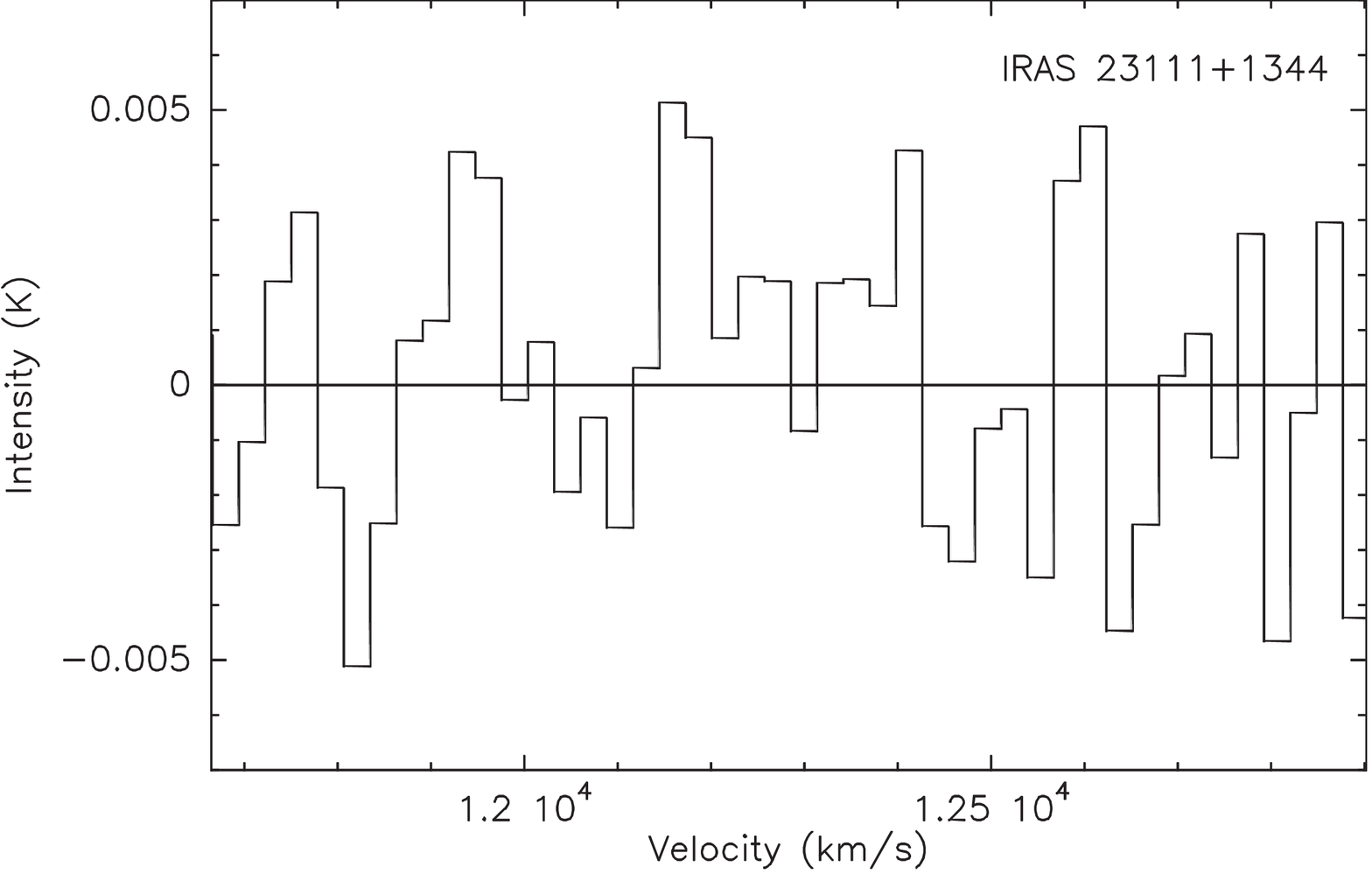} \\[20pt]

\includegraphics[scale=.22, angle = -90]{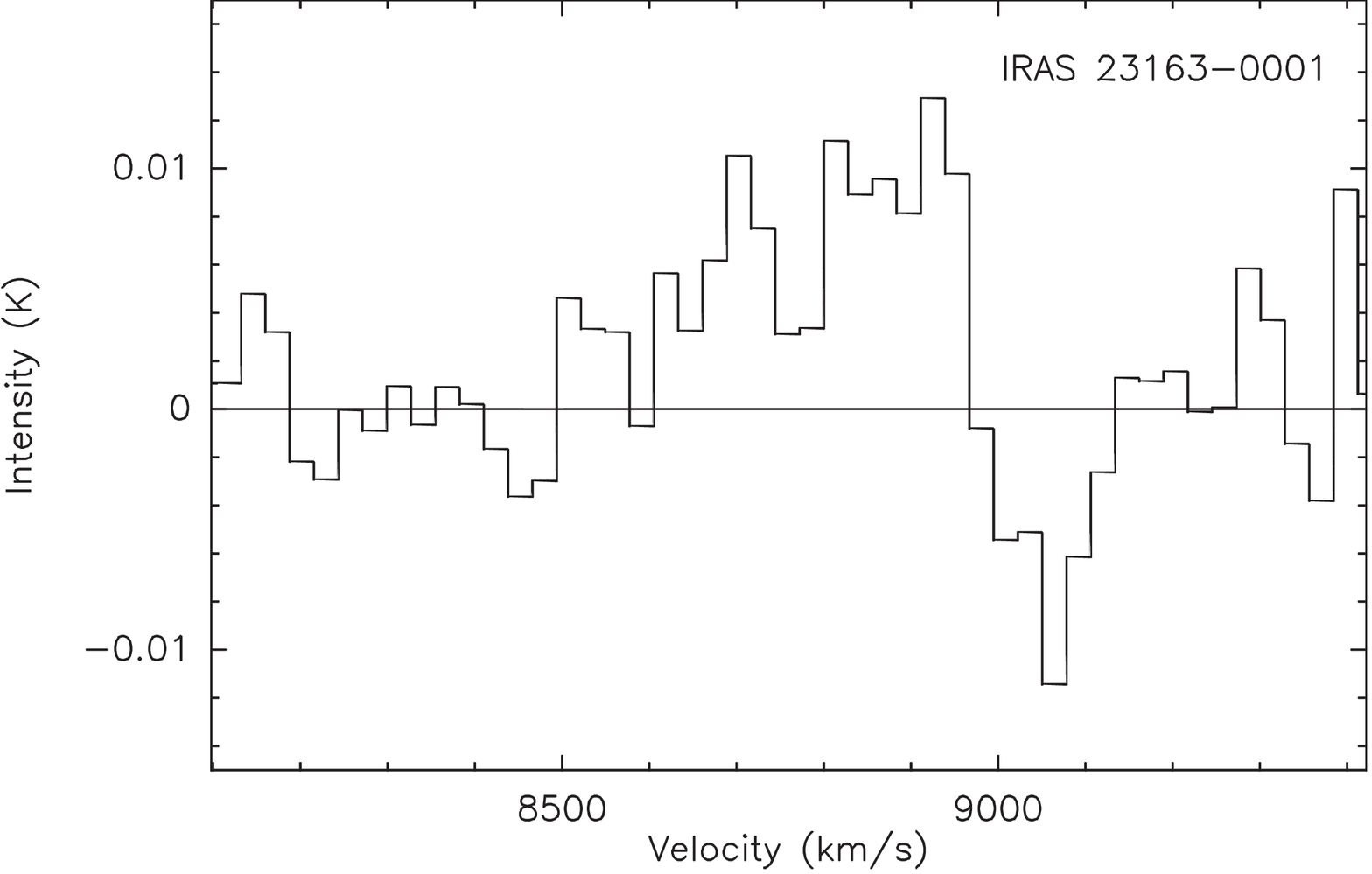}\hspace{.1cm}
\includegraphics[scale=.22, angle = -90]{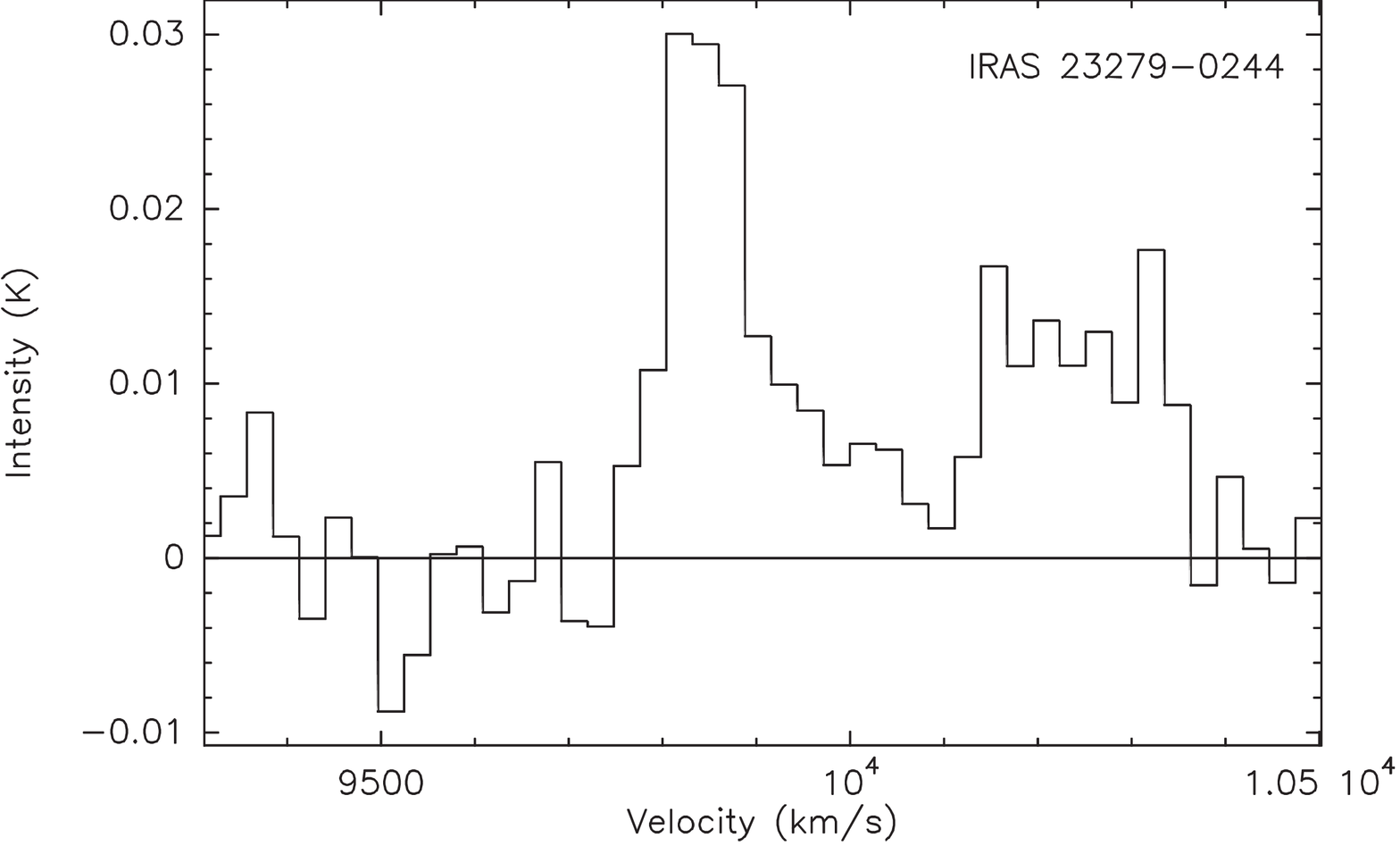}\hspace{.1cm}
\includegraphics[scale=.22, angle = -90]{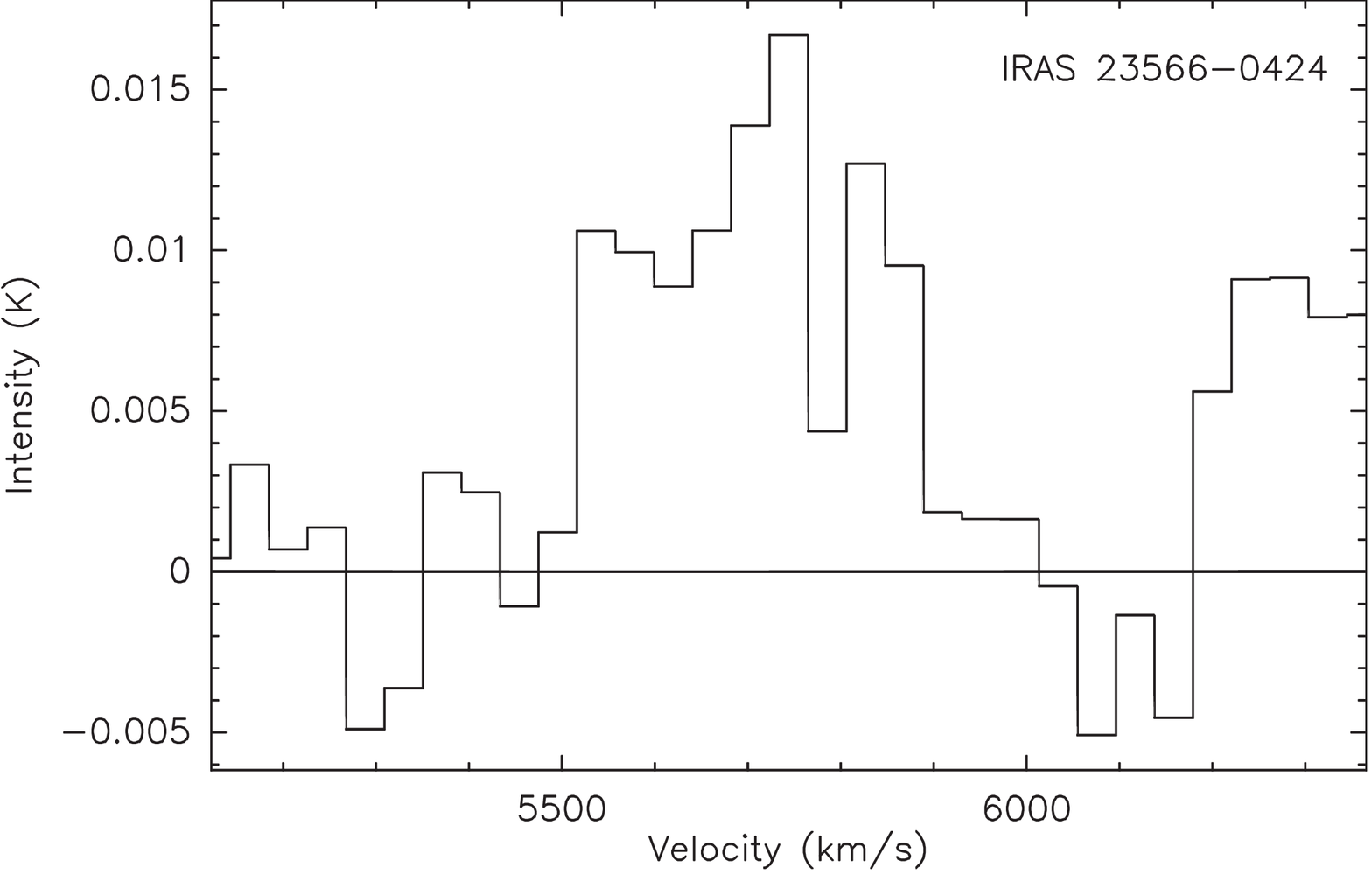}
\caption{continued}
\end{figure*}

\clearpage

\begin{figure}[h!]
%\resizebox{9cm}{!}{
\center
\includegraphics[angle=0,scale=.75]{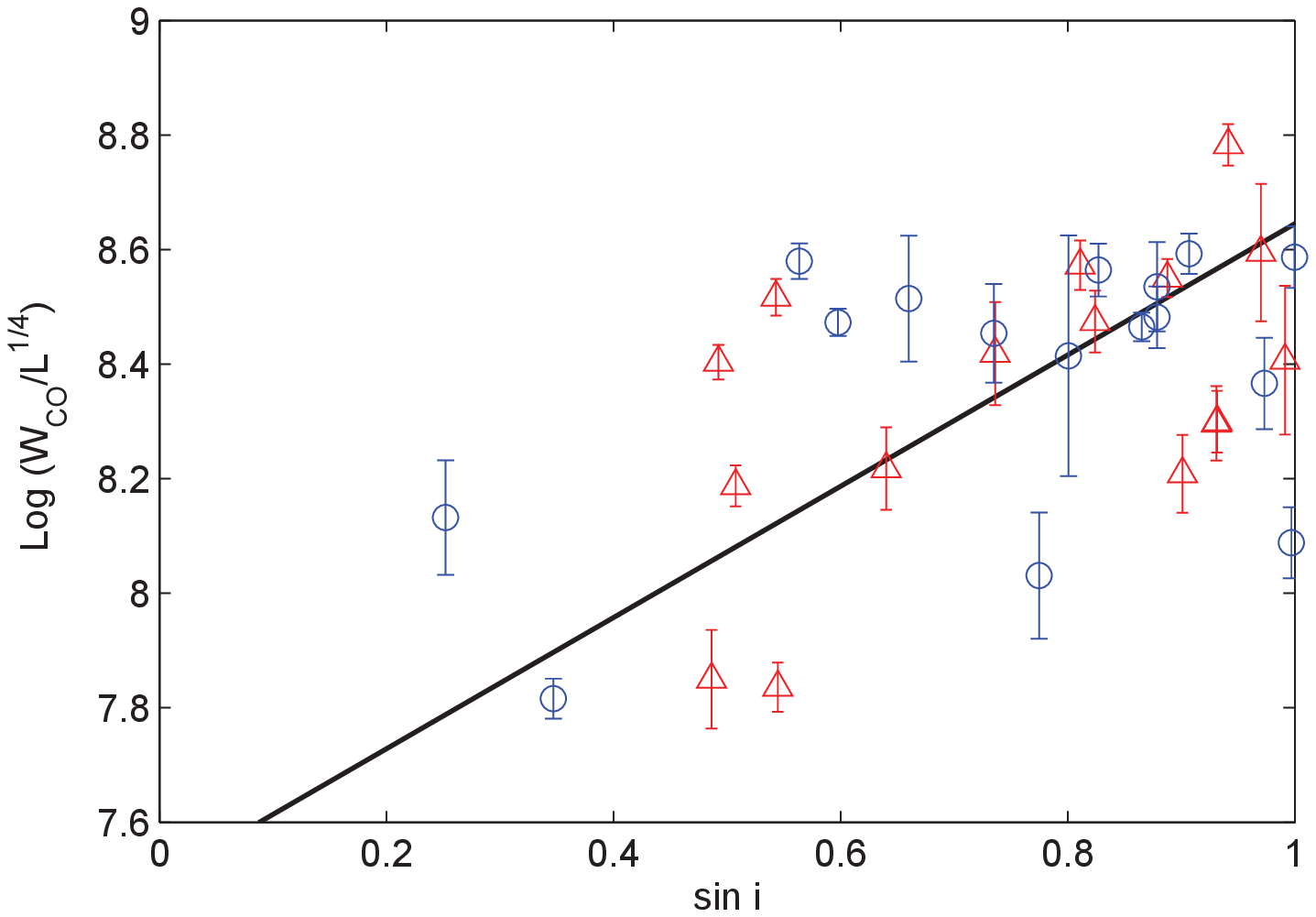}%Wco_sini_Lcorr_errors.eps
\caption{\label{CO line width} CO line width $[km \cdot s^{-1}]$ blue luminosity corrected versus the sine of the galaxy inclination for the observed sample, Sy1s (circles) and Sy2s (triangles). The CO line width and sin i are correlated with a 0.49 factor and a 0.21 of null probability (see table \ref{tbl-3}). The straight line represents the least squares fit applied to the whole sample (both Sy1s and Sy2s)}
\label{Wco-sini}
\end{figure}

\clearpage

\begin{figure}[b!]
\center
\centering
%\vspace{1cm}
\includegraphics[scale=.55]{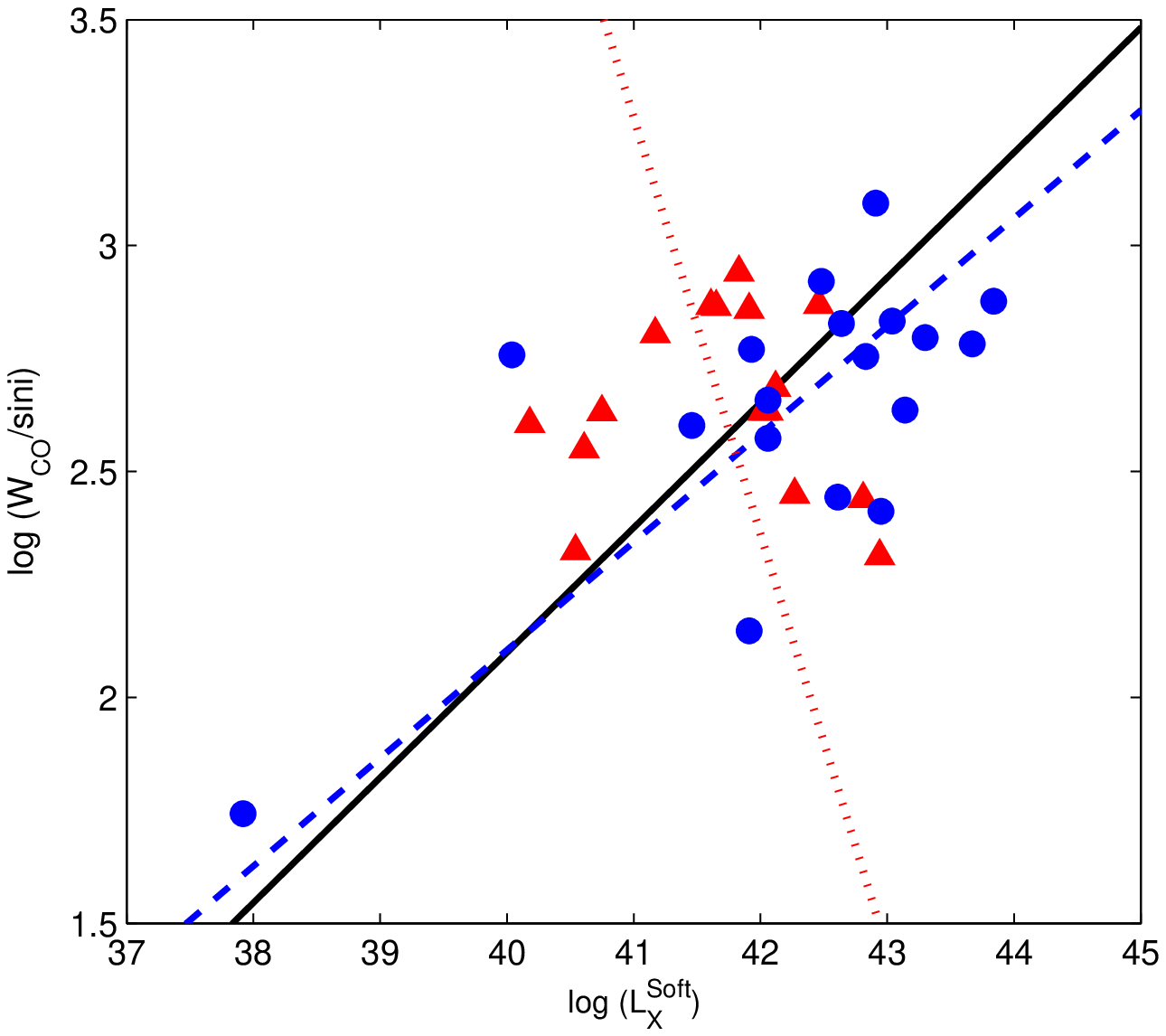}\hspace{.01cm}
\includegraphics[scale=.55]{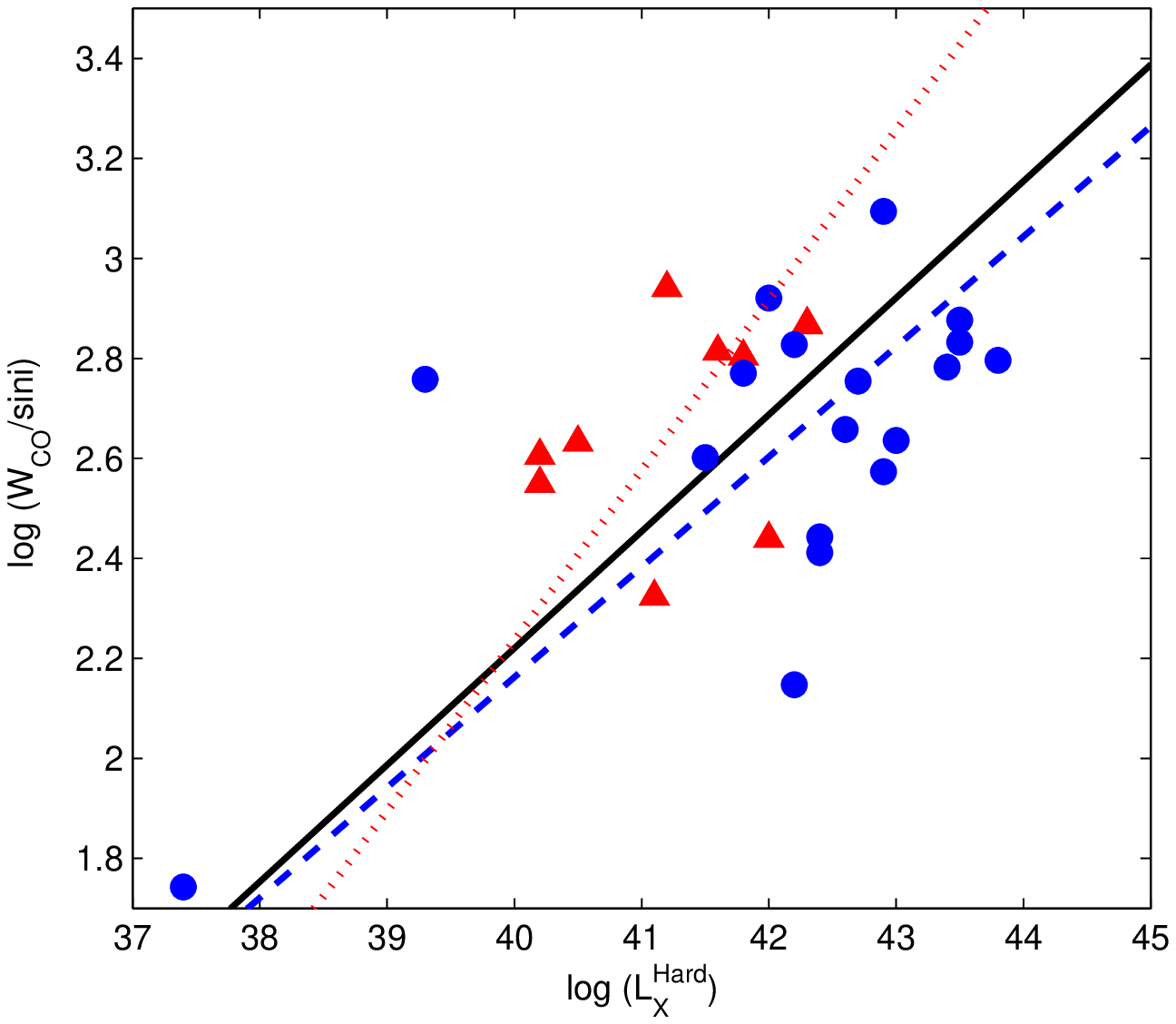}\\
\caption{\label{Wco-lx}(Left) CO line width corrected for inclination $[km \cdot s^{-1}]$ versus soft (0.1--2.4 keV) X--ray luminosity $[ergs \cdot s^{-1}]$ for the detected Sy1s (circles) and Sy2s (triangles).(Right) CO line width corrected for inclination $[km \cdot s^{-1}]$ versus hard (0.3--8 keV) X--ray luminosity $[ergs \cdot s^{-1}]$ for the detected Sy1s (circles) and Sy2s (triangles). Least-squares fits to the data are indicated by the blue dashed (Sy1s), red dotted (Sy2s) lines and solid (both Sy1s and Sy2s). The correlation factor for each group and null probability are shown in Table~\ref{tbl-3}.}

\end{figure}
\clearpage

\begin{figure}[hb!]
%\resizebox{9cm}{!}{
\center
\includegraphics[angle=0,scale = 0.75]{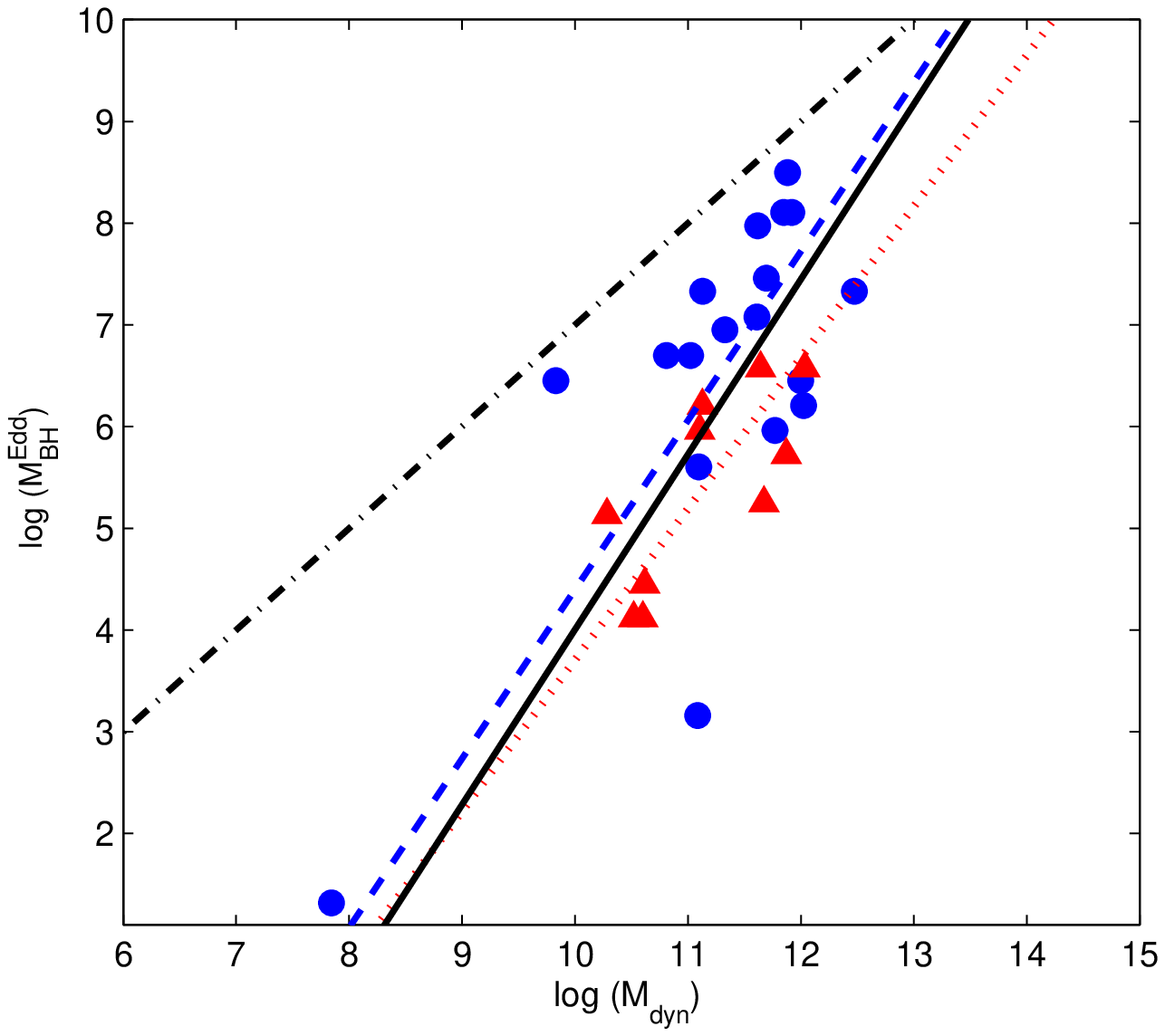}%Mbh_Mdyn_size.eps}}
\caption{\label{masses} Black hole mass M$^{Edd}_{BH}$ [M$_{\odot}]$ versus the dynamical mass of the galaxy M$_{dyn}$ [M$_{\odot}]$ for both type 1 (filled circles) and type 2 (triangles) Seyfert galaxies.  Least-squares fits to the data are indicated by the blue dashed (Sy1s), red dotted (Sy2s) lines and solid (both Sy1s and Sy2s). The correlation factor for each group and null probability are shown in Table~\ref{tbl-3}. The M$_{BH}$-M$_{bulge}$ relation for normal galaxies, i.e.   $\left\langle M_{BH}/M_{bulge}\right\rangle$ $\sim$ 0.001, is indicated by the dot-dashed line. }
\label{masses}
\end{figure}
\clearpage
\begin{figure}[h!]
\center
%\resizebox{9cm}{!}{
\includegraphics[angle=0,scale = 1]{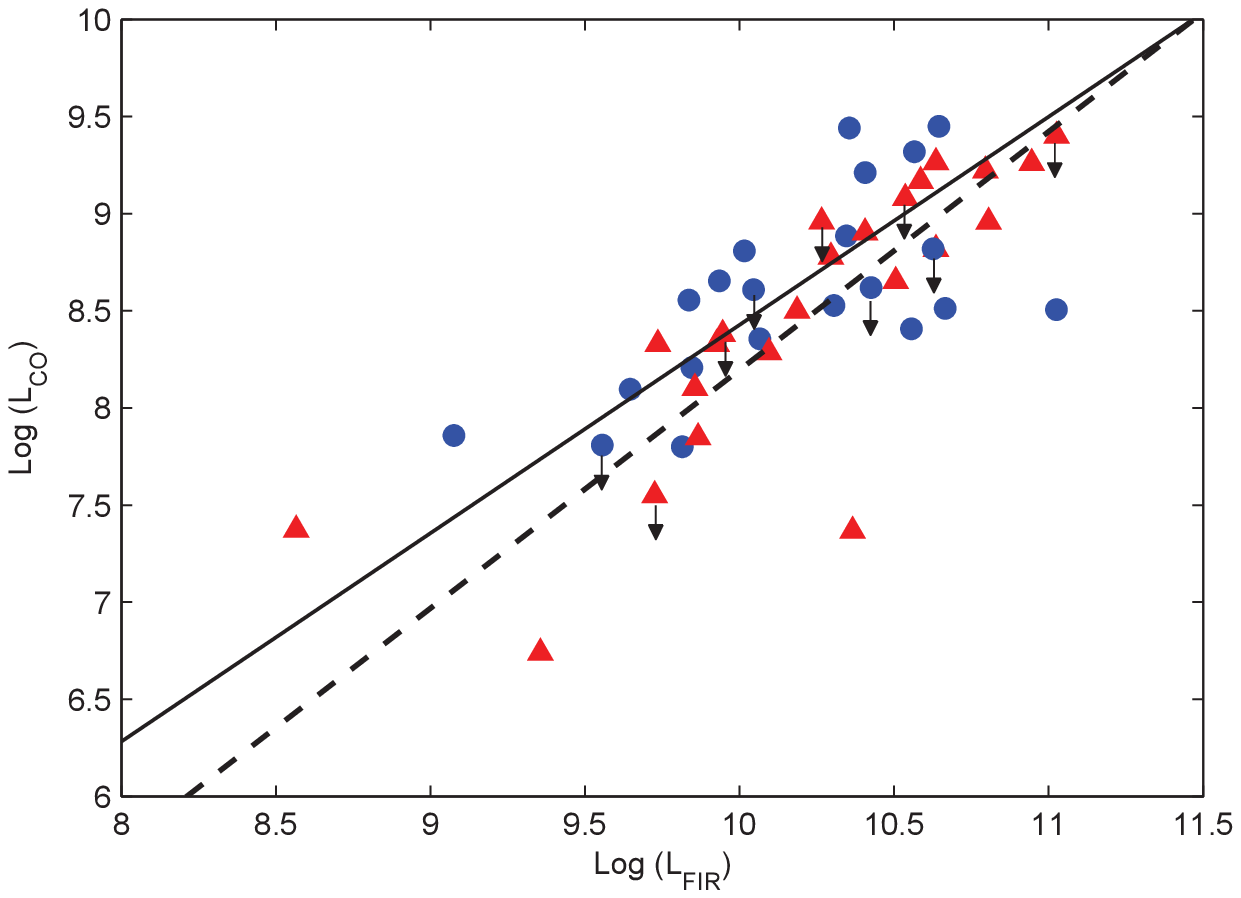}
\caption{\label{Lco-Lfir} Log $L_{CO}$ $[K \cdot km \cdot s^{-1} \cdot pc^{2}]$ vs $Log ~L_{FIR}$ $[L_{\odot}]$ for the observed Sy1s (circles) and Sy2s (triangles). Least-squares fits to the data are indicated by the solid (Sy1s) and dashed (Sy2s) lines. Upper limit of the non-detection are plotted as well. The correlation factor for each group and probability of not being correlated are shown in Table \ref{tbl-3}.}
\end{figure}

\clearpage

\begin{figure}[t!]
%%\resizebox{9cm}{!}{
\center
\includegraphics[angle=0,scale = 0.55]{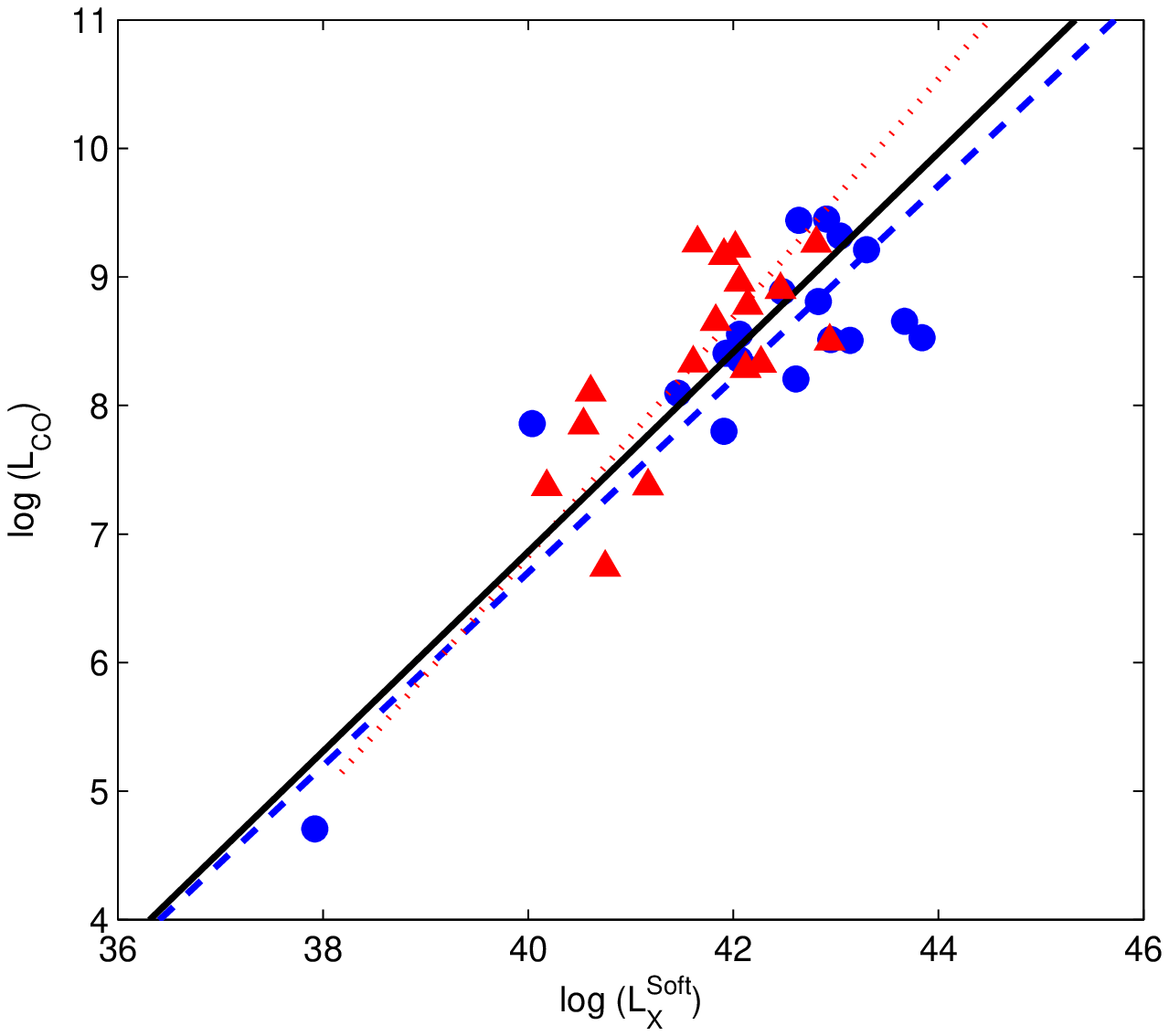}
\includegraphics[angle=0,scale = 0.55]{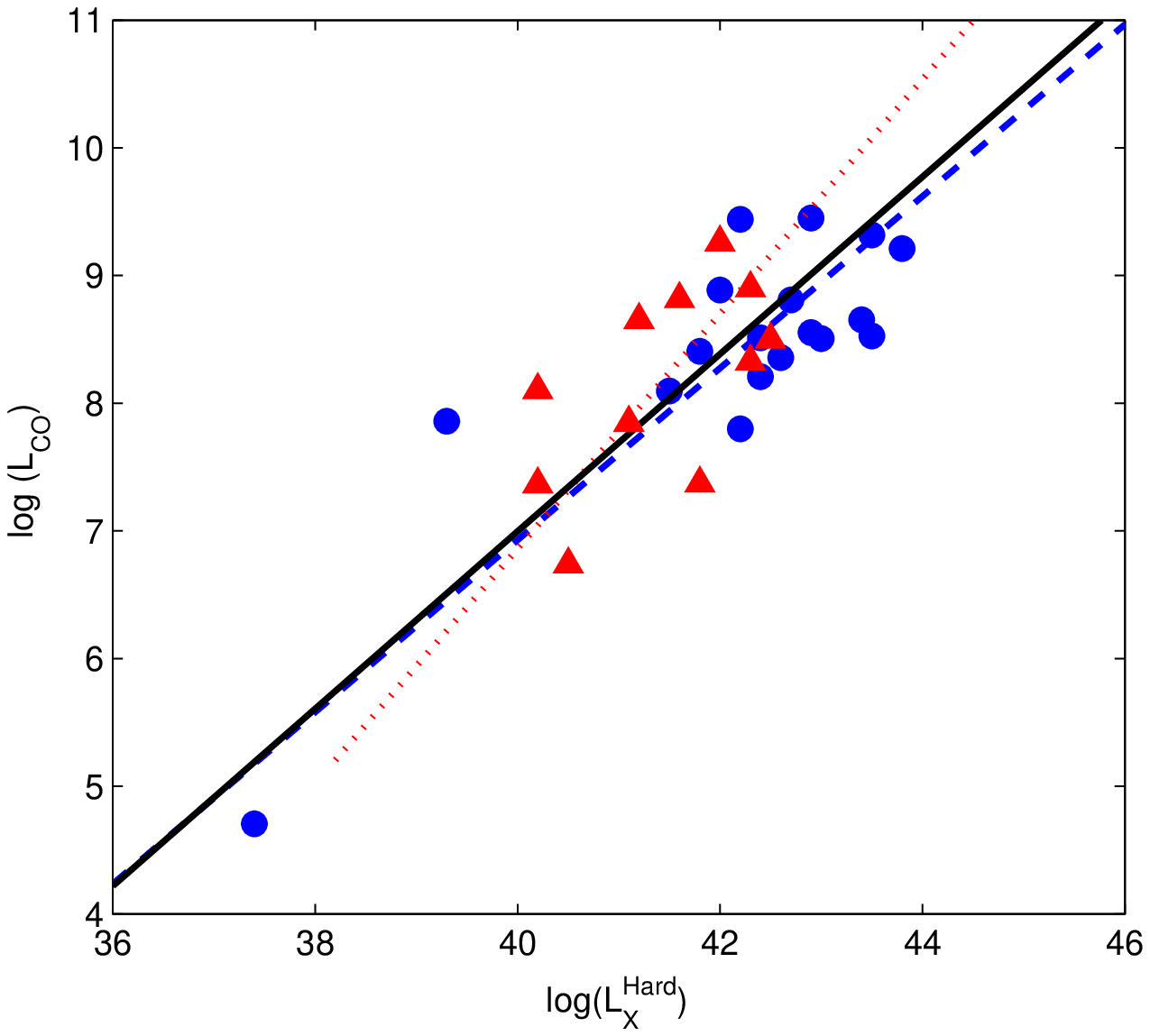}
\caption{\label{Lco_Lx}(Left) CO luminosity [K $\cdot km \cdot s^{-1} \cdot pc^{2}]$ versus soft (0.1--2.4 keV) X--ray luminosity $[erg \cdot s^{-1}]$ for the observed Sy1s (circles) and Sy2s (triangles). (Right) CO luminosity [K $\cdot km \cdot s^{-1} \cdot pc^{2}]$ versus hard (0.3--8 keV) X--ray luminosity $[erg \cdot s^{-1}]$. Least-squares fits to the data are indicated by the blue dashed (Sy1s), red dotted (Sy2s) lines and solid (both Sy1s and Sy2s). The correlation factor for each group and null probability are shown in Table~\ref{tbl-3}.}
\end{figure}

%\clearpage

\begin{deluxetable}{ccrccl}
\tabletypesize{\scriptsize}
%\rotate
\tablecaption{Summary of Galaxies \label{tbl-1}}
\tablewidth{421pt}
\tablehead{
\colhead{Galaxy}&\colhead{Name}& \colhead{Type} & \colhead{Velocity} & \colhead{Redshift} &\colhead{i$^{a}$}
\\
 &  &     & (km/s)     &         & $(~^{\circ})$
}
\startdata
IRAS 00076-0459 & MRK 0937&SBa/b, HII, Sy2& 8846 & 0.029507 & 29.1\\[2pt]
IRAS 00317-2142& ESO 540- G 001&SB(r1)bc, Sy1.8 &  8048 & 0.026845&33\\[2pt]
IRAS 04260+6444&NGC 1569&IBm;Sbrst, Sy1&-104&-0.000347&64.7\\[2pt]
IRAS 04565+0454&UGC 03223 &SBa  Sy1 & 4683 & 0.015621   & 61.5\\[2pt]
IRAS 05128+5308&CGCG 258-006 & S, Sy1.8 & 8482 & 0.028293 &  68.6\\[2pt]
IRAS 05497-0728&NGC 2110 &SAB0-, Sy2 & 2335 & 0.007789 &  46.4\\[2pt]
IRAS 06280+6342&UGC 03478 & Sb, Sy1.2 & 3828 & 0.012769 &  76.7\\[2pt]
IRAS 07388+4955& UGC 03973& SBb, Sy1.2 & 6652  &0.022189& 36.7\\[2pt]
IRAS 08331-0354& NGC 2617& Sy1.8 &  4261 & 0.014213&39.8\\[2pt]
IRAS 09585+5555&NGC 3079 &SB(s)c;LINER, Sy2&1116&0.00372&82.5\\[2pt]
IRAS 10126+7339&NGC 3147&SA(rs)bc;HII, Sy2&2820&0.009407&29.5\\[2pt]
IRAS 10291+6517&NGC 3259 & SAB(rs)bc:BLAGN, Sy1 & 1686 & 0.005624  &60.7\\[2pt]
IRAS 10295-1831&2MASX J10315733-1846333 &Sp, Sy1 & 12070 & 0.040261 &20.3\\[2pt]
IRAS 10589-1210& MCG -02-28-039&  Sy1.5&7720&  0.025751 & 59.9\\[2pt]
IRAS 11033+7250& NGC 3516& (R)SB(s): Sy1.5&2649&  0.008836& 37.0\\[2pt]
IRAS 11083-2813&ESO 438- G 009 &RSB(rl)ab  Sy1.5 & 7199 & 0.024013 &50.8\\[2pt]
IRAS 11112+0951&IC 2637 &E+, pec, Sy1.5 & 8763 & 0.029230   &35\\[2pt]
IRAS 11210-0823&NGC 3660 &SB(r)bc Sy2 & 3683 & 0.012285  &32.9\\[2pt]
IRAS 11376+2458& NGC 3798& SB0, Sy1&3552 &  0.011848 & 90\\[2pt]
IRAS 11395+1033&NGC 3822 &Sb Sy2 & 6138 & 0.020474   &76.0\\[2pt]
IRAS 11500-0455& MCG -01-30-041& (R)SB(rs)ab, Sy1.8 &  5641 & 0.018816&55.5\\[2pt]
IRAS 12373-1120& MESSIER 104&SA(s)a;LINER, Sy1.9&1024&0.003416&78.5\\[2pt]
IRAS 12393+3520& NGC 4619&SB(r)b pec? Sy1&6927&0.023106&14.6\\[2pt]
IRAS 12409+7823& 2MASX J12423600+7807203&Sy1.9&6625&0.022100&75.7\\[2pt]
IRAS 12495-1308&NGC 4748&Sa Sy1&4386&0.014630&53.2\\[2pt]
IRAS 13218-1929& 2MASX J13243528-1945114&$(R'_1)$SB(l:)a, Sy1.9& 5284  & 0.017625&$\cdots$\\[2pt]
IRAS 14060+7207& NGC 5643 & Sy 1.9 &  10251 &0.034194& 54.2\\[2pt]
IRAS 14105+3932&NGC 5515 &  Sab, Sy1.9 & 7719  & 0.025749&62.6\\[2pt]
IRAS 14156+2522&NGC 5548 &  (R')SA(s)0/a   Sy1.5 & 5149  & 0.017175&41.4\\[2pt]
IRAS 14294-4357& NGC 5643 & SAB(rs)c, Sy2 &  1199 &0.003999& 30.5\\[2pt]
IRAS 15361-0313& CGCG 022-021&HII, Sy1.9 &  7137 &0.023806& 68.7\\[2pt]
IRAS 15564+6359& CGCG 319-034& Sy1.9 & 9023  &0.030097& 54.2\\[2pt]
IRAS 16277+2433& VV 807& Sy1.9 & 11241  &0.037496& 57.2\\[2pt]
IRAS 17020+4544& B3 1702+457 & SBab;Sy1,Sy2 & 18107  &0.060400& 59.9\\[2pt]
IRAS 17023-0128& UGC 10683 & S0+ pec: Sy1.5 & 9149  & 0.030518& 41.3\\[2pt]
IRAS 18001+6638 &NGC 6552 &SB?, Sy2  & 7942  & 0.026492&47.4\\[2pt]
IRAS 19399-1026& NGC 6814& SAB(rs)bc;HII, Sy1.5&  1563 &0.005214& 85.6\\[2pt]
IRAS 20069+5929&CGCG 303-017& Sbrst, Sy2  & 11132 &0.037132&48.8\\[2pt]
IRAS 20437+5929&MRK 0896& SBb, Sy1& 7922 &0.026424&64.0\\[2pt]
IRAS 22062-2803&NGC 7214  &SB(s)bc pec: Sy1.2 & 6934 & 0.023128  &47.3 \\[2pt]
IRAS 22330-2618& NGC 7314& SAB(rs)bc, Sy1.9 &  1428  & 0.004763&70.3\\[2pt]
IRAS 22595+1541&NGC 7465 & $(R')$SB(s): Sy2 &  1968 &0.006565&63.4 \\[2pt]
IRAS 23111+1344& NGC 7525& E, Sy1.5 & 12262  & 0.040900&30.6 \\[2pt]
IRAS 23163-0001&NGC 7603 & SA(rs)b: pec, Sy1.5 &   8851& 0.029524&61.5\\[2pt]
IRAS 23279-0244& UM 163&   SB(r)b pec:;HII, Sy1&10022   & 0.033430&65.1\\[2pt]
IRAS 23566-0424& IC 1490 & SA(rs)b pec:, Sy1 &  5744  & 0.019160&55.8\\[2pt]
\enddata
%% Text for table notes should follow after the \enddata but before
%% the \end{deluxetable}. Make sure there is at least one \tablenotemark
%% in the table for each \tablenotetext.
\tablenotetext{a}{Inclination to the line of sight obtained from the Hyperleda database.}
\end{deluxetable}

\clearpage

\begin{deluxetable}{cccccccccc}
\tabletypesize{\scriptsize}
\rotate
\tablecaption{Observational Parameters and Gaussian Line Fits \label{tbl-2}}
\tablewidth{0pt}%542pt}
\tablehead{
\colhead{Source}& \colhead{Size$^{a}$}& \colhead{$\Delta V^{b}$} & \colhead{$I^{c} =\int T^{*}_{A} dV$} & \colhead{$Intensity ^{d}$} & \colhead{$S_{CO}^{e}$} &\colhead{Log $L_{CO}^{f}$}& \colhead{Log $L_{FIR}^{g}$} & \colhead{$Log L_{x}^{h}$}& \colhead{$S_{X}^{i}$} \\
\\
 &(kpc)& (km $s^{-1}$) &  (K km $s^{-1}$)& (erg $cm^{-2}$ $s^{-1}$ $sr^{-1}$)& (Jy km $s^{-1}$)&(K km $s^{-1} pc^{2}$)& ($L_{\odot}$) &(ergs $s^{^-1}$)&(erg s$^{-1}$ cm$^{-2}$)
}
\startdata
IRAS 00076-0459 &16.71&100$\pm 20$&1.18 $\pm 0.2$ &2.15 $\cdot10^{-8}$$\pm3.6 \cdot10^{-9}$&73.1$\pm 12$&8.77$\pm 0.0059$&10.18&42.94 &$\cdots$\\[2pt]
IRAS 00317-2142&15.20&150$\pm 15$ & 5.45$\pm 0.9$&9.93$\cdot10^{-8}$$\pm 1.6 \cdot10^{-8}$&335.01$\pm 54$&9.35$\pm 0.0068$&10.94&42.81&7.49$\cdot10^{-13}$\\[2pt]
IRAS 04260+6444&0.19&50$\pm 7$&1.01$\pm 0.18$&1.85$\cdot10^{-8}$ $\pm 3.3 \cdot10^{-9}$&	 57.69	 $\pm 10$&	4.8 $\pm 0.075$&$\cdots$&37.92&1.62$\cdot$10$^{-13}$\\[2pt]
IRAS 04565+0454&8.85&400 $\pm 50$&1.68$\pm 0.2$&3.06$\cdot10^{-8}$ $\pm 3.6 \cdot10^{-9}$	 &99.82$\pm 12$	&	8.35$\pm 0.013$&10.06&	 42.06&$\cdots$\\[2pt]
IRAS 05128+5308&16.02&560 $\pm 60$&4.09$\pm 0.7$&	7.46$\cdot10^{-8}$  $\pm 1.3 \cdot10^{-8}$&	 252.84$\pm 44$&9.28$\pm 0.0067$&10.81&42.06&$\cdots$\\[2pt]
IRAS 05497-0728&4.41&$\cdots$&$\leq 1.08$&$\cdots$&$\cdots$&$\leq 7.55$&9.72&40.99&1.04$\cdot$10$^{-11}$\\[2pt]
IRAS 06280+6342&7.23&270 $\pm 50$&1.79$\pm 0.17$&3.26$\cdot10^{-8}$ $\pm 3.1 \cdot10^{-9}$ &105.46$\pm 10$&8.21$\pm 0.013$&9.84&42.61&$\cdots$\\[2pt]
IRAS 07388+4955&12.56& 450 $\pm 25$&1.21$\pm 0.16$	&2.19$\cdot10^{-8}$$\pm 2.9 \cdot10^{-9}$ &	 73.06	$\pm 9.7$&8.52$\pm 0.0079$&10.31&43.84&3.18$\cdot$10$^{-11}$\\[2pt]
IRAS 08331-0354&8.05&180 $\pm 30$ &2.22$\pm 0.65$	&4.04$\cdot10^{-8}$	$\pm 1.2 \cdot10^{-8}$ &131.4$\pm 39$&8.39$\pm 0.036$&9.73&42.27&$\cdots$\\[2pt]
IRAS 09585+5555&2.11&400 $\pm 120$&3.21$\pm 0.3$&	5.84$\cdot10^{-8}$ $\pm 5.4 \cdot10^{-9}$ &	 184.15$\pm 17$ &7.37$\pm 0.034$&10.36	 &40.18&6.39$\cdot10^{-13}$\\[2pt]
IRAS 10126+7339&5.33&430$\pm 30$&9.33$\pm 1.1$	&1.70$\cdot10^{-7}$ $\pm 2.0 \cdot10^{-8}$ &	 544.49 $\pm 64$&8.65$\pm 0.024$&10.51&41.83&1.05$\cdot10^{-12}$\\[2pt]
IRAS 10291+6517&3.18&500$\pm 40$&4.28$\pm 0.8$&7.81$\cdot10^{-8}$	$\pm 1.5 \cdot10^{-8}$ &247.45 $\pm 48$	&7.86 $\pm 0.083$&9.07&40.04&3.31$\cdot$10$^{-14}$\\[2pt]
IRAS 10295-1831&22.80&150$\pm 12$&1.45$\pm 0.15$&	2.64$\cdot10^{-8}$ $\pm 2.7 \cdot10^{-9}$	 &92.88$\pm 9.5$&9.08$\pm 0.002$&11.02&43.14&$\cdots$\\[2pt]
IRAS 10589-1210&14.58& 510$\pm 30$&1.86$\pm 0.2$&	3.40$\cdot10^{-8}$ $\pm 3.6 \cdot10^{-9}$&114.40$\pm 12$&8.85$\pm 0.0048$&10.55&41.93&$\cdots$\\[2pt]
IRAS 11033+7250&5.00&$\cdots$&$\leq 1.53$&$\cdots$&$\cdots$&$\leq 7.81$&9.55&41.73&6.21$\cdot$10$^{-11}$\\[2pt]
IRAS 11083-2813&13.59&300$\pm 50$&1.95$\pm 0.3$&	3.56$\cdot10^{-8}$ $\pm 5.5 \cdot10^{-9}$	 &119.13$\pm 18$&8.81$\pm 0.0078$&10.66&42.95&$\cdots$\\[2pt]
IRAS 11112+0951&16.55&700$\pm 50$&6.34$\pm 0.9$&	1.15$\cdot10^{-7}$	$\pm 1.6 \cdot10^{-8}$	 &391.90$\pm 54$&9.49$\pm 0.005$&10.65&42.91&$\cdots$\\[2pt]
IRAS 11210-0823&6.96&450 $\pm 30$ &2.57$\pm 0.28$&	4.68$\cdot10^{-7}$ $\pm 5.1 \cdot10^{-9}$&	 151.25$\pm 65$&	8.33$\pm 0.063$&9.92&	 41.61&7.60$\cdot$10$^{-12}$\\[2pt]
IRAS 11376+2458&6.71&400$\pm 50$&1.63$\pm 0.3$&	2.97$\cdot10^{-8}$ $\pm 5.5 \cdot10^{-9}$&	 95.84$\pm 18$&8.09$\pm 0.029$&9.64&41.46&$\cdots$\\[2pt]
IRAS 11395+1033&11.59&700$\pm 200$&6.24$\pm 0.6$&	1.14$\cdot10^{-7}$ $\pm 1.1 \cdot10^{-8}$&	 376.55$\pm 36$&9.17$\pm 0.0065$&10.58&41.91&$\cdots$\\[2pt]
IRAS 11500-0455&10.65& 400$\pm 50$&1.16$\pm 0.16$ &	2.12$\cdot10^{-8}$ $\pm 2.9 \cdot10^{-9}$	 &69.82$\pm 9.6$&8.36$\pm 0.011$&10.09	 &42.12&$\cdots$\\[2pt]
IRAS 12373-1120&1.93&420$\pm 50$&0.88$\pm 0.2$&1.61$\cdot10^{-8}$ $\pm 2.2 \cdot10^{-9}$&50.89$\pm 5.0$&6.73$\pm 0.042$&9.35&40.75&1.61$\cdot10^{-12}$\\[2pt]
IRAS 12393+3520&13.08& 210$\pm 50$&2.52$\pm 0.23$&	4.60$\cdot10^{-8}$ $\pm 4.2 \cdot10^{-9}$&	 153.57$\pm 14$&8.88$\pm 0.005$&10.09&42.12&$\cdots$\\[2pt]
IRAS 12409+7823&12.51& $\cdots$&$\leq 0.86$&$\cdots$&$\cdots$&$\leq 8.38$&9.94	 &41.87&$\cdots$\\[2pt]
IRAS 12495-1308&8.28&300$\pm 150$&3.24$\pm 0.4$&5.91$\cdot10^{-8}$ $\pm 7.3 \cdot10^{-9}$&	 192.42$\pm 24$&8.58$\pm 0.014$&9.83&	42.06&$\cdots$\\[2pt]
IRAS 13218-1929&9.98& 360$\pm 30$&5.01$\pm 0.49$&9.11$\cdot10^{-8}$ $\pm 8.9 \cdot10^{-9}$ &299.28$\pm 29$&8.93$\pm 0.0084$&10.29&	42.14&$\cdots$\\[2pt]
IRAS 14060+7207&19.36&$\cdots$&$\leq 1.33$&$\cdots$&$\cdots$&$\leq 8.96$&10.26&	 42.15&$\cdots$\\[2pt]
IRAS 14105+3932&14.58&650$\pm 50$&4.83$\pm 0.66$	&8.81$\cdot10^{-8} $	 $\pm 1.2 \cdot10^{-8}$&296.26$\pm 40$&9.26$\pm 0.0062$&10.63&41.65&$\cdots$\\[2pt]
IRAS 14156+2522&9.73& 400$\pm 100$&2.72$\pm 0.66$	&	4.96$\cdot10^{-8}$	 $\pm 1.2 \cdot10^{-8}$&162.68$\pm 39$&8.65$\pm 0.0022$&9.93&43.67&5.39$\cdot$10$^{-11}$\\[2pt]
IRAS 14294-4357&2.26&180$\pm 15$&15.88$\pm 1.04$ &	2.89$\cdot10^{-7}$ $\pm 1.9 \cdot10^{-8}$&	 911.72$\pm 60$&8.13$\pm 0.024$&9.85	 &40.61&5.31$\cdot10^{-13}$\\[2pt]
IRAS 15361-0313&13.48&400$\pm 50$&6.40$\pm 1.17$&1.17$\cdot10^{-7}$ $\pm 2.1 \cdot10^{-8}$ &389.84$\pm 70$&9.31$\pm 0.0094$&10.79&	42.02&$\cdots$\\[2pt]
IRAS 15564+6359&17.04& 600$\pm 60$&2.90$\pm 0.67$&5.29$\cdot10^{-7}$ $\pm 1.2 \cdot10^{-8}$&180.10 $\pm 41$&9.18$\pm 0.0078$&10.41	&42.46&1.12$\cdot$10$^{-12}$\\[2pt]
IRAS 16277+2433& 21.23&$\cdots$&$\leq 1.45$&$\cdots$&$\cdots$&$\leq 9.08$&10.53	 &42.02&2.25$\cdot$10$^{-12}$\\[2pt]
IRAS 17020+4544&34.20& $\cdots$&$\leq 1.11$&$\cdots$&$\cdots$&$\leq 9.40$&11.02	 &44.06&1.30$\cdot$10$^{-11}$\\[2pt]
IRAS 17023-0128&17.28& $\cdots$&$\leq 0.76$&$\cdots$&$\cdots$&$\leq 8.62$&10.42	 &42.95&$\cdots$\\[2pt]
IRAS 18001+6638 &15.00&500$\pm 100$&1.72$\pm 0.5$&	3.14$\cdot10^{-8}$ $\pm 9.1 \cdot10^{-9}$	 &105.99 $\pm 31$&8.84$\pm 0.013$&10.63	 &$\cdots$&2.94$\cdot$10$^{-13}$\\[2pt]
IRAS 19399-1026&2.95& 140$\pm 20$&4.36$\pm 0.68$&7.95$\cdot10^{-8}$ $\pm 1.2 \cdot10^{-8}$&251.69 $\pm 38$&7.80$\pm 0.048$&9.81	&41.91&3.08$\cdot$10$^{-11}$\\[2pt]
IRAS 20437-0259&14.96& $\cdots$&$\leq 0.97$&$\cdots$&$\cdots$&$\leq 8.61$&10.04	 &43.31&1.06$\cdot$10$^{-11}$\\[2pt]
IRAS 22062-2803&13.09&500$\pm 100$& 6.97$\pm 0.7$&	1.27$\cdot10^{-7}$ $\pm 1.3 \cdot10^{-8}$&	 423.65$\pm 43$&9.33$\pm 0.0056$&10.56&43.04&$\cdots$\\[2pt]
IRAS 22330-2618&2.69& 140$\pm 50$&1.91$\pm 0.3$&	3.47$\cdot10^{-8}$	$\pm 5.5 \cdot10^{-9}$&109.76$\pm 17$&7.37$\pm 0.052$&9.56&41.17&1.40$\cdot10^{-11}$\\[2pt]
IRAS 22595+1541&3.72&190$\pm 30$&3.07$\pm 0.6$&	5.59$\cdot10^{-8}$ 	$\pm 1.1 \cdot10^{-8}$ &	 177.79$\pm 35$&7.85$\pm 0.055$&9.86&40.54&1.54$\cdot10^{-12}$\\[2pt]
IRAS 23111+1344&23.16& $\cdots$ &$\leq 0.65$&$\cdots$&$\cdots$&$\leq8.83$&10.62&	 42.81&$\cdots$\\[2pt]
IRAS 23163-0001&16.71&550$\pm 100$&3.22$\pm 0.4$&	5.86$\cdot10^{-8}$ $\pm 7.3 \cdot10^{-9}$ &	 199.27$\pm 25$	&9.21$\pm 0.0045$&10.41	 &43.3&3.65$\cdot$10$^{-11}$\\[2pt]
IRAS 23279-0244&18.93&610$\pm 50$&4.75$\pm 0.6$&	8.64$\cdot10^{-8}$ $\pm 1.1 \cdot10^{-8}$&	 297.36$\pm 38$&	9.49$\pm 0.0036$&10.35&	42.64&$\cdots$ \\[2pt]
IRAS 23566-0424& 10.85&470$\pm 50$&3.35$\pm 0.6$&	6.09$\cdot10^{-8}$ $\pm 1.1 \cdot10^{-8}$&	 201.05$\pm 36$&	8.83$\pm 0.014$&10.01&	 42.83&$\cdots$\\[2pt]
\enddata
%% Text for table notes should follow after the \enddata but before
%% the \end{deluxetable}. Make sure there is at least one \tablenotemark
%% in the table for each \tablenotetext.
\tablenotetext{a}{Beam size at each source.}
\tablenotetext{b}{Width of the line at 20 \% of the peak intensity.}
\tablenotetext{c}{Integrated intensity in $T^{*}_{A}$ corrected for beam efficiency. For the non-detection an upper limit of 3$\sigma$ has been used.}
\tablenotetext{d}{Intensity from the Rayleigh-Jeans relations.}
\tablenotetext{e}{Flux.}
\tablenotetext{f}{CO luminosity, see Appendix A for derivation.}
\tablenotetext{g}{Far-infrared luminosity in the wavelength range $40 \mu m \leq\lambda\leq120\mu m$.}
\tablenotetext{h}{Soft (0.1 - 2.4 keV) X--ray luminosity from ROSAT data.}
\tablenotetext{i}{Hard absorption-corrected X--ray flux (0.3 - 8 keV) from Chandra and XMM-Newton data.}
\end{deluxetable}

\clearpage

\begin{deluxetable}{ccccc}
\tabletypesize{\scriptsize}
%\rotate
\tablecaption{Correlations \label{tbl-3}}
\tablewidth{0pt}
\tablehead{
\colhead{Relation}& \colhead{$\chi^{2}$ }& \colhead{Probability} & \colhead{$r^{a}$} & \colhead{Null Probability$^{b}$}\\
&			&		(\%)&	&(\%)
}
\startdata
$W_{CO}/L^{1/4}$ \textit{vs} \textit{sin (i)} & 22.75& 82.55 &0.49 & 0.21 \\[6pt]
\hline
$log(W_{CO}/\textit{sin (i)})$ \textit{vs} log(L$^{Soft}_{X})$ ~~~~  \\ [1pt]
~~i) Sy1s& 8.80& 92.05 & 0.67 & 0.12\\[1pt]
~~ii) Sy2s & 13.99& 45.00 & 0.02 & 46.88\\[1pt]
~~iii)Sy1s \& Sy2s& 24.77& 81.50 & 0.47 & 0.23\\[4pt]
\hline
$log(W_{CO}/\textit{sin (i)})$ \textit{vs} log(L$^{Hard}_{X})$ ~~~~  \\ [1pt]
~~i) Sy1s& 9.90& 87.17 & 0.62 & 0.32\\[1pt]
~~ii) Sy2s & 6.65& 57.48 & 0.41 & 11.92\\[1pt]
~~iii)Sy1s \& Sy2s& 18.62& 85.22 & 0.53 & 0.18\\[4pt]
\hline
log(L$_{CO}$) \textit{vs} log(L$^{Soft}_{X})$ \\[1pt]
~~i) Sy1s& 4.27 & 99.84 & 0.86 & 2.9$\cdot$$10^{-4}$\\[1pt]
~~ii) Sy2s& 7.39& 94.59 &0.71& 0.066\\[1pt]
~~iii)Sy1s \& Sy2s& 12.55& 99.95 & 0.79 & 1.0$\cdot10^{-6}$ \\[4pt]
\hline
$log(L_{CO})$ \textit{vs} log(L$^{Hard}_{X})$ \\[1pt]
~~i) Sy1s& 4.64& 99.73 & 0.84 & 5.7$\cdot10^{-4}$\\[1pt]
~~ii) Sy2s& 5.70& 76.94 &0.60& 2.40\\[1pt]
~~iii)Sy1s \& Sy2s& 10.19& 99.84 & 0.79 & 1.83$\cdot10^{-5}$ \\[4pt]
\hline
$log(M_{dyn})$ \textit{vs} $log(M_{BH})$$^{c}$ \\[1pt]
~~i)  Sy1s & 7.60& 95.86 & 0.72 & 3.54$\cdot10^{-2}$\\[1pt]
~~ii) Sy2s & 3.20& 92.05 & 0.77 & 0.40\\[1pt]
~~iii)Sy1s \& Sy2s& 13.19& 98.22 & 0.70 & 1.58$\cdot10^{-2}$\\[4pt]
\hline
$log(L_{FIR})$ \textit{vs} $log(L_{CO})$ \\[1pt]
~~i)  Sy1s & 9.58& 84.49& 0.60 & 0.54\\[1pt]
~~ii) Sy2s & 5.82& 98.99 & 0.79 & 3.7$\cdot10^{-3}$\\[1pt]
~~iii)Sy1s \& Sy2s& 15.99& 99.44 & 0.71 & 6.08$\cdot10^{-5}$\\[4pt]
\hline
%$log(L_{FIR})$ \textit{vs} $log(L_{X})$ \\[1pt]
%~~i)  Sy1s & 11.15& 67.36 & 0.45 & 3.99\\[1pt]
%~~ii) Sy2s & 11.75& 62.64 & 0.40 & 6.19\\[1pt]
%~~iii)Sy1s \& Sy2s& 26.01& 67.46 & 0.36 & 2.01\\[1pt]
%\hline
%$log(M_{dyn})$ \textit{vs} $log(M_{BH})$$^{c}$ \\[1pt]
%~~i)  Sy1s & 11.08& 67.92 & 0.46 & 3.79\\[1pt]
%~~ii) Sy2s & 12.97& 44.99 & 0.04 & 43.66\\[1pt]
%~~iii)Sy1s \& Sy2s& 27.34& 55.33 & 0.24 & 9.16\\[4pt]
%% Text for table notes should follow after the \enddata but before
%% the \end{deluxetable}. Make sure there is at least one \tablenotemark
%% in the table for each \tablenotetext.
\enddata
\tablenotetext{a}{The linear correlation coefficient:~$r\equiv\frac{N\sum{x_{i}y_{i}}-\sum{x_{i}\sum{y_{i}}}}{[N\sum{x^{2}_{i}}-(\sum{x_{i}})^{2}]^{1/2}[N \sum{y^{2}_{i}}-(\sum{y_{i}})^{2}]^{1/2}]}$,
where \textit{r} ranges from 0, when there is no correlation to 1 when there is complete correlation.}
\tablenotetext{b}{Null probability of the correlation (in percentage), indicates the probability that the observed data could have come from an uncorrelated parent population. A small value of the probability implies that the observed variables are probably correlated.}
%\tablenotetext{c}{Dynamical mass calculated assuming a rotating disk R, of 5 kpc.}
\tablenotetext{c}{Dynamical mass calculated assuming a rotating disk R, equal to the telescope beam FWHM at each source.}
\end{deluxetable}

\end{document}